\documentclass{aa}

\usepackage{graphicx}
\usepackage{txfonts}
\usepackage[colorlinks,citecolor=blue,]{hyperref}

\usepackage{amsmath}	% Advanced maths commands
\usepackage{verbatim,xspace,pifont,array,multirow,bigints}
\usepackage{subfigure}
\usepackage[utf8]{inputenc}

\usepackage[dvipsnames]{xcolor}
\usepackage{comment} 
\usepackage{multirow}
\usepackage{siunitx}

\usepackage{lipsum}
\usepackage{booktabs}

\newcommand{\msol}{\,M$_{\odot}$\xspace}

\newcommand{\ergs}{\,erg\,s$^{-1}$\xspace}

\newcommand{\kev}{\,keV\xspace}

\newcommand{\gev}{\,GeV\xspace}

\newcommand{\ledd}{\,L$_{\rm Edd}$\xspace}

\newcommand{\ptde}{\,$P_{\mathrm{TDE}}$\xspace}
\newcommand{\rdisk}{\,$R_{\rm disk}$\xspace}
\newcommand{\ldisk}{\,$L_{\rm disk}$\xspace}
\newcommand{\rcor}{\,$R_{\rm cor}$\xspace}
\newcommand{\tcor}{\,$T_{\rm cor}$\xspace}
\newcommand{\taucor}{\,$\tau_{\rm cor}$\xspace}

\newcommand{\gr}{$\gamma$-ray\xspace}
\newcommand{\grs}{$\gamma$\,rays\xspace}
\newcommand{\g}{$\gamma$\xspace}
\newcommand{\pg}{p$\gamma$\xspace}
\newcommand{\fermi}{\textit{Fermi}/LAT\xspace}

\newcommand{\dsg}{AT2019dsg\xspace}
\newcommand{\ehz}{AT2019ehz\xspace}
\newcommand{\ali}{ASASSN--14li\xspace}

\newcommand{\CASS}{The Center for Astrophysics and Space Science, NYUAD Research Institute, New York University Abu Dhabi, PO Box 129188, Abu Dhabi, UAE \email{dimitrios.kantzas@nyu.edu}}

\newcommand{\IA}{Institute of Astrophysics, FORTH, GR-71110 Heraklion, Greece}

\newcommand{\UoC}{Department of Physics, University of Crete, GR-70013 Heraklion, Greece}

\newcommand{\astrobiologyUoC}{Institute of Astrobiology, University of Crete, GR-70013 Heraklion, Greece}

\newcommand{\INAFR}{INAF-Istituto di Radioastronomia, via Gobetti 101, 40129, Bologna, Italy}

\begin{document}

   \title{Hidden engines in tidal disruption events:
multimessenger signatures in a lepto-hadronic framework}
   \titlerunning{TDEs in gamma rays and neutrinos}

   \author{D.~Kantzas
          \inst{1}
          \and
          D.A.~Langis
          \inst{2, 3}
          \and
        I.~Liodakis
          \inst{2, 4}
          \and
        A.~Paggi
        \inst{2, 5}
          }
    \authorrunning{Kantzas et al. }
   \institute{
   \CASS
   \and
   \IA
   \and
   \UoC
   \and
   \astrobiologyUoC
   \and
    \INAFR
             }
   \date{Received --; accepted --}

% \abstract{}{}{}{}{} 
% 5 {} token are mandatory
 
  \abstract
   { 
Tidal disruption events (TDEs) provide a unique laboratory for studying transient accretion, outflows, particle acceleration, and the possible connection between stellar disruptions and high-energy neutrinos. Motivated by recent candidate TDE--neutrino associations and by the forthcoming sensitivity of multimessenger facilities, we investigate under which physical conditions TDEs can be detected in MeV, GeV, and TeV gamma rays, and in high-energy neutrinos. We develop a lepto-hadronic emission framework and apply it to three representative sources: AT2019dsg, AT2019ehz, and ASASSN-14li. For each source we explore a set of physically motivated models that vary the proton and electron spectral indices, the minimum Lorentz factor of non-thermal protons, the particle energy partition, the acceleration efficiency, the magnetic compactness, and the presence of a compact corona. We compute the broadband spectral energy distributions, the associated muon-neutrino spectra, and compare them with reference sensitivities for COSI, Fermi/LAT, CTAO, and KM3NeT/ARCA. We find that hard proton spectra and/or large minimum proton Lorentz factors are the most favourable conditions for TeV gamma-ray and neutrino detectability, while GeV-bright models can be predominantly leptonic and neutrino quiet. AT2019dsg and AT2019ehz can reach the CTAO and KM3NeT benchmarks in hadronically favourable configurations, whereas ASASSN-14li is more naturally bright in the MeV--GeV band and constrained by its radio synchrotron emission. Our results highlight the importance of combining MeV, GeV, TeV, radio, and neutrino observations to distinguish between transparent, hidden, and lepton-dominated TDE emission regimes.

}

   \maketitle
%
%-------------------------------------------------------------------

\section{Introduction}

Tidal disruption events (TDEs) occur when a star is scattered close enough to a massive black hole that the tidal force exceeds the star's self-gravity.  A fraction of the stellar debris remains bound, returns to pericenter, circularizes, and can power a luminous flare from the otherwise faint nucleus of a galaxy \citep{Rees:1988bf}.  Because the debris initially falls back at highly super-Eddington rates for many systems, TDEs provide a transient view of accretion, outflows, shocks, and non-thermal particle acceleration around massive black holes.  Wide-field optical surveys and rapid multiwavelength follow-up have shown that TDEs are not a single observational class: some events are dominated by optical/UV emission, some are X-ray bright, some launch radio-emitting outflows or jets, and only a few have been discussed as possible high-energy neutrino emitters \citep{Langis:2025btl}.

The origin of the optical/UV emission remains one of the central open questions in TDE physics. In the simplest picture, the observed luminosity would trace efficient accretion of the fallback material, but the radiated energy and characteristic temperatures of optical TDEs are often difficult to reconcile with a bare, compact, steady thin disc \citep{Mummery:2025lgo}. Several alternatives have therefore been proposed.  The optical/UV flare may be powered, at least partly, by shocks produced during stream self-intersection and circularization \citep{Shiokawa_2015,Piran:2015gha}; by reprocessing of inner-disc or coronal radiation in an extended photosphere, wind, or debris layer; or by a combination of accretion, shocks, and outflow emission whose relative importance changes with time and viewing angle. Magnetic fields can further modify this picture by affecting the returning debris stream and the subsequent circularization flow \citep{Guillochon:2016ith,Bonnerot:2016krr}.  Recent optical polarimetric observations have strengthened the case that at least some TDEs require non-axisymmetric geometries and shock-powered emission during disc formation \citep{Liodakis:2022wak,Koljonen:2024wzm,Koljonen:2025jpr,Floris:2025fhp,Floris:2026whn}. These uncertainties are important for high-energy modelling because they determine the photon fields, magnetic fields, and compactness that control both particle cooling and \gr absorption.

TDEs have also been proposed as sources of cosmic rays and high-energy neutrinos. Early studies focused on relativistic TDE jets and considered whether protons accelerated in internal shocks or external outflows could produce neutrinos through \pg and pp interactions \citep{Farrar:2008ex,
Wang:2011ip,Farrar:2014yla,WangLiu2016,DaiFang2017,Senno2017,LunardiniWinter2017,Guepin:2017abw,Biehl:2017hnb}. Subsequent work showed that the same basic idea can be realised in several physical environments, including choked or off-axis jets, coronae, radiatively inefficient accretion flows, disc winds, and debris-interaction regions \citep{HayasakiYamazaki2019,Murase:2020lnu,Liu:2020isi,Fang:2020bkm, Winter:2020ptf, Matsumoto:2021qqo,WinterLunardini2023,Zheng:2022afy,Mukhopadhyay:2023mld,Wu:2024qjw, Li:2024qcp, Yuan:2024daj}. These models differ in geometry, target photon field, baryon loading, and internal opacity, but they share a common observational challenge: neutrino production is accompanied by electromagnetic cascades, and the escaping \gr emission must remain consistent with X-ray, MeV, GeV, and TeV constraints.

The possible association of \dsg with IceCube-191001A brought this question into sharper focus \citep{Stein:2020xhk}. Further candidate associations involving accretion flares or TDE-like nuclear transients, such as AT2019fdr and AT2019aalc, have strengthened the motivation for considering TDEs as neutrino sources \citep{Reusch:2021ztx,vanVelzen:2021zsm}. At the same time, the statistical significance of individual associations remains debated.  Updated IceCube event catalogues, \fermi non-detections, and population-level correlation studies all indicate that current TDE--neutrino associations should be treated as suggestive rather than established \citep{Peng2022FermiTDEs,IceCube:2025uzh,2025ApJ...990...18L,Langis:2026dyj,Mukhopadhyay:2026qlq}. This motivates a question that is broader than the interpretation of any single event: under which physical conditions should a TDE be detectable in \grs and neutrinos at all?

A particularly useful diagnostic is the energy band in which the non-thermal power escapes. If the source is transparent, pion-decay emission and inverse-Compton radiation can appear in the GeV--TeV range and may be constrained by \fermi or detected by CTAO. If the source is compact and photon rich, the same hadronic activity can be hidden at TeV energies and reprocessed into the hard X-ray or MeV band through pair production and electromagnetic cascades \citep{Murase:2020lnu,YuanWinter2023}. The MeV range is therefore a key bridge between thermal X-ray/UV emission and GeV--TeV searches. Future observations with the Compton Spectrometer and Imager (COSI) will be especially valuable for testing hidden or partially hidden TDE scenarios \citep{Tomsick:2023aue}, while CTAO and KM3NeT/ARCA will probe complementary TeV \gr and TeV--PeV neutrino signatures \citep{Carosi2021CTA,KM3NeT:2024arca}.

In this paper we explore the detectability of non-thermal emission from TDEs using a lepto-hadronic emission framework based on the steady-state multi-zone \textsc{bhjet} model and its hadronic extension \citep{Lucchini:2021scp,Kantzas:2020mld,Kantzas:2023tzw}.  
Our aim is not to obtain a unique fit to each source, nor to claim that any individual TDE must have produced a particular neutrino alert. Instead, each model is a representative or reference realization chosen to isolate a specific physical regime, such as MeV--GeV brightness, TeV transparency, neutrino production, or radio synchrotron consistency. We use these representative models to identify the combinations of particle acceleration, proton spectral shape, minimum proton Lorentz factor, energy partition, magnetic field, photon compactness, and coronal opacity that make a TDE visible to COSI, \fermi, CTAO, and KM3NeT. This parameter-space approach is designed to connect the intrinsic physics of the source to practical multimessenger detectability. The overplotted multiwavelength data are used to check whether the model normalization and spectral shape are broadly compatible with the observed emission during the selected epoch, but the models are not selected by minimizing a global likelihood or chi-square.

We apply the model to three TDEs that represent different observational regimes. \dsg is used as a neutrino-motivated benchmark because of its proposed association with IceCube-191001A and evidence for non-thermal activity. \ehz is treated as a prompt optical/UV outburst case with X-ray follow-up, allowing us to test whether high-energy emission could accompany the main TDE flare even without an established neutrino association. Finally, \ali provides an X-ray- and radio-bright benchmark. Its radio data constrain the low-energy synchrotron component and allow us to test whether primary electrons, secondary pairs, or proton synchrotron can contribute to the observed non-thermal emission.

The paper is organized as follows. In Sect.~\ref{Sec: Model}, we describe the lepto-hadronic model and the photon attenuation processes included in the calculations. In Sect.~\ref{Sec: Multiwavelength observations}, we summarize the multiwavelength data used for the three sources. In Sect.~\ref{Sec: SED and neutrinos of TDEs}, we present the model families, spectral energy distributions, neutrino predictions, and benchmark-proximity ratios for COSI, \fermi, CTAO, and KM3NeT. We discuss the physical interpretation and comparison with previous work in Sect.~\ref{Sec: discussion}, and we summarize our conclusions in Sect.~\ref{Sec: summary and conclusions}.

\section{Model}\label{Sec: Model}
We use a semi-analytical, steady-state lepto-hadronic model to calculate the intrinsic photon and neutrino emission from compact dissipation regions in TDE environments. The model follows the philosophy of \texttt{bhjet}, where the dynamical properties of the emitting plasma are coupled to radiative transfer, but it is adapted here to the geometry expected during the circularization and early accretion phase of a TDE rather than to a scale-invariant compact jet \citep{Lucchini:2021scp}. The hadronic extension follows the treatment developed in \citet{Kantzas:2020mld}, where relativistic protons are accelerated together with electrons and interact with gas and radiation fields, producing secondary photons, pairs, and neutrinos. The goal of the present work is not to fit a unique solution for each source, but to identify the physical conditions under which the GeV-to-TeV \gr and neutrino emission can become observable.

\subsection{Geometry and dynamical assumptions}\label{sec: model geometry}
We consider a TDE around a black hole of mass $M_{\rm BH}$ and define the gravitational radius as $R_g=GM_{\rm BH}/c^2$. During the circularization of the stellar debris, shocks are expected to form in the accretion flow and/or in the stream-intersection region \citep{Shiokawa_2015,Steinberg:2022rqq}. We describe the system with an accretion disk in the equatorial plane and use the angle $\theta$ between the disk axis and the line of sight as the viewing angle. Thus, $\theta \simeq 0^{\circ}$ corresponds to a face-on system, whereas $\theta \simeq 90^{\circ}$ corresponds to an edge-on system. The disk is truncated at a characteristic radius \rdisk and radiates with luminosity \ldisk, with the thermal component peaking in the optical/UV band for the parameter range explored here.

In addition to the disk radiation, we include a compact thermal corona. The corona is described by its radius \rcor, temperature \tcor, and Thomson optical depth \taucor. The corresponding number density is
$n_{\rm cor} = \tau_{\rm cor}/\sigma_T R_{\rm cor}$,
where $\sigma_T$ is the Thomson cross section. Unless stated otherwise, we assume a charge-neutral plasma with equal numbers of thermal electrons and protons. The thermal electrons Comptonize the disk photons and provide an additional target field for inverse Compton scattering and photohadronic interactions. In the numerical examples below the coronal temperature should be interpreted as an electron thermal energy, $k_{\rm B}T_{\rm cor}$, and is therefore quoted in energy units.

The dissipation region and the corona should not be interpreted as two independent acceleration zones. In the present implementation, non-thermal electrons and protons are accelerated only in the compact shock region located at radius $R$. The corona, when included, acts as a thermal photon field and as a reservoir of thermal target protons for Compton scattering, photohadronic interactions, and pp interactions. Its characteristic radius is denoted by $R_{\rm cor}$. Thus, models with $R<R_{\rm cor}$ correspond to dissipation embedded within the coronal environment, whereas models with $R>R_{\rm cor}$ place the shock outside the nominal coronal scale. This distinction affects the target photon density, the internal opacity, and the efficiency of hadronic interactions, but protons are not separately accelerated throughout the corona.

We assume that a fraction of the power released during the TDE is dissipated in shocks located at a characteristic radius $R$ from the black hole. This radius also sets the escape length of particles from the acceleration region. The dissipated power available to non-thermal particles and magnetic fields is denoted by \ptde. We parametrize the partition between particle and magnetic pressure through the plasma beta, $\beta = P_{\rm particles}/P_{\rm magnetic}$. Low values of $\beta$ correspond to magnetically dominated shocks, while high values of $\beta$ correspond to particle-dominated shocks. We further use the parameter $\eta_e$ to control the relative amount of non-thermal power carried by electrons and protons. This parameter is important for the high-energy output because the GeV emission can be dominated by leptonic inverse Compton scattering, whereas the TeV and neutrino emission require an efficient hadronic component.

The emitting region is treated as quasi-static in the source frame. Therefore, the calculations presented here do not include quantities associated with a relativistic jet. The framework is intended to describe compact disk/corona, circularization, or mildly outflowing shock regions rather than a fully relativistic jetted TDE. 

\subsection{Particle acceleration}\label{sec: particle acceleration}
At the shocks, a fraction of the thermal particles is accelerated to non-thermal energies. We do not impose a specific microscopic acceleration mechanism, but instead parametrize the accelerated distributions as power laws with exponential cutoffs. For each particle species $i=e,p$, we use
\begin{equation}\label{equation: non thermal particle distribution}
    \frac{{\rm d}N_i}{{\rm d}E_i}
    = N_{0,i}\, E_i^{-p_i}\exp\left(-\frac{E_i}{E_{i,{\rm max}}}\right),
    \qquad E_i \geq E_{i,{\rm min}},
\end{equation}
where $p_i$ is a free parameter and $N_{0,i}$ is fixed by the non-thermal power assigned to that species. Unless stated otherwise, the minimum electron energy is tied to the peak of the relativistic Maxwell-J\"uttner distribution, while for protons we set $\gamma_{p,{\rm min}}=1$. Below $E_{i,{\rm min}}$, the electrons follow the thermal Maxwell-J\"uttner distribution and the protons are assumed to remain cold.

The maximum particle energy is calculated by comparing the acceleration time-scale with the relevant cooling and escape time-scales. We use the acceleration time-scale \citep{jokipii1987rate}
\begin{equation}\label{eq: acceleration timescale}
    t_{\rm acc}(E_i)=\frac{4E_i}{3f_{\rm sc}ecB},
\end{equation}
where $e$ is the elementary charge, $B$ is the characteristic magnetic field strength in the acceleration region, and $f_{\rm sc}\lesssim1$ is the acceleration efficiency parameter. The value of $E_{i,{\rm max}}$ is obtained from
\begin{equation}
    t_{\rm acc}^{-1}(E_{i,{\rm max}})=\sum_j t_{j}^{-1}(E_{i,{\rm max}}),
\end{equation}
where the sum includes synchrotron cooling, inverse Compton scattering, inelastic hadronic collisions, photohadronic losses, Bethe--Heitler losses, and escape, whenever these processes are relevant for the particle species under consideration.

For the parameter space explored in this work, the maximum electron energy is typically set by synchrotron cooling,
\begin{equation}
    t_{\rm acc}^{-1}(E_{e,{\rm max}})\simeq t_{\rm syn}^{-1}(E_{e,{\rm max}})
    = \frac{\sigma_T B^2 E_{e,{\rm max}}\beta_e^2}{6\pi m_e^2 c^3},
\end{equation}
where $\beta_e c$ is the electron velocity. By contrast, the maximum proton energy is often limited by escape from the acceleration region,
\begin{equation}
    t_{\rm acc}^{-1}(E_{p,{\rm max}})\simeq t_{\rm esc}^{-1}(E_{p,{\rm max}})
    = \frac{\beta_pc}{R},
\end{equation}
although photohadronic and Bethe--Heitler losses can dominate in sufficiently compact and luminous radiation fields. The most important parameters controlling the highest photon and neutrino energies are therefore $B$, $R$, $f_{\rm sc}$, \ldisk, \taucor, and \tcor.

To calculate the normalization of the power-law of non-thermal particles, we assume that a fraction $\varepsilon_i$ of the injected power is injected to the non-thermal particles, with $i=e$ for electrons and $i=p$ for protons. Typically, the fraction $\varepsilon_i$ is 10\% for other astrophysical sources \citep{Petropoulou:2015upa, Murase:2020lnu,Kantzas:2020mld, kantzas2023possible}, hence, we adopt this value here as well unless we explicitly mention another value.

\subsection{Radiative processes}\label{sec: radiative processes}
The accelerated electrons emit synchrotron radiation in the local magnetic field. For the full synchrotron emissivity we use the analytical expressions of \cite{Blumenthal:1970gc} for both electrons and protons, although the proton synchrotron component is generally subdominant for the parameters considered here \citep{Muecke:2002bi}.

The same non-thermal electrons inverse-Compton scatter the synchrotron photons, disk photons, and coronal photons. We include the transition from the Thomson to the Klein--Nishina regime in the calculation of the emitted spectrum and in the cooling time-scale. This leptonic channel is particularly important in the GeV band, especially when the magnetic field is weak enough for inverse Compton cooling to compete with synchrotron cooling.

Relativistic protons interact with both gas and radiation. Inelastic proton-proton collisions occur on the thermal protons of the corona, disk atmosphere, or debris environment, while photohadronic interactions occur on the synchrotron, disk, and coronal photon fields. These interactions produce charged and neutral pions. Neutral pions decay into \grs, while charged pions decay into neutrinos and secondary pairs. We use the semi-analytical prescription of \citet{Kelner:2006tc}, with the updated cross sections of \citet{Kafexhiu:2014cua}, for pp interactions and the prescription of \citet{Kelner:2008ke} for p\g interactions. The resulting secondary electrons and positrons are followed radiatively, allowing electromagnetic cascades to redistribute part of the hadronic power to lower photon energies.

The neutrino spectrum is calculated consistently with the charged-pion component of the pp and p\g interactions. At production, the flavor composition is dominated by the pion-decay chain, while at Earth we assume standard flavor mixing, leading approximately to equal fluxes of electron, muon, and tau neutrinos. For a source at luminosity distance $d_L$, the all-flavor neutrino energy flux is obtained from the intrinsic neutrino luminosity through the usual geometric dilution. To estimate the number of detectable events in KM3NeT/ARCA, the model flux can then be folded with the declination- and energy-dependent effective area,
\begin{equation}
    N_{\nu}=\Delta t\sum_{\alpha}\int {\rm d}E_{\nu}\,
    A_{{\rm eff},\alpha}(E_{\nu},\delta)\,\phi_{\nu_\alpha}(E_{\nu}),
\end{equation}
where $\Delta t$ is the duration of the flare, $\delta$ is the source declination, $\phi_{\nu_\alpha}$ is the neutrino flux per flavor, and the sum runs over the relevant event channels \citep{KM3NeT:2024arca}. This expression makes explicit that neutrino detectability depends not only on the intrinsic hadronic luminosity, but also on the flare duration, source distance, sky position, and spectral hardness.

\subsection{Illustrative parameter dependence}\label{sec: illustrative parameter dependence}
\begin{figure*}[!ht]
    % \resizebox{\hsize}{!}
    \centering
    \subfigure{
        \hspace{-1cm}
        \includegraphics[width=0.9\columnwidth]{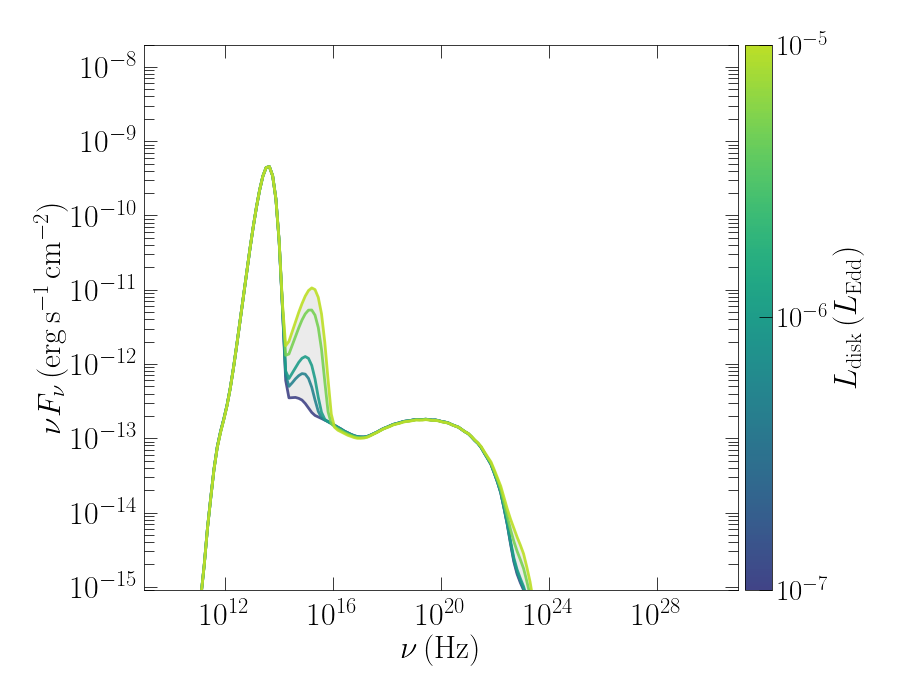}
    }
    \subfigure{
        \hspace{-.5cm}
        \includegraphics[width=0.9\columnwidth]{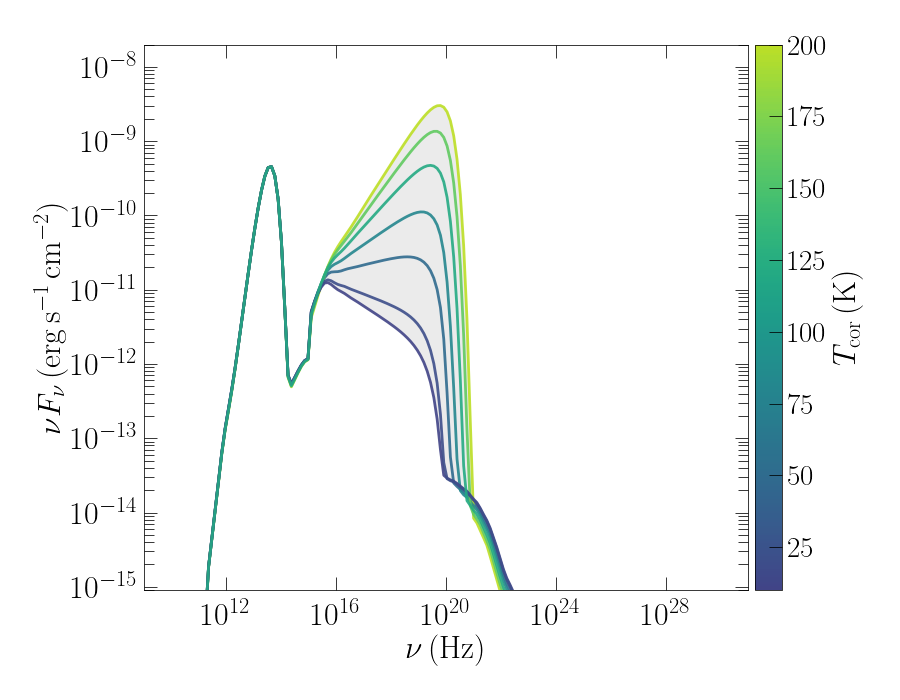}
    }
    \caption{
    Intrinsic spectral flux density $F_{\nu}$ per photon frequency $\nu$ for representative model realizations. Both panels include a dusty torus that peaks in the IR and a non-thermal electron component that contributes in the X-ray band. In the left panel we vary the disk luminosity, while in the right panel we include a thermal corona that Comptonizes the disk photons. The colored lines show different values of the parameter indicated by the color bar, and the shaded regions highlight the full range spanned by that parameter. The remaining model parameters are described in Sec.~\ref{sec: illustrative parameter dependence}.
 }   
    \label{fig: seds for disk and corona}
\end{figure*}

\begin{figure*}[!ht]
    % \resizebox{\hsize}{!}
    \centering

    \subfigure{
        \hspace{-1cm}
        \includegraphics[width=0.9\columnwidth]{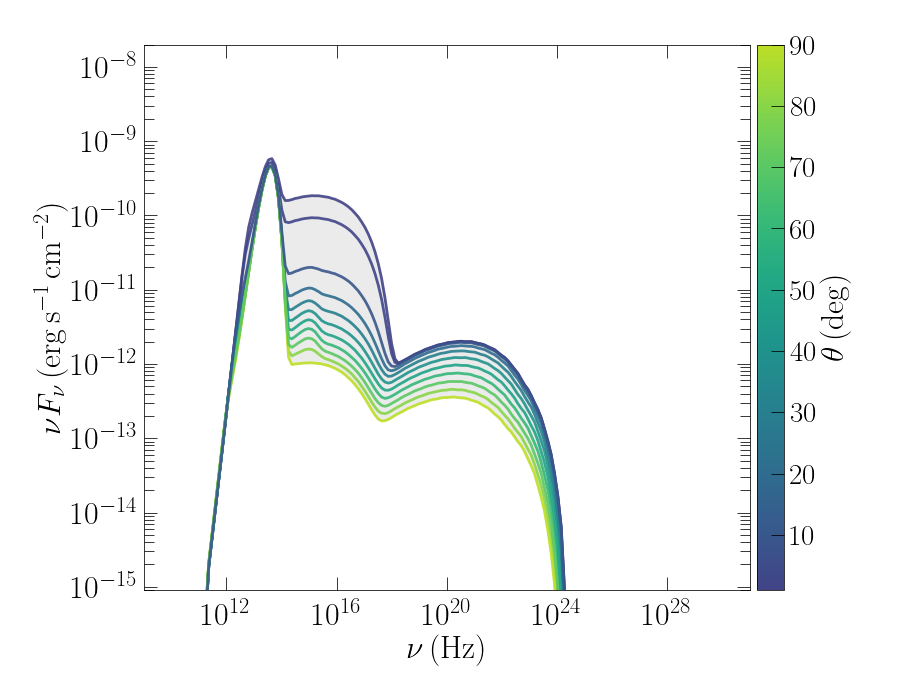}
    }
        \vspace{-.75cm}
    \subfigure{
        \hspace{-.70cm}
        \includegraphics[width=0.9\columnwidth]{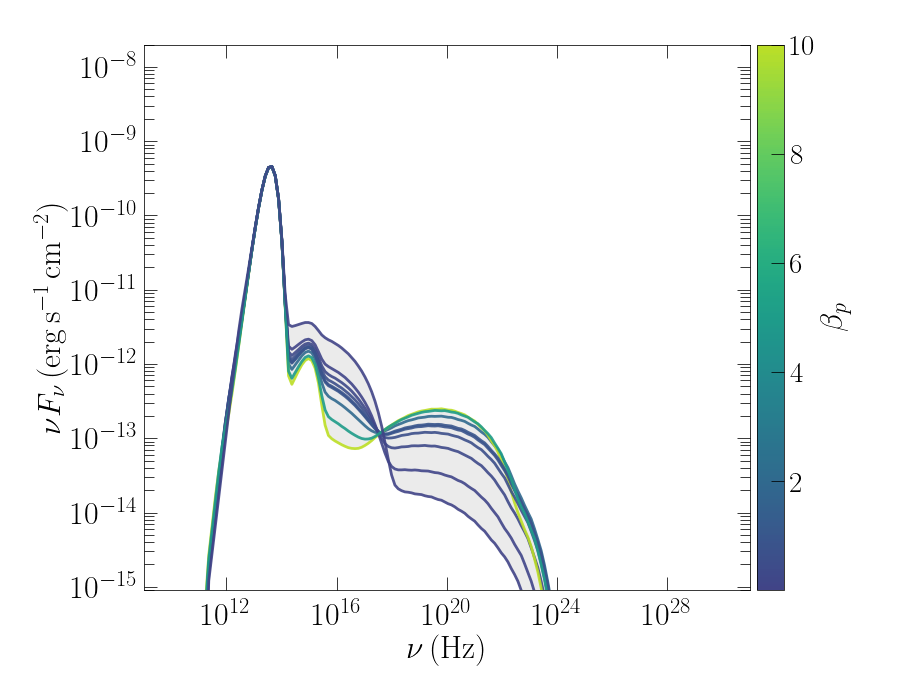}
    }    
    \vspace{-.5cm}
    \subfigure{
        \hspace{-1cm}
        \includegraphics[width=0.9\columnwidth]{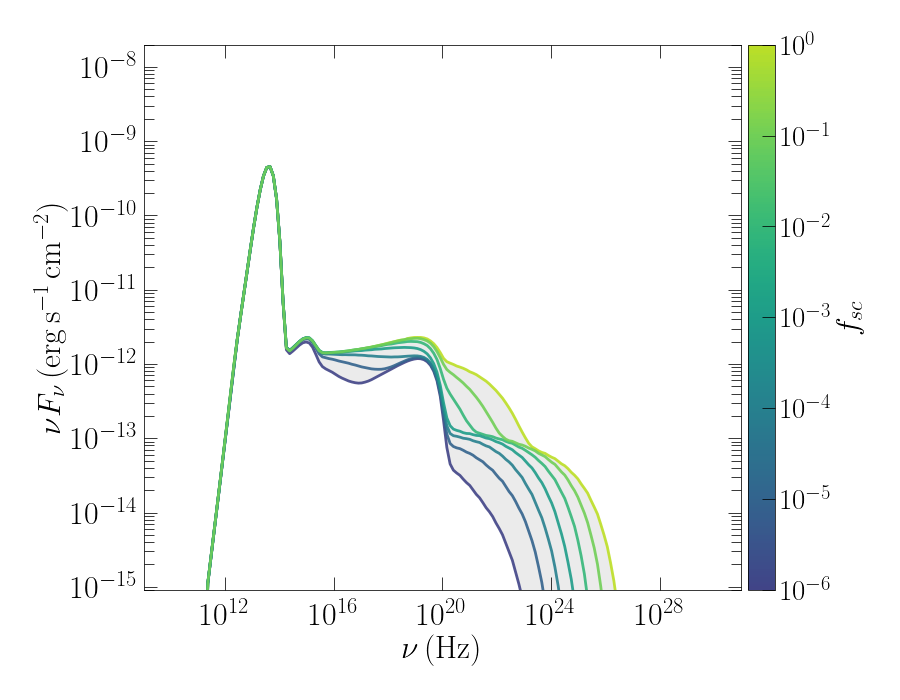}        
    }
    \subfigure{
         \hspace{-0.70cm}
        \includegraphics[width=0.9\columnwidth]{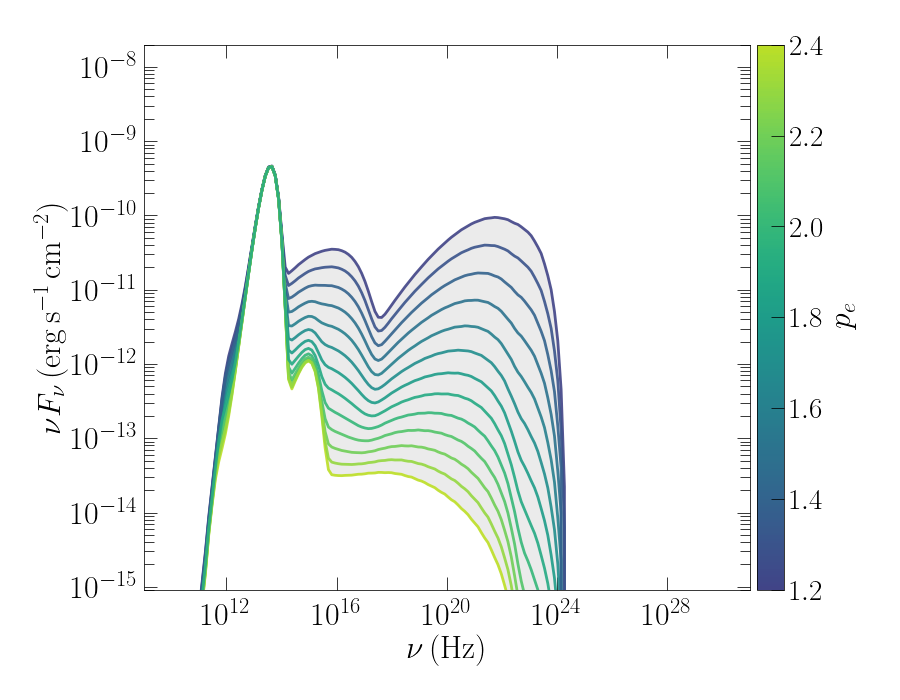}
    }
    \caption{
    Intrinsic spectral flux density $F_{\nu}$ per photon frequency $\nu$ for leptonic models without a thermal corona. In each panel, we vary one model parameter, as indicated by the color bar, while the other parameters are fixed to $\theta=60^{\circ}$, $\beta=1$, $f_{\rm sc}=10^{-6}$, and $p_e=2.0$, unless that parameter is varied in the panel. The remaining model parameters are described in Sec.~\ref{sec: illustrative parameter dependence}.
}   
    \label{fig: seds for non-thermal electrons without corona}
\end{figure*}

\begin{figure*}[!ht]
    \centering
        \includegraphics[width=0.9\columnwidth]{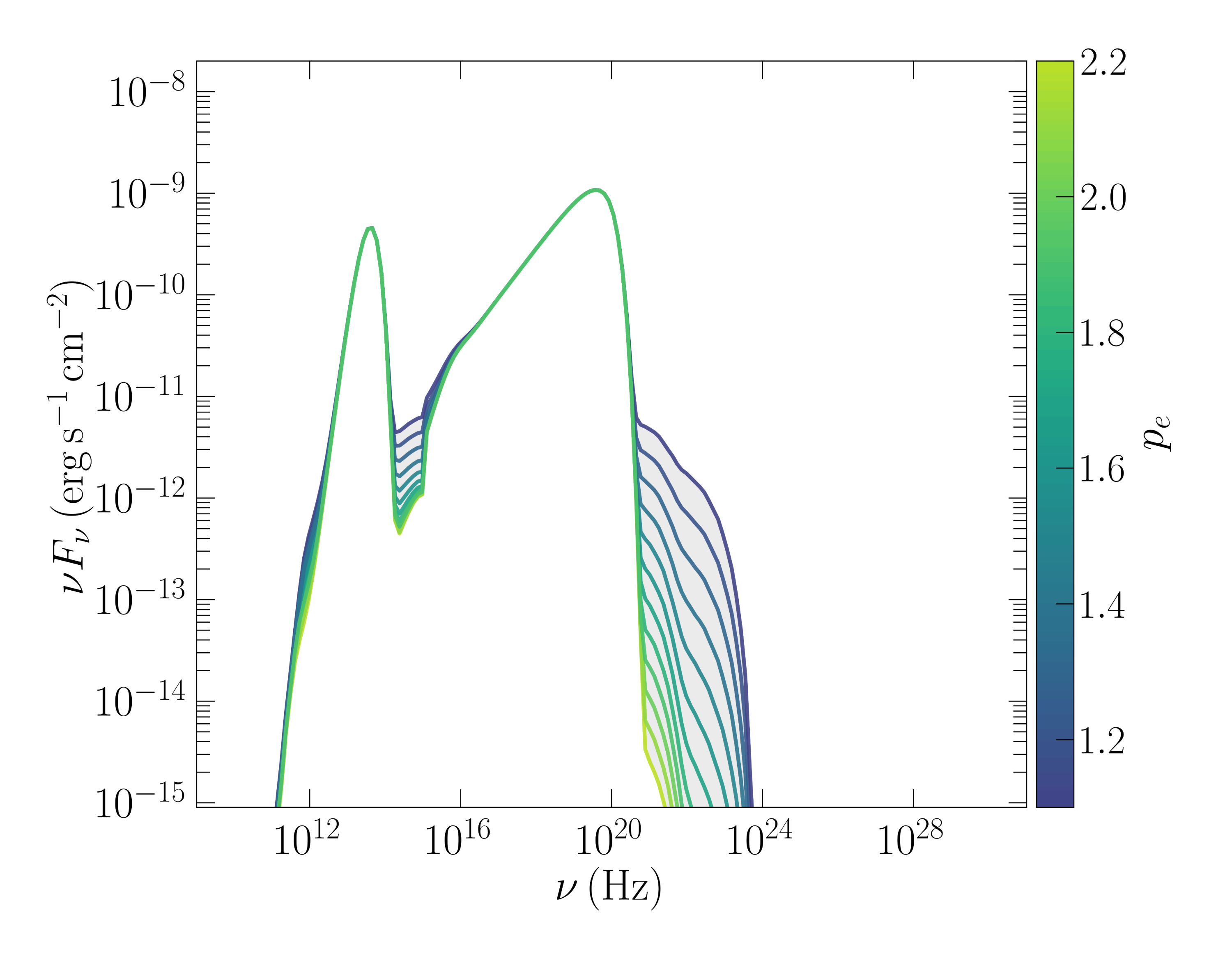}
        \includegraphics[width=0.9\columnwidth]{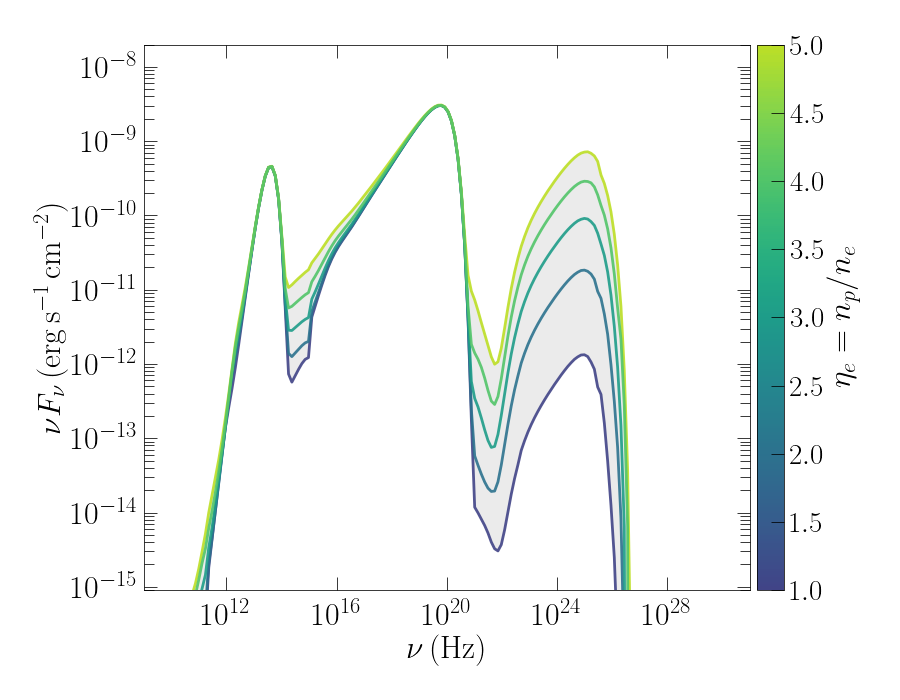}
    \caption{
    Intrinsic spectral flux density $F_{\nu}$ per photon frequency $\nu$ for models including a thermal corona with $\tau_{\rm cor}=0.5$ and $k_{\rm B}T_{\rm cor}=100\,\kev$. In the upper panel, only electrons are accelerated and we vary the electron power-law index $p_e$. In the lower panel, protons are also accelerated with $p_p=1.5$ while $p_e=2$, and we vary the parameter controlling the relative non-thermal power in electrons and protons. The remaining model parameters are described in Sec.~\ref{sec: illustrative parameter dependence}.
 }   
    \label{fig: seds for non thermal electrons and protons with corona}
\end{figure*}

\begin{figure*}[!ht]
    \centering
        \includegraphics[width=0.67\columnwidth]{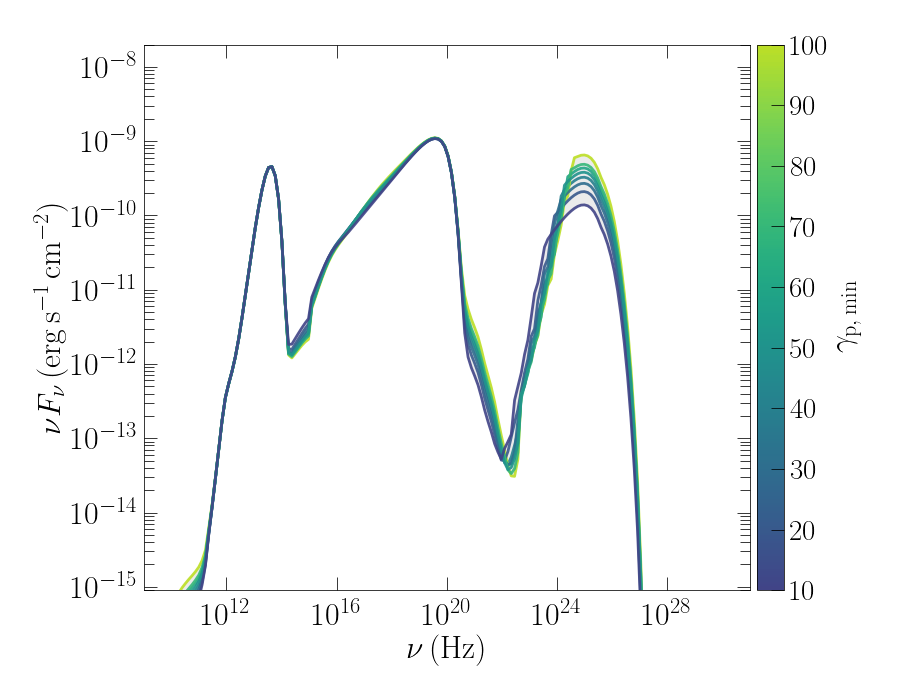}  
        \hspace{-.4cm}
        \includegraphics[width=0.67\columnwidth]{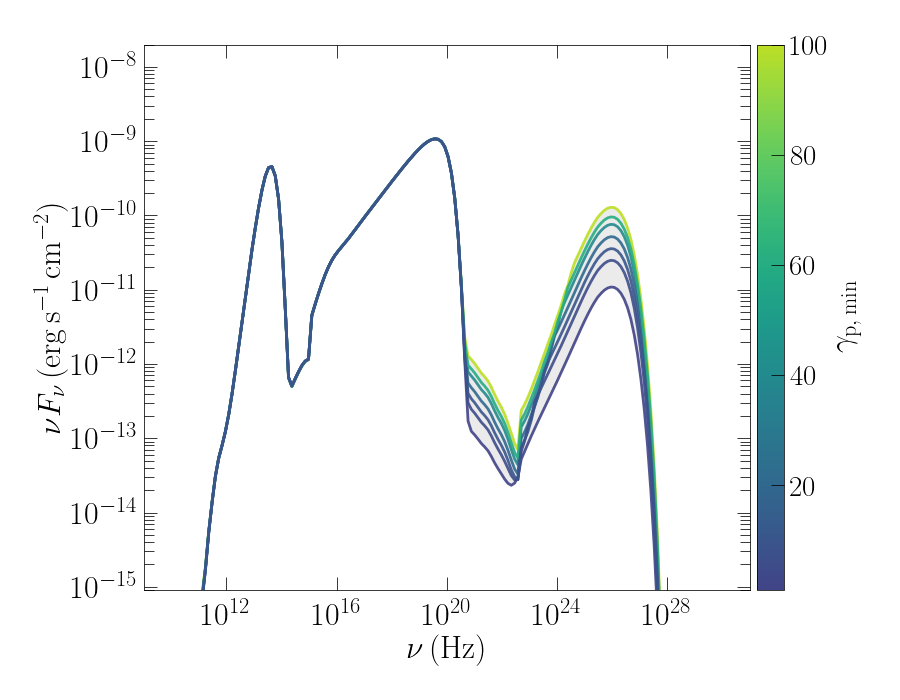} 
        \hspace{-.3cm}\includegraphics[width=0.67\columnwidth]{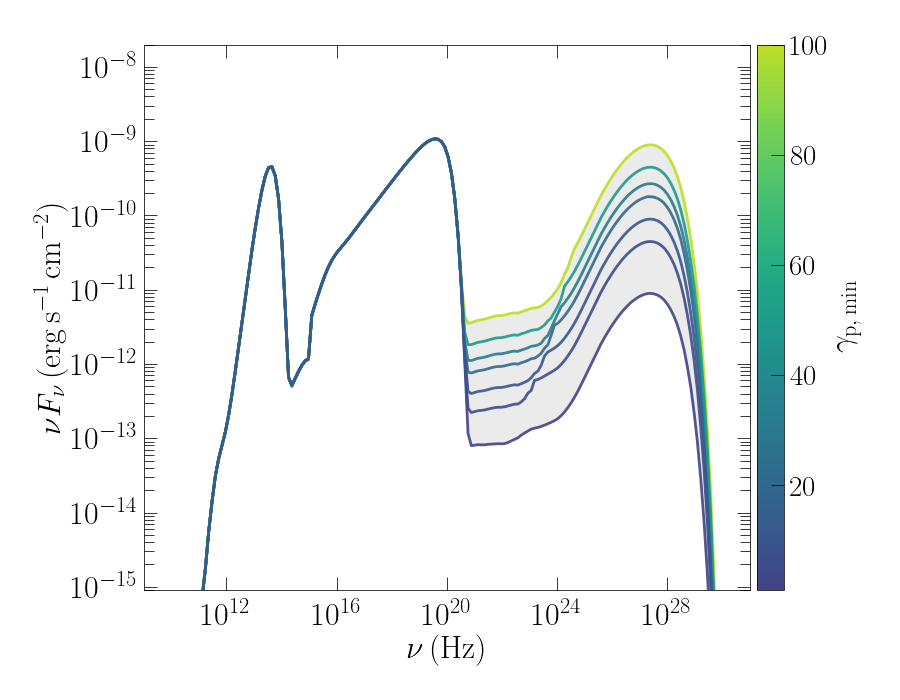} 
    \caption{
    Intrinsic spectral flux density $F_{\nu}$ per photon frequency $\nu$ for hadronic models including a thermal corona with $\tau_{\rm cor}=0.5$ and $k_{\rm B}T_{\rm cor}=150\,\kev$. The color bar shows the minimum Lorentz factor of the accelerated protons. We assume $p_p=1.5$ and $p_e=2$. From top to bottom, the panels correspond to $(f_{\rm sc},\eta_e)=(10^{-6},3)$, $(10^{-5},1)$, and $(10^{-3},1)$. The remaining model parameters are described in Sec.~\ref{sec: illustrative parameter dependence}.
 }   
    \label{fig: seds for non thermal protons with corona}
\end{figure*}

In Figs.~\ref{fig: seds for disk and corona}--\ref{fig: seds for non thermal protons with corona}, we show representative intrinsic spectral energy distributions (SEDs) to illustrate how the main model parameters affect the emitted radiation before attenuation. These curves are not intended as fits to a specific TDE. They are used to identify the regions of parameter space in which TDEs may become detectable in the GeV--TeV band and in neutrinos.

Figure~\ref{fig: seds for disk and corona} shows the effect of the thermal photon fields. In the left panel, we vary the disk luminosity while keeping the disk truncation radius fixed at $10R_g$. A more luminous disk naturally increases the optical/UV emission and also enhances the external photon field available for inverse Compton scattering and p\g interactions. In the right panel, we include a thermal corona with $\tau_{\rm cor}=0.5$ and vary $k_{\rm B}T_{\rm cor}$. A hotter corona upscatters disk photons to higher energies and increases the density of X-ray target photons. The remaining parameters are fixed to a fiducial model in which a dusty torus with luminosity $6\times10^{39}$\ergs dominates the IR band, non-thermal electrons are injected from a Maxwell-J\"uttner distribution peaking at 50\kev, the electron power-law index is $p_e=2$, the shocks are located at $20R_g$, $f_{\rm sc}=10^{-6}$, and $\beta=1$.

Figure~\ref{fig: seds for non-thermal electrons without corona} isolates the leptonic component in the absence of a corona. The viewing angle affects the observed normalization because the emission is preferentially released along the axis perpendicular to the disk plane. The plasma beta controls the balance between synchrotron and inverse Compton radiation: near equipartition, the stronger magnetic field enhances synchrotron emission, whereas larger $\beta$ values suppress synchrotron cooling and can increase the relative importance of inverse Compton scattering. The acceleration efficiency $f_{\rm sc}$ controls the maximum particle energy and therefore the high-energy cutoff. Finally, harder electron spectra, corresponding to smaller $p_e$, produce harder and brighter GeV spectra.

Figure~\ref{fig: seds for non thermal electrons and protons with corona} demonstrates the role of the corona and of the lepton-to-hadron power ratio. In the upper panel, only electrons are accelerated and the corona provides both an X-ray component and an additional cooling channel. Compared to the corona-free case, the stronger cooling suppresses the highest-energy inverse Compton emission. In the lower panel, relativistic protons are also included with $p_p=1.5$. The p\g interactions on coronal photons generate a TeV component, and the strength of this component increases as a larger fraction of the non-thermal power is placed in protons.

Figure~\ref{fig: seds for non thermal protons with corona} shows the effect of the minimum proton Lorentz factor. This parameter controls how much of the proton power is placed close to the interaction thresholds and how much is available at the highest energies. For fixed total proton power, increasing $\gamma_{p,{\rm min}}$ can harden the hadronic output and shift the cascade emission, but it also removes lower-energy protons that contribute to pp interactions. The minimum proton energy is therefore one of the key parameters for connecting the TeV \gr output to the neutrino spectrum.

\subsection{Photon attenuation}\label{Sec: photon attenuation optical depths}
The SEDs shown above are intrinsic. High-energy photons produced close to the black hole may be attenuated before escaping the source, while TeV photons can also be attenuated by extragalactic background light during propagation to Earth. We include the dominant local attenuation channels: photon-photon pair production, Compton scattering on thermal electrons, pair production in the electric fields of nuclei, and photoelectric absorption. The observed photon flux is then
\begin{equation}
    F_{\nu,{\rm obs}}(E_{\gamma}) = F_{\nu,{\rm int}}(E_{\gamma})\,\exp[-\tau_{\rm eff}(E_{\gamma})],
\end{equation}
where $F_{\nu,{\rm int}}$ is the intrinsic model spectrum.

\subsubsection{Photon-photon pair production}
Electron-positron pairs are produced when a photon of energy $E_{\gamma}$ collides with a target photon of energy $\epsilon$ and satisfies
\begin{equation}
E_\gamma\,\epsilon\,(1-\mu)
 \ge 2\,(m_e c^{2})^{2},
\label{eq:threshold}
\end{equation}
where
\begin{equation}
\mu \equiv \cos\phi,
\label{eq:mu_def}
\end{equation}
and $\phi$ is the collision angle, with $\phi=\pi$ corresponding to a head-on collision. The Lorentz factor of the center-of-momentum frame is
\begin{equation}
\gamma_{\rm cm}^{2}
 = \frac{E_\gamma\,\epsilon\,(1-\mu)}
        {2\,(m_e c^{2})^{2}},
\label{eq:gcm2}
\end{equation}
with
\begin{equation}
\beta_{\rm cm}
 = \left(1 - \gamma_{\rm cm}^{-2} \right)^{1/2}.
\label{eq:beta_cm}
\end{equation}

The optical depth of a photon of energy $E_{\gamma}$ propagating through a target photon field $n_{\gamma\gamma}(r,\epsilon)$ is
\begin{equation}
\tau_{\gamma\gamma}(E_\gamma)
 =
 \int_{\gamma\gamma} {\rm d}r
 \int_{-1}^{1} {\rm d}\mu\,\frac{1-\mu}{2}
 \int_{\epsilon_{\rm min}}^{\epsilon_{\rm max}} {\rm d}\epsilon\;
   n_{\gamma \gamma}(r, \epsilon)\,
   \sigma_{\gamma\gamma}(E_\gamma,\epsilon,\mu),
\label{eq:tau photopair gamma gamma}
\end{equation}
where the path integral is evaluated over the region occupied by the target photon field. The cross section is \citep{Nikishov:1962rmq,Gould:1967zza,Blumenthal:1970nn,Chodorowski1992reaction,coppi1997}
\begin{equation}
\begin{aligned}
    \sigma_{\gamma\gamma}(E_\gamma,\epsilon,\mu)
 &=
 \frac{3}{16}\,\sigma_T\,(1-\beta_{\rm cm}^{2})\, \times 
 \\[0.5em]
 &
 \left[
   (3-\beta_{\rm cm}^{4})
   \ln\left(\frac{1+\beta_{\rm cm}}{1-\beta_{\rm cm}}\right)
   - 2\,\beta_{\rm cm}\,(2-\beta_{\rm cm}^{2})
 \right].
\end{aligned}
\label{eq:sigma_gg}
\end{equation}

\begin{figure}
    \centering
    \includegraphics[width=0.9\linewidth]{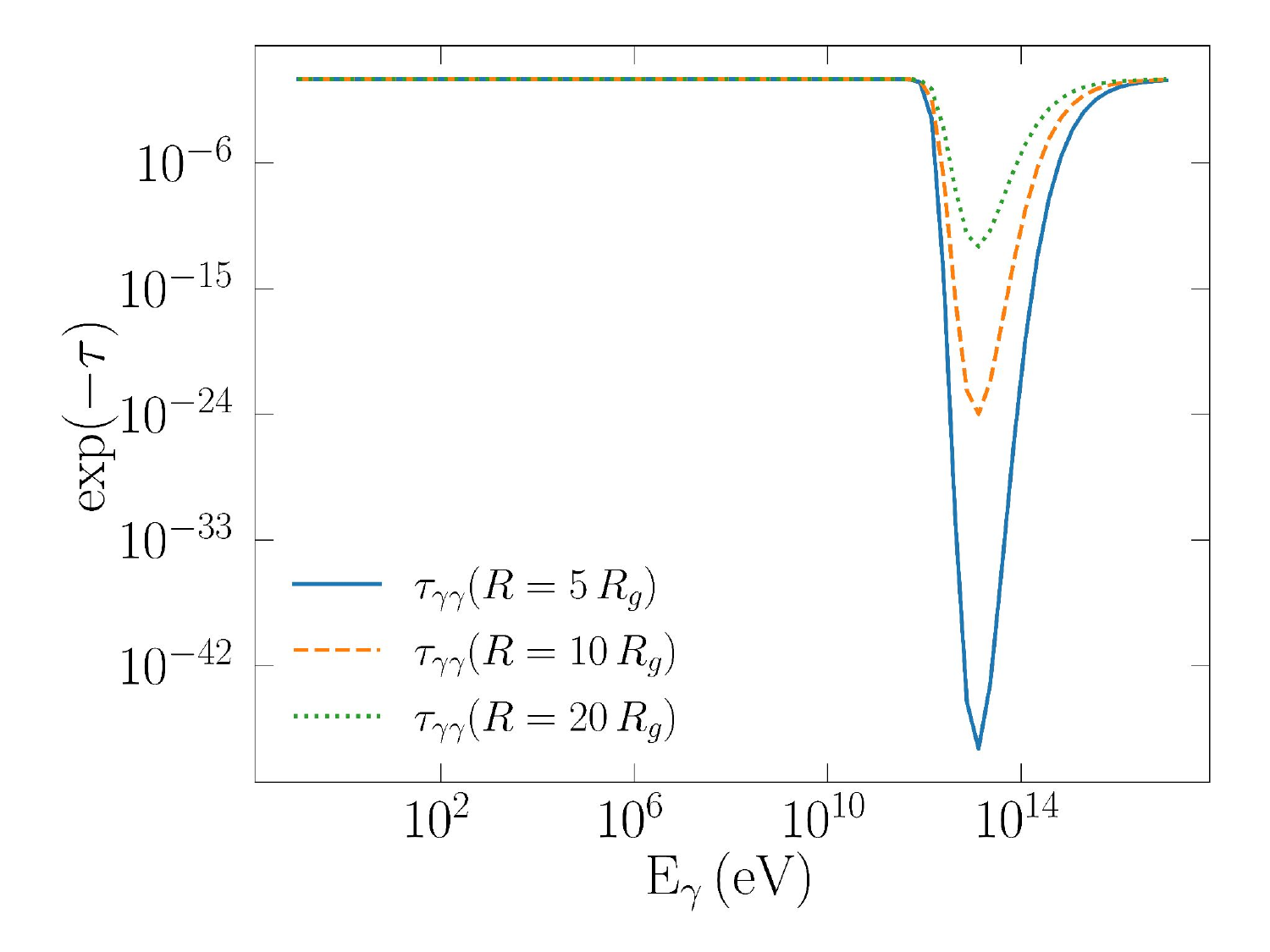}
    \caption{Photon-photon pair-production optical depth per photon energy for different radii $R$ of the dissipation region, assuming an IR photon field from the dusty torus.}
    \label{fig: tau optical depth gamma gamma gg photopair for different radii}
\end{figure}

\begin{figure}
    \centering
    \includegraphics[width=0.9\linewidth]{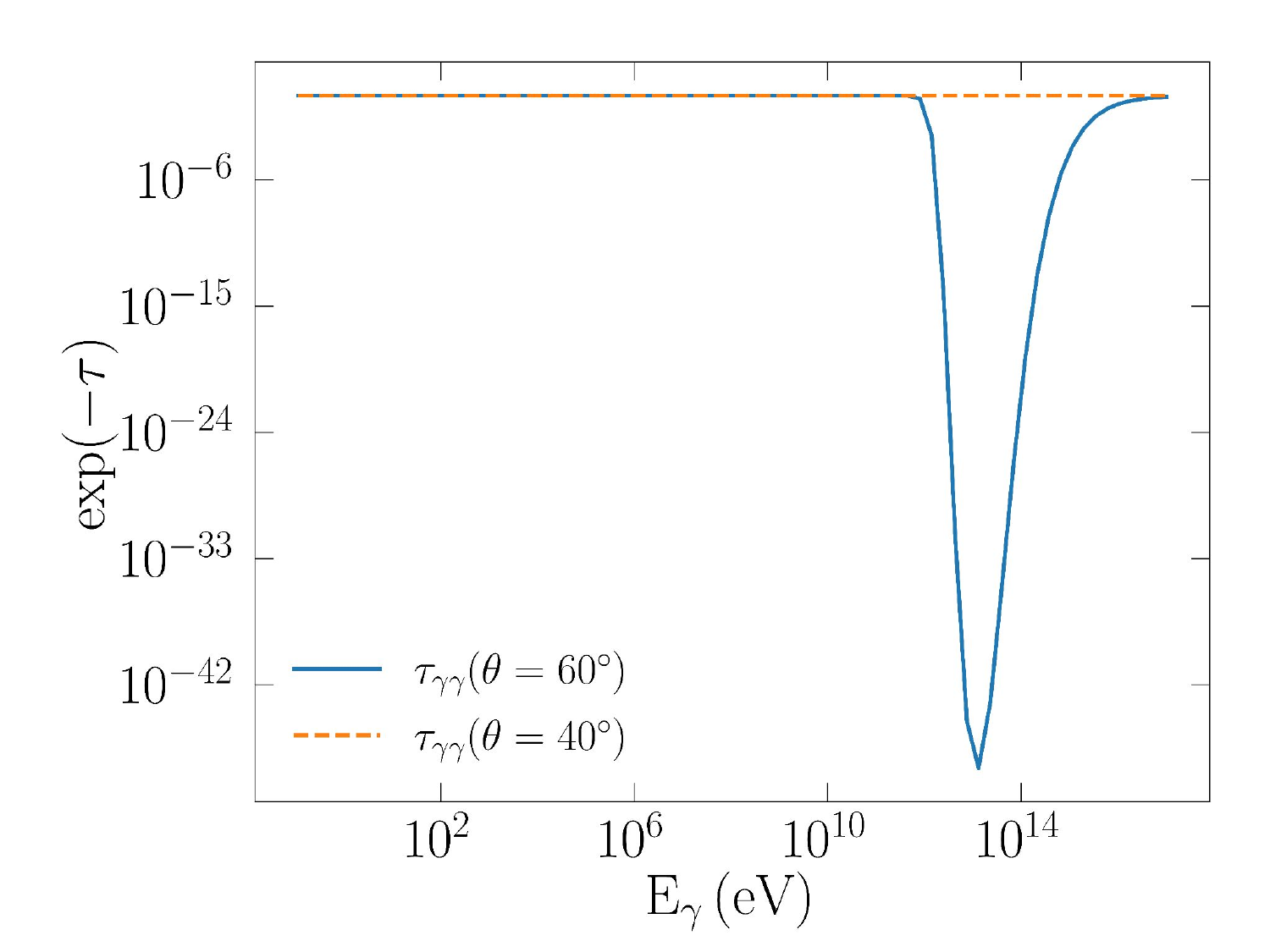}
    \caption{Photon-photon pair-production optical depth as a function of photon energy for different viewing angles $\theta$, as shown in the legend.}
    \label{fig: tau optical depth gamma gamma gg photopair for different angles}
\end{figure}

Photon-photon absorption is most important when the target photon energy is close to the pair-production threshold. In practice, TeV photons are efficiently attenuated by IR photons from the dusty torus, while GeV photons are more sensitive to optical/UV and X-ray target fields. In Fig.~\ref{fig: tau optical depth gamma gamma gg photopair for different radii}, we show the dependence of the optical depth on the radius of the dissipation region. In Fig.~\ref{fig: tau optical depth gamma gamma gg photopair for different angles}, we show the dependence on viewing angle. Photons escaping close to the polar direction cross a smaller column of disk and torus photons, whereas photons emitted closer to the equatorial plane experience stronger attenuation.

\subsubsection{Compton scattering}
High-energy photons may also scatter on the thermal electrons of the corona. The relevant cross section decreases in the Klein--Nishina regime. For a photon with $x=E_{\gamma}/m_ec^2$, the total Klein--Nishina cross section is \citep{klein1929streuung,Rybicki:2004hfl}
\begin{equation}
\begin{aligned}
\sigma_{\rm KN}(x)
 &=
  \frac{3\sigma_T}{4}
 \Bigg[
   \frac{1+x}{x^{3}}
   \left(
     \frac{2x(1+x)}{1+2x}
     - \ln(1+2x)
   \right)
 \\[0.5em]
 &\qquad
   + \frac{1}{2x}\ln(1+2x)
   - \frac{1+3x}{(1+2x)^2}
 \Bigg].
\end{aligned}
\label{eq:KN_single_number}
\end{equation}
For $x\ll1$, $\sigma_{\rm KN}\simeq\sigma_T$, whereas for $x\gg1$ the cross section decreases approximately as $\ln(2x)/x$.

\begin{figure}
    \centering
    \includegraphics[width=0.9\linewidth]{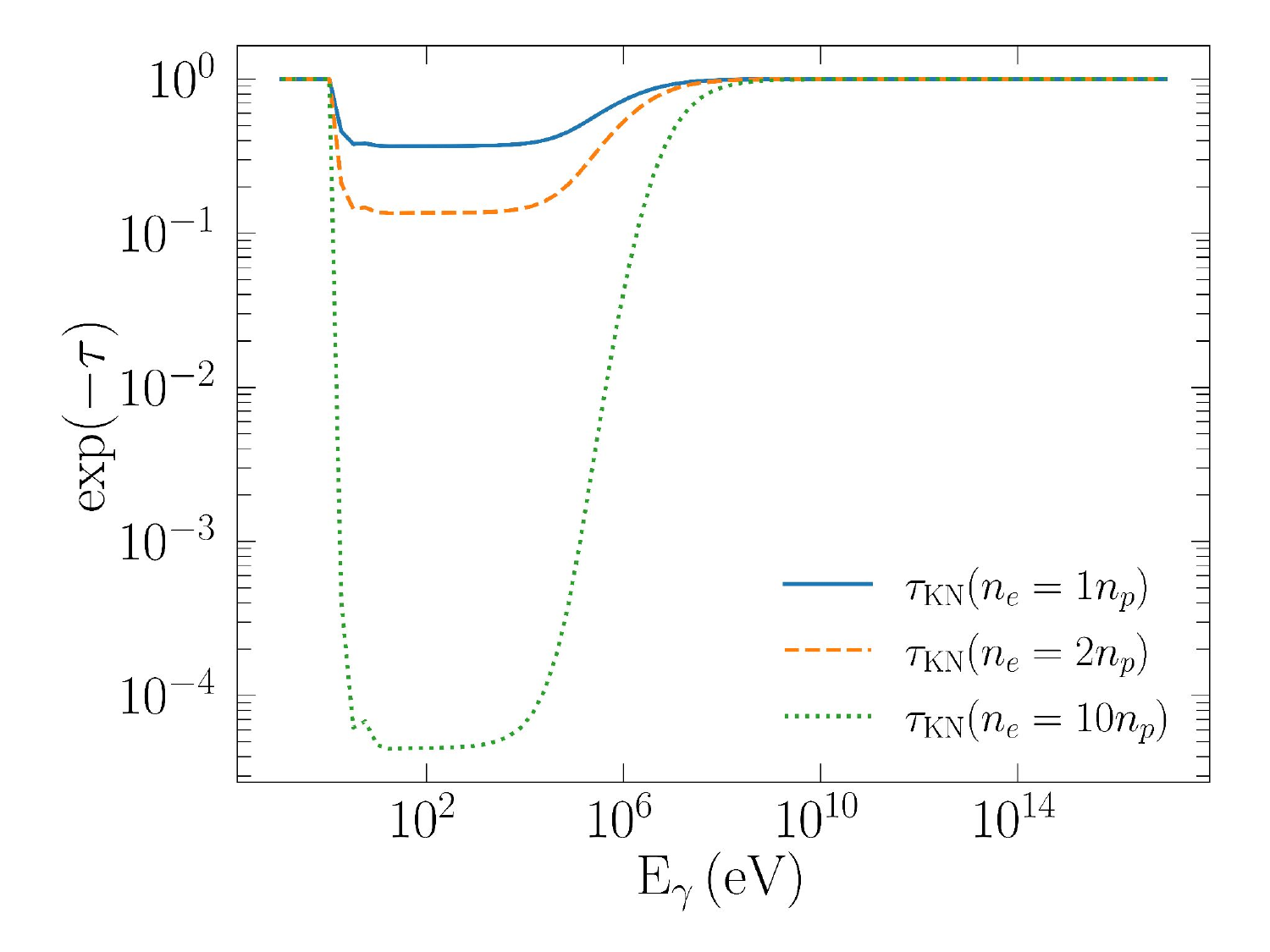}
    \caption{Compton-scattering optical depth of photons propagating through the thermal electrons of the corona. The different curves correspond to different target electron densities relative to the baseline coronal density, as shown in the legend. The feature around $1\,$eV is caused by the disk radiation field, while the remaining attenuation is dominated by the corona.}
    \label{fig: tau optical depth ICS KN}
\end{figure}

The corresponding optical depth is
\begin{equation}
\tau_{\rm KN}(E_\gamma)
 =
 \int_{\rm KN} {\rm d}r\; n_e(r)\,\sigma_{\rm KN}(E_\gamma),
\label{eq:tau klein nishina kn}
\end{equation}
where $n_e(r)$ is the density of thermal electrons along the photon path. In Fig.~\ref{fig: tau optical depth ICS KN}, we show this optical depth for different coronal electron densities. This attenuation is most relevant from the UV/X-ray band up to the energy range where the Klein-Nishina suppression becomes important.

\subsubsection{Bethe--Heitler pair production}
\begin{figure}
    \centering
    \includegraphics[width=0.9\linewidth]{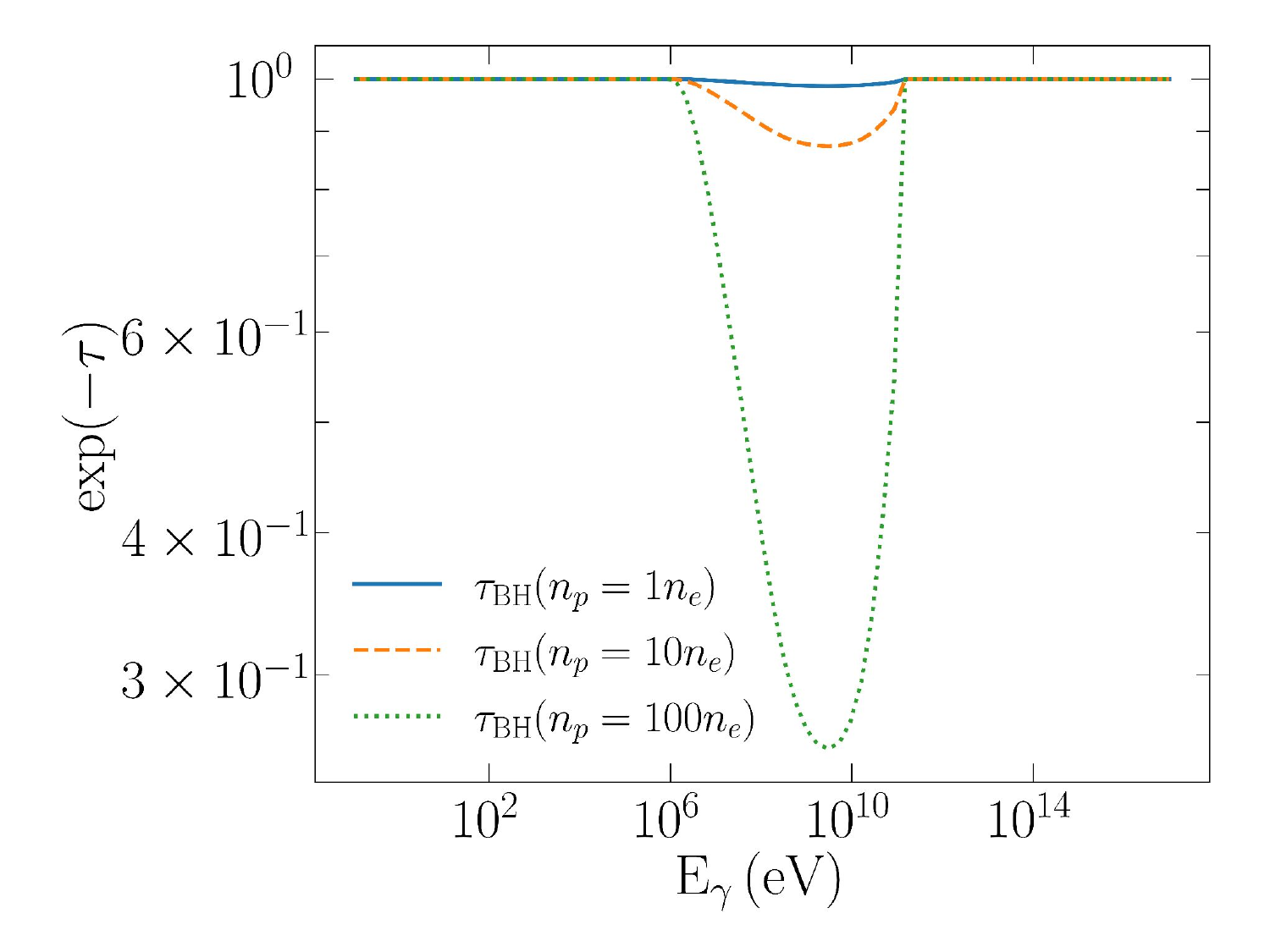}
    \caption{Bethe--Heitler photon attenuation optical depth as a function of photon energy. The different curves correspond to different target proton densities relative to the baseline coronal density.}
    \label{fig: tau optical depth BH}
\end{figure}

Bethe--Heitler pair production enters the calculation in two ways. First, it acts as a cooling channel for relativistic protons interacting with photons,
\begin{equation}\label{eq: bethe heitler bh threshold}
    \gamma_p \epsilon \ge 2m_ec^2,
\end{equation}
where $\gamma_p$ is the proton Lorentz factor. Second, high-energy photons can be attenuated by pair production in the electric field of nuclei in the cold gas. In the high-energy limit, the photon-nucleus pair-production cross section can be approximated by \citep{Bethe:1934za,heitler1984quantum}
\begin{equation} \label{eq: sigma cross section bethe heitler BH}
    \sigma_{\rm BH} \simeq \frac{28}{9} \alpha r^2_e Z^2 \left[ \ln(2x)  - \frac{109}{42} \right],
\end{equation}
where $x=E_{\gamma}/m_ec^2$, $\alpha$ is the fine-structure constant, $r_e$ is the classical electron radius, and $Z$ is the nuclear charge. In the numerical calculation, we use the tabulated attenuation coefficients from the \textit{NIST XCOM} database between 1\kev and 100\gev \citep{berger1987xcom}. The photon optical depth is
\begin{equation}
\tau_{\rm BH}(E_\gamma)
 =
 \int_{\rm BH} {\rm d}r\; n_Z(r)\,\sigma_{\rm BH}(E_\gamma,Z),
\label{eq: bethe heither BH tau}
\end{equation}
where $n_Z(r)$ is the number density of the relevant nuclei in the disk, corona, debris, or dusty torus. In Fig.~\ref{fig: tau optical depth BH}, we show the resulting attenuation for different target densities.

\subsubsection{Photoelectric absorption}
\begin{figure}
    \centering
    \includegraphics[width=0.9\linewidth]{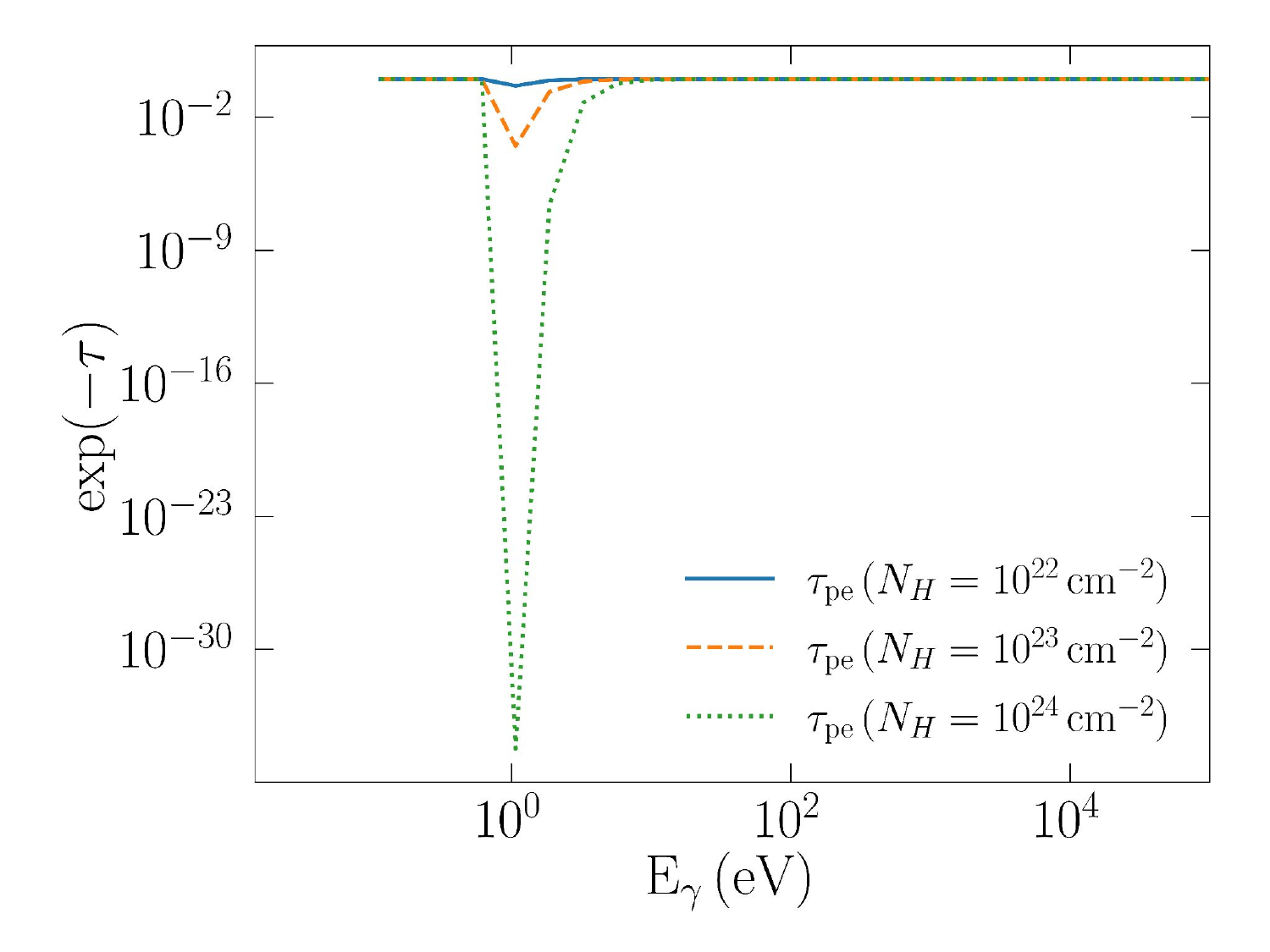}
    \caption{Photoelectric optical depth as a function of photon energy for different column densities, as indicated in the legend.}
    \label{fig: tau optical depth photoelectric pe}
\end{figure}

Photoelectric absorption dominates the opacity of neutral material in the soft X-ray band and can also represent an effective opacity from atomic and dust absorption at lower energies. We use the approximate power-law form \citep{Morrison1983interstellar,Wilms:2000ez}
\begin{equation}
    \sigma_{\rm pe} \simeq \sigma_0 \left( \frac{E_\gamma}{E_0} \right)^{-s},
\end{equation}
where $E_0=1\,$keV, $s\lesssim3$, and $\sigma_0\simeq10^{-22}\,{\rm cm^2}$ per H atom at 1\kev. For a column density $N_H$, the optical depth is
\begin{equation}
    \tau_{\rm pe}(E_\gamma)=N_H\,\sigma_{\rm pe}(E_\gamma),
\label{eq: photoelectric pe tau}
\end{equation}
with the more general path-integral form obtained by replacing $N_H$ with $\int n_H(r){\rm d}r$. In Fig.~\ref{fig: tau optical depth photoelectric pe}, we show the resulting optical depth for different column densities of cold material associated with the dusty torus.

\subsubsection{Total effective optical depth}\label{subsection: total effective optical depth}
The total local optical depth is the sum of the individual contributions,
\begin{equation}\label{eq:tau_eff}
    \tau_{\rm eff}(E_\gamma) = \tau_{\gamma\gamma}(E_\gamma) + \tau_{\rm KN}(E_\gamma) + \tau_{\rm BH}(E_\gamma) + \tau_{\rm pe}(E_\gamma).
\end{equation}
The relative importance of these terms depends on photon energy, source geometry, viewing angle, coronal density, and the amount of cold gas along the line of sight. The same physical ingredients that enhance neutrino production, namely high gas density and high photon density, can therefore suppress the escaping \gr flux. This tension is central to the detectability problem addressed in this work: TDEs are most efficient neutrino sources when the hadronic interaction depth is high, but they are most promising TeV \gr sources when the pair-production and matter opacities are low enough for high-energy photons to escape.
The optical depth in Eq.~(\ref{eq:tau_eff}) describes attenuation local to the TDE environment. In the observed SEDs we also include attenuation by the
extragalactic background light during propagation to Earth.

\section{Multiwavelength observations}
\label{Sec: Multiwavelength observations}

\subsection{Data preparation}
We test the newly developed model against three typical optical TDEs, \dsg, \ehz, \ali, which have publicly available multi-wavelength observations spanning from radio to X-rays. For the infrared, optical, and UV observations, we utilised data from TDECat\footnote{\url{https://github.com/dlangis/TDECat}} \citep{Langis:2025btl}, while for the radio observations, we used data from the Open multiwavelength Transient Event Repository\footnote{\url{https://otter.idies.jhu.edu/}} \citep[OTTER;][]{2026ApJ...999..243F}. In all cases, we utilised Swift BAT and UVOT \citep{2005SSRv..120...95R, Burrows2005SwiftXray}, Wide-field Infrared Survey Explorer \citep[WISE;][]{Wright2010WISE} and Near-Earth Object WISE Reactivation \citep[NEOWISE;][]{Mainzer2011NEOWISE,Mainzer2014NEOWISE} data, while for \dsg and \ehz we also included Zwicky Transient Facility \citep[ZTF;][]{Graham2019ZTFScObj, Bellm2019ZTFSys, Masci2019ZTFDataPr} g- and r-band light curves. For AT2019dsg, the Fermi/LAT upper limits reported by \citet{Peng2022FermiTDEs} are above the predicted GeV fluxes of all models considered here; therefore, the explored configurations are not ruled out by the archival LAT non-detection.

\begin{figure}
    \centering
    \includegraphics[width=\linewidth]{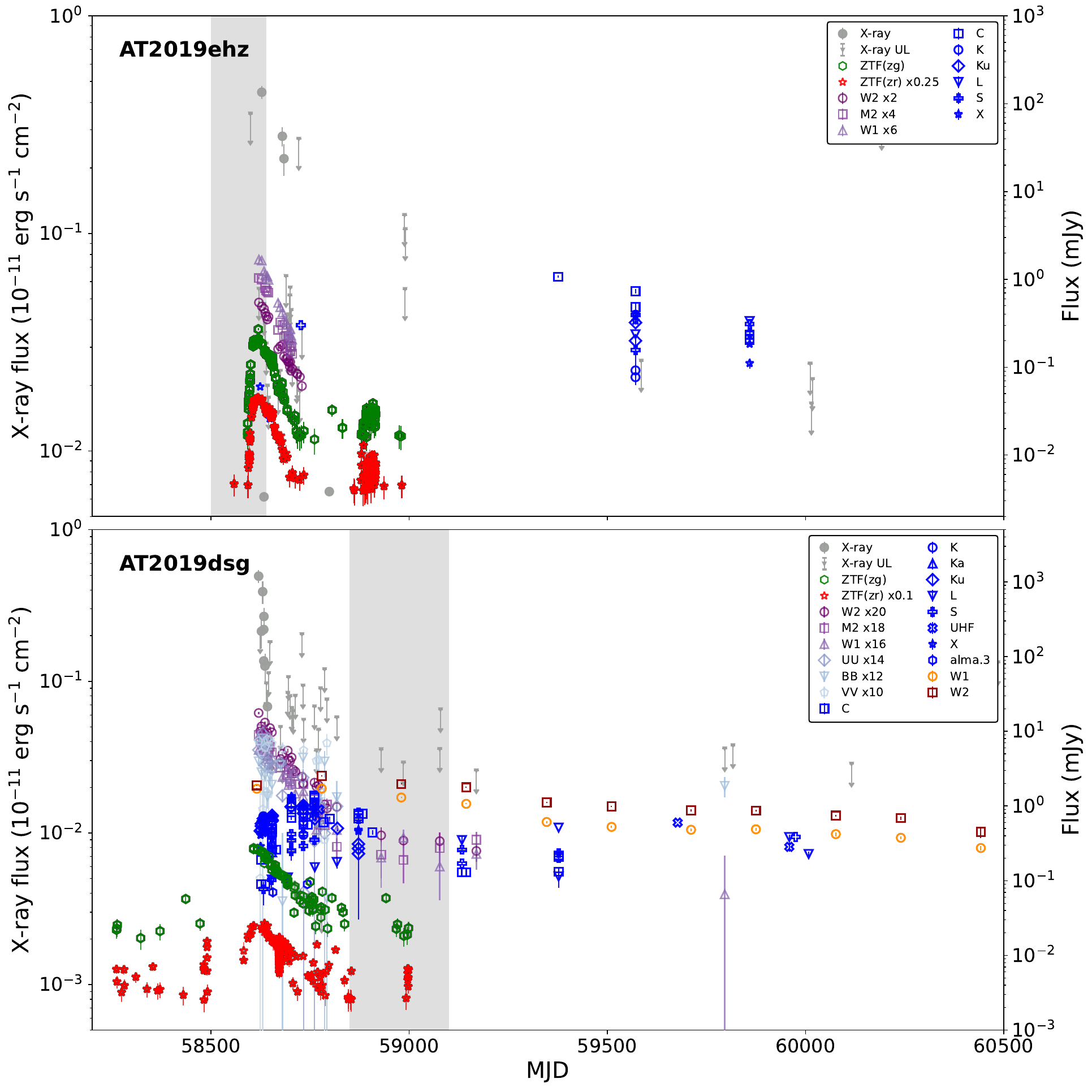}
    \caption{The lightcurves of \ehz (upper) and \dsg (lower) for the different instruments and bands as shown in the legend. The shaded grey region indicates the dates we use to explore the multiwavelength emission below. }
    \label{fig:lightcurves for dsg and ehz}
\end{figure}

\begin{figure}
    \centering
    \includegraphics[width=\linewidth]{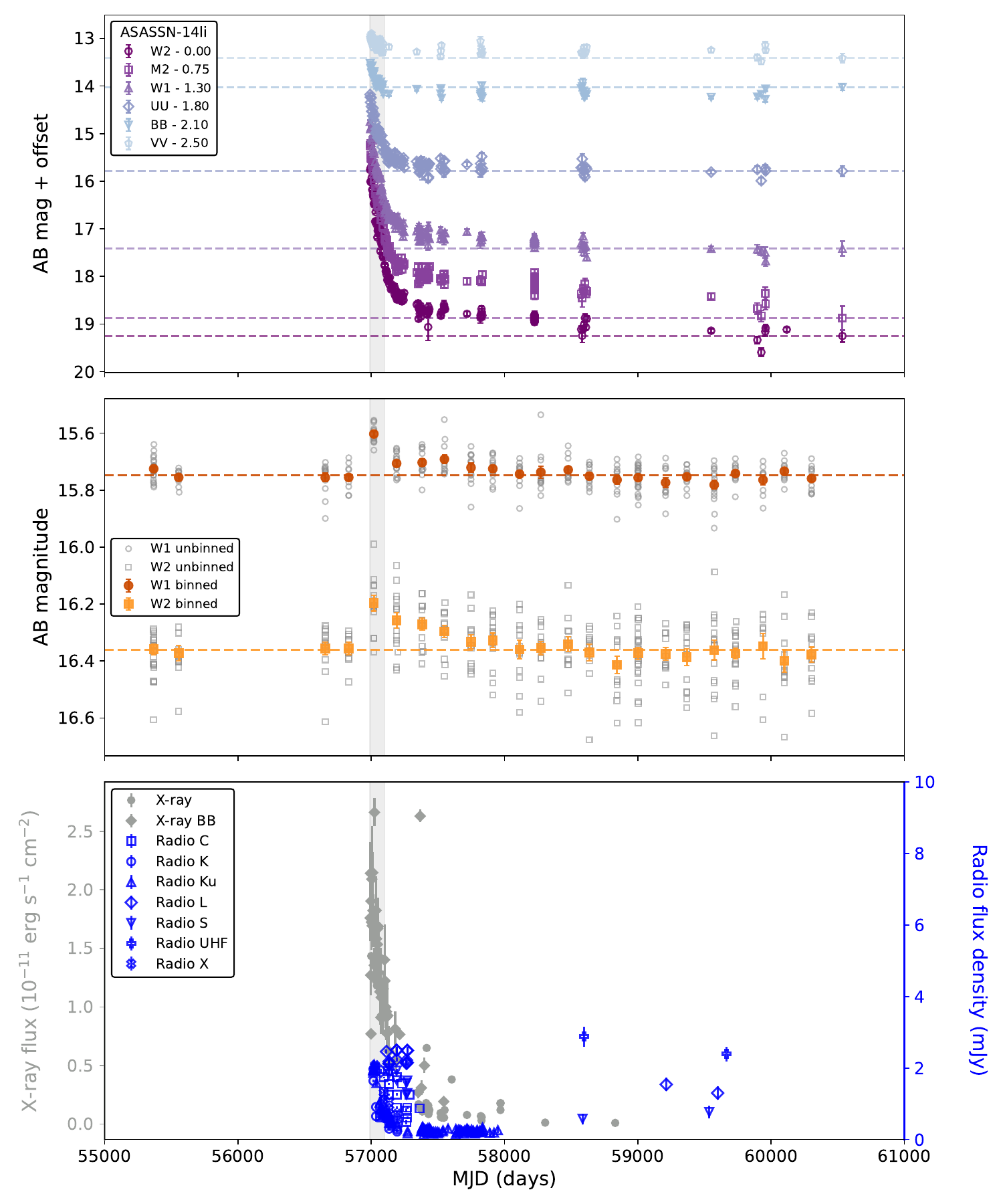}
    \caption{The lightcurve of \ali for the different bands and instruments, as shown in the legends. The middle panel shows the detected flux after subtracting the background emission from the host galaxy. The shaded grey region shows the range of dates we use to explore the multiwavelength emission. }
    \label{fig:lightcurves for ali}
\end{figure}

All data are shown in Fig.~\ref{fig:lightcurves for dsg and ehz} for \dsg and \ehz, and Fig.~\ref{fig:lightcurves for ali} for \ali. 

For all three TDEs, we correct the IR, optical and UV light curves for the host-galaxy contribution by converting from magnitude to flux, calculating the mean of the pre-flare light curve for the infrared and optical, and utilising one post-flare period for the UVOT light curves. This value is then subtracted from the full light curve. This process is shown for \ali in Fig. \ref{fig:lightcurves for ali}, where the UVOT and infrared pre-host-subtraction light curves are shown in the upper and lower panels respectively. The horizontal lines show the host contribution considered in each case. We note that after the host subtraction, we considered as detections the observations that had a signal-to-noise ratio larger than 3.

The X-ray SEDs for \ehz, and \ali were built from the reduced spectra following the steps described in Appendix B of \cite{Langis:2025btl}. We utilized Swift-XRT OBSIDs 00011397002 and 00011397011 for \ehz, and OBSIDs 00033539001--00033539012, 00033539016, 00033539017, 00033539019--00033539027, 00033539029, 00033539030, 00033539032--00033539034 for \ali. Source and background spectra were extracted for each observation, together with appropriate auxiliary response files (ARFs) and redistribution matrix files (RMFs). The spectra were then fitted in XSPEC \citep{1996ASPC..101...17A} with two alternative spectral models, a black body and a power-law, both including an absorption component fixed to the Galactic value at the source coordinates  \citep{2005A&A...440..775K}. The preferred model was selected using the Akaike information criterion \citep{2007MNRAS.377L..74L}, and the corresponding unfolded, model-dependent SED was obtained using XSPEC's eeufspec for the best-fitting model. We note that the X-ray SED data presented in this study for both \ehz and \ali were binned for the epochs inside the time interval considered.

\subsection{Different observational regimes}
The three TDEs considered here probe different observational regimes. For \dsg, we focus on a delayed neutrino-motivated phase, MJD $58850$--$59100$. This source is the only object in our sample with a published high-energy neutrino association, namely IceCube-191001A, detected about $154$\,d after the optical/UV peak. Radio observations of \dsg indicate non-thermal activity from a central engine, making it a natural benchmark for testing whether TDEs can produce detectable neutrino emission together with GeV-to-TeV gamma rays.

For \ehz, we instead model the prompt optical/UV outburst phase, MJD $58500$--$58640$. The source was classified as a TDE and was observed to be variable in Swift/XRT, but it does not have an established high-energy neutrino association. Sparse radio detections occur at much later epochs and therefore do not directly constrain the particle distribution during the interval modelled here. We therefore use \ehz as a test case for whether \grs and neutrino emission could accompany the main TDE outburst.

Finally, \ali is used as an X-ray/radio-bright benchmark. We model the interval MJD $56991$--$57100$, during which the source was bright in both X-rays and radio. Unlike \dsg, \ali is not associated with a reported high-energy neutrino event, but its rich multiwavelength coverage provides the strongest constraints on the low-energy synchrotron component.

\section{Multimessenger emission of \dsg, \ehz and \ali}
\label{Sec: SED and neutrinos of TDEs}
In this section we apply the model described in Sec.~\ref{Sec: Model} to three TDEs and examine the conditions under which they can become observable at high energies. The data shown in the SEDs are the quasi-simultaneous multi-wavelength points described in Sec.~\ref{Sec: Multiwavelength observations}.  

Since the plotted SEDs combine heterogeneous multiwavelength data, host-subtracted photometry, binned X-ray spectra, and non-simultaneous measurements within finite time windows, we do not use a global chi-square minimization to define the models. Such a statistic would be sensitive to band-dependent systematics, source variability, and the model-dependent construction of the unfolded X-ray SEDs. Nevertheless, to make the comparison with the data explicit, we report a reduced chi-square-like diagnostic, $\chi^2_{\rm red}$, for each model using the plotted data points and their reported uncertainties. We use this quantity only as an a posteriori consistency measure, not as a model-selection criterion.

The high-energy sensitivity curves shown in the figures are used only as observational benchmarks. In particular, we use the predicted sensitivity of the Compton Spectrometer and Imager (COSI) from \citet{Tomsick:2023aue}. 
For neutrinos, the grey curves correspond to the nominal ten-year KM3NeT/ARCA230 point-source sensitivity \citep{KM3NeT:2024arca}.  
Since a TDE is transient, these curves should not be interpreted as a direct discovery threshold; the relevant quantity for detectability is the expected number of events during the assumed active period.

Our models calculate epoch-averaged emission states rather than time-dependent light curves. Microscopic particle timescales (acceleration, cooling, escape) are far shorter than the light-curve evolution windows (Figs.~\ref{fig:lightcurves for dsg and ehz} and \ref{fig:lightcurves for ali}), hence, we treat the particle distributions as quasi-steady. Each model represents average active-region conditions rather than a point-by-point fit. Explicit time-dependent modeling (evolving $L_{\rm disk}$, coronal compactness, shock parameters, and non-thermal injection) could modify estimated \gr and neutrino fluences by factors of a few to order-of-magnitude levels depending on variability and flare behavior. Consequently, while our framework checks if time-averaged fluences match event-level expectations, explaining time delays—such as IC191001A arriving $\sim 154$ days post-peak in \dsg—requires full time-dependent modeling of particle injection, photon field evolution, and opacity.

The model families below are exploratory and source-tailored rather than a uniform parameter grid for all three TDEs. The three sources present different observational constraints and different motivating questions: \dsg is treated as a neutrino-motivated delayed-phase case, \ehz as a prompt-outburst high-energy detectability case, and \ali as a radio/X-ray-bright system with strong constraints on the low-energy synchrotron component. We therefore choose the parameters in Tables~\ref{tab: at2019dsg models}--\ref{tab: asassn14li models} to probe the physical regimes most relevant to each source.

\subsection{AT2019dsg}
\label{Sec: AT2019dsg model description}
We first focus on \dsg, which was proposed as the possible electromagnetic counterpart of a high-energy IceCube neutrino \citep{Stein:2020xhk}. Because this association remains debated, we do not attempt to infer a unique neutrino flare model. Instead, we use \dsg as a benchmark system and explore a small set of representative lepto-hadronic configurations. The \dsg sequence is designed to test whether a delayed, neutrino-motivated phase can produce a detectable high-energy counterpart, and therefore varies the proton spectral slope, $\gamma_{p,\min}$, non-thermal proton fraction, and acceleration efficiency around a hard-proton reference case.

We consider five exploratory models for \dsg, listed in Table~\ref{tab: at2019dsg models}.  All models assume a disc inner radius of $r_{\rm in}=300\,R_{\rm g}$ and disk power of $3\times 10^{-5}$\ledd, which keep the thermal peak compatible with the optical/UV data while providing a photon field for photohadronic interactions after Compton scattering. We assume a viewing angle of 60$^\circ$ due to lack of accurate measurements, a mass of the black hole $5.46\times 10^6$\msol, a luminosity distance of 224\,Mpc, a redshift of 0.0512 \citep{Cannizzaro:2020xzc},  the injected power to be 0.01\ledd, the electron temperature to be 100\,keV, the plasma $\beta=1$, and a thermal corona of temperature 150\,keV and optical depth $\tau = 0.75$. We use Model~A as the reference hard-proton configuration for \dsg. This model is not a best-fit solution, but a representative case chosen to test whether a hard proton distribution with a moderate minimum Lorentz factor can produce CTAO- and KM3NeT-relevant emission while remaining broadly compatible with the observed optical/UV and X-ray SED. We use a hard proton distribution with $p_{\rm p}=1.7$, and a moderate minimum proton Lorentz factor of $\gamma_{p,\rm min}=20$. This choice is motivated by the fact that a hard proton spectrum can enhance the multi-TeV neutrino and \gr output without requiring the most extreme hadronic normalization. Models~B--D test alternative ways of obtaining a comparable high-energy signal: a softer proton spectrum with a much larger minimum proton Lorentz factors, a larger fraction of power injected into non-thermal protons, and a higher acceleration efficiency. Model~E is qualitatively different: it is optimized for GeV \gr emission, has no corona, and uses a hard electron spectrum together with a softer proton spectrum. It therefore serves as a Fermi/LAT-oriented, neutrino-quiet comparison case.

\begin{table*}[!ht]
    \centering
    \setlength{\tabcolsep}{4.5pt}
    \renewcommand{\arraystretch}{1.5}
    \caption{Exploratory models used for \dsg.
    The parameters are defined in Sec.~\ref{Sec: Model}.
    $\varepsilon_{p}$ denotes the fraction of the available non-thermal proton power injected in the power-law component.}
    \label{tab: at2019dsg models}
    \begin{tabular}{lcccccclp{10.2cm}}
        \hline\hline
        Model &  $p_{\rm p}$ & $p_{\rm e}$ &  $\gamma_{p, \rm min}$ & $f_{\rm sc}$ & $\varepsilon_p$ & $\eta$ & $\chi^2_{\rm red}$ &
        Physical role \\
        \hline
        A &  1.7 & 1.7 &  20  & $10^{-4}$       & 0.1 & 1 & 28.86 & Reference hard-proton model for CTAO/KM3NeT detectability. \\
        B &  2.0 & 2.0 &  200 & $10^{-3}$ & 0.1 & 1 & 28.48 & Soft-proton model requiring a large $\gamma_{p, \rm min}$. \\
        C &  2.0 & 2.0 &  20  & $10^{-3}$ & 0.8  & 1 & 28.64 & Large non-thermal proton fraction. \\
        D &  2.0 & 2.0 &  2   & $10^{-4}$       & 0.1  & 4    & 36.90 & Low-$\gamma_{p,\min}$ model with more efficient acceleration. \\
        E &  2.2 & 1.6  & 1 & $2\times10^{-4}$ & 0.1 & 1 & 36.44 & Hard-electron, no-corona model optimized for Fermi/LAT emission. \\
        \hline
    \end{tabular}
\end{table*}

\subsection{\ehz}
\label{Sec: ehz model description}
We next apply the same exploration to \ehz.  
The five models are listed in Table~\ref{tab: at2019ehz models}.  
As for \dsg, these models are not intended to define a unique best fit. They are designed to isolate the physical conditions that make the source bright in the GeV band, the TeV band, and in neutrinos. 
The \ehz sequence is chosen to contrast photon-rich and more transparent prompt-outburst environments, and therefore emphasizes the presence or absence of a cold corona, the proton spectral shape, $\gamma_{p,\min}$, and the acceleration efficiency.

For all models, we assume a viewing angle of $60^\circ$ due to lack of better constraints, a mass of the black hole $6.6\times 10^6$\msol, a luminosity distance of 346\,Mpc, and a redshift of 0.074 \citep{Ramsden2022bulge}.

We use Model~A as the reference cold-corona, hard-proton configuration for \ehz. It combines a hard proton spectrum, $p_{\rm p}=1.7$, with $\gamma_{p,\rm min}=50$ and a cold corona with $k_{\rm B}T_{\rm cor}=15\,{\rm keV}$ and $\tau_{\rm cor}=0.55$.  This is a useful reference case because the source remains embedded in a photon-rich environment that can drive photohadronic interactions, rather than relying on a completely transparent, no-corona geometry. 

Model~B removes the corona and increases the acceleration efficiency, producing a more transparent hard-proton solution that can be relevant for Fermi/LAT, CTAO, and KM3NeT simultaneously. Model~C keeps the cold corona but uses a softer proton spectrum and a high proton low-energy cutoff. Model~D tests a proton-rich configuration in which a large fraction of the non-thermal proton power is placed in the power-law component. Finally, Model~E is a Fermi/LAT-oriented model with no corona, low disc luminosity, inefficient scattering, and a low proton minimum energy; it is included to illustrate a GeV-bright but comparatively neutrino-quiet regime.

\begin{table*}[!ht]
    \centering
    \setlength{\tabcolsep}{2.0pt}
    \renewcommand{\arraystretch}{1.4}
    \caption{Exploratory models used for \ehz.  The column labelled ``Corona'' 
    indicates the existence of a thermal Corona of $k_{\rm B}T_{\rm cor}=15$keV and $\tau_{\rm cor}=0.55$.  A dash indicates that no thermal corona is included.}
    \label{tab: at2019ehz models}
    \begin{tabular}{lcccccccccp{8.5cm}}
        \hline\hline
        Model & $p_{\rm p}$ & $\gamma_{p,\rm min}$ & Corona & $L_{\rm disk}$/\ledd & $L_{\rm shock}$/\ledd & $f_{\rm sc}$ & $\eta$ & $\varepsilon_p$ & $\chi^2_{\rm red}$ & Physical role \\
        \hline
        A & 1.7 & 50  & $\checkmark$  & $2\times10^{-5}$ & $10^{-2}$ & $10^{-3}$ & 1 & 0.1 & 51.41 & Reference hard-proton, cold-corona model for CTAO/KM3NeT. \\
        B & 1.7 & 10  & --                  & $9\times10^{-6}$ & $2\times10^{-3}$ & $10^{-3}$ & 4 & 0.1 & 4.60 & Hard-proton rich model for Fermi/LAT, CTAO, and KM3NeT. \\
        C & 2.0 & 500 & $\checkmark$  & $2\times10^{-5}$ & $10^{-2}$ & $10^{-3}$ & 1 & 0.1 & 4.60 &  Soft-proton model with high $\gamma_{p,\rm min}$. \\
        D & 2.0 & 20  & $\checkmark$  & $2\times10^{-5}$ & $10^{-2}$ & $10^{-4}$ & 1 & 0.8 & 65.00 & Large non-thermal, soft-proton model. \\
        E & 2.0 & 3   & --                  & $2\times10^{-6}$ & $10^{-2}$ & $10^{-6}$ & 4 & 0.1 & 4.64 & Proton-rich, no-corona model optimised for Fermi/LAT emission and IR secondary synchrotron. \\
        \hline
    \end{tabular}
\end{table*}

\subsection{\ali}
\label{Sec: ali model description}
We finally apply the same framework to \ali.  
This source is useful as a third benchmark because the explored parameter space does not naturally lead to the same TeV-bright behaviour found in some of the \dsg and \ehz models. 
Instead, 
the \ali sequence is constrained by the radio/X-ray-bright nature of the source and is designed to test whether compact non-thermal shocks can contribute to the radio-to-MeV/GeV SED and neutrino output without requiring a robust CTAO-bright counterpart.
We separate \ali into three hadronic configurations that can be relevant for Fermi/LAT and KM3NeT, but remain faint for CTAO, and one primarily leptonic configuration that illustrates a neutrino-quiet GeV-emitting state. We therefore keep four representative models, listed in Table~\ref{tab: asassn14li models}. 

All \ali models assume a mass of the black hole of $2\times 10^6$\msol, a viewing angle of $60^\circ$, a luminosity distance of 90.3\,Mpc, a redshift of 0.0206 \citep{Holoien2016six}, an emission-region radius of $R_0=10^3\,R_{\rm g}$, and a disc with $L_{\rm disc}=4\times10^{-4}\,L_{\rm Edd}$, and $r_{\rm in}=4000\,R_{\rm g}$. No active corona is included in these models. We use Model~A as the reference hot, hard-particle configuration for \ali. This model was selected to test whether the source can be bright in the MeV--GeV band and produce an observable neutrino flux without becoming a CTAO-bright source, while remaining broadly consistent with the observed X-ray and radio emission. Model~A combines a hard proton spectrum, $p_{\rm p}=1.6$, with a hard electron spectrum, $p_{\rm e}=1.8$, a hot thermal electron population, and a moderate $\gamma_{p,\rm min}$. This model is designed to test whether \ali can be bright in the MeV--GeV band and produce an observable neutrino flux without becoming a CTAO source. Models~B and C explore two alternative hadronic routes: a large non-thermal proton fraction at lower jet power, and a magnetically dominated, high-hadronic-weighting case with a softer proton spectrum. Model~D is a lepton-dominated comparison model, with a large non-thermal electron fraction and weak hadronic normalization.

\begin{table*}[!ht]
    \centering
    % \scriptsize
    \setlength{\tabcolsep}{3.0pt}
    \renewcommand{\arraystretch}{1.4}
    \caption{Exploratory models used for \ali.  
    The parameters $\varepsilon_{\rm e}$ and $\varepsilon_{\rm p}$ denote the fractions of the available non-thermal electron and proton power injected in the power-law component. }
    \label{tab: asassn14li models}
    \begin{tabular}{lccccccccccp{7.5cm}}
        \hline\hline
        Model & $p_{\rm p}$ & $p_{\rm e}$ & $kT_{\rm e}/\mathrm{keV}$ & $L_{\rm shock}/L_{\rm Edd}$ & $f_{\rm sc}$ & $\beta$ & $\varepsilon_{\rm e}$ & $\varepsilon_{\rm p}$ & $\gamma_{p, \rm min}$ & $\chi^2_{\rm red}$ & Physical role \\
        \hline
        A & 1.6 & 1.8 & $3000$ & $2.5\times10^{-2}$ & $2\times10^{-3}$ & 1    & 0.1 & 0.1 & 60  & 57.58 & Reference hot, hard-particle model for COSI/Fermi and KM3NeT. \\
        B & 1.7 & 2.0 & $511$ & $2.5\times10^{-3}$ & $10^{-2}$        & 1    & 0.1 & 0.8 & 5   & 58.51 & Proton-rich, low-power hadronic model. \\
        C & 2.0 & 1.7 & $51$ & $5.0\times10^{-2}$ & $5\times10^{-3}$ & 0.1  & 0.1 & 0.1 & 500 & 58.51 & Low-$\beta$, high-min Lorentz factor. \\
        D & 2.0 & 2.0 & $511$ & $5.0\times10^{-2}$ & $10^{-2}$        & 10   & 0.8 & 0.1 & 1   & 57.74 &  Lepton-dominated Fermi/LAT-orientated model. \\
        \hline
    \end{tabular}
\end{table*}

\subsection{Spectral energy distributions}
\label{Sec: SEDs of all three sources}
\begin{figure}[!ht]
    \centering
    \vspace{-.4cm}
    \includegraphics[width=\columnwidth]{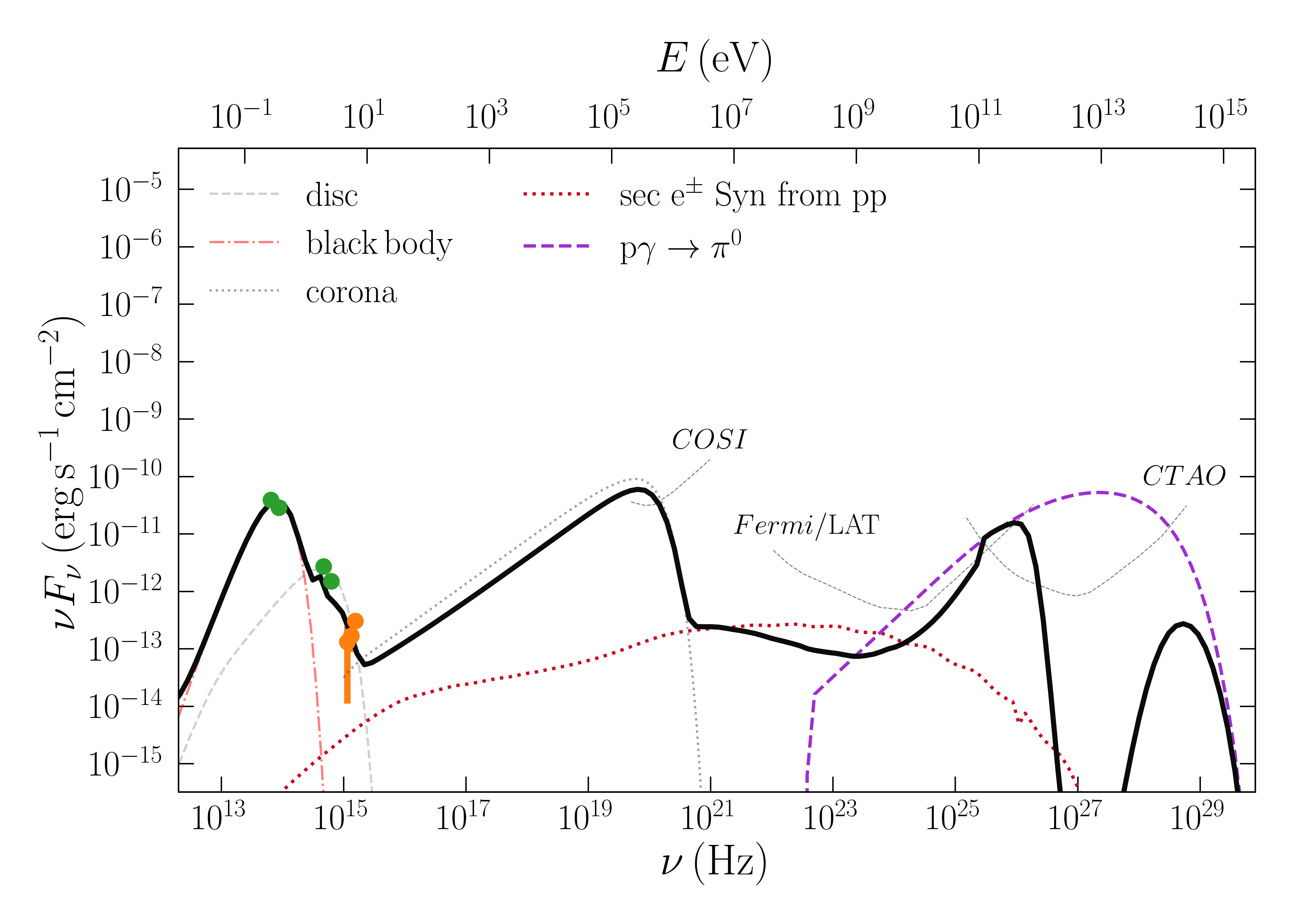}
    \vspace{-.3cm}
    \includegraphics[width=\columnwidth]{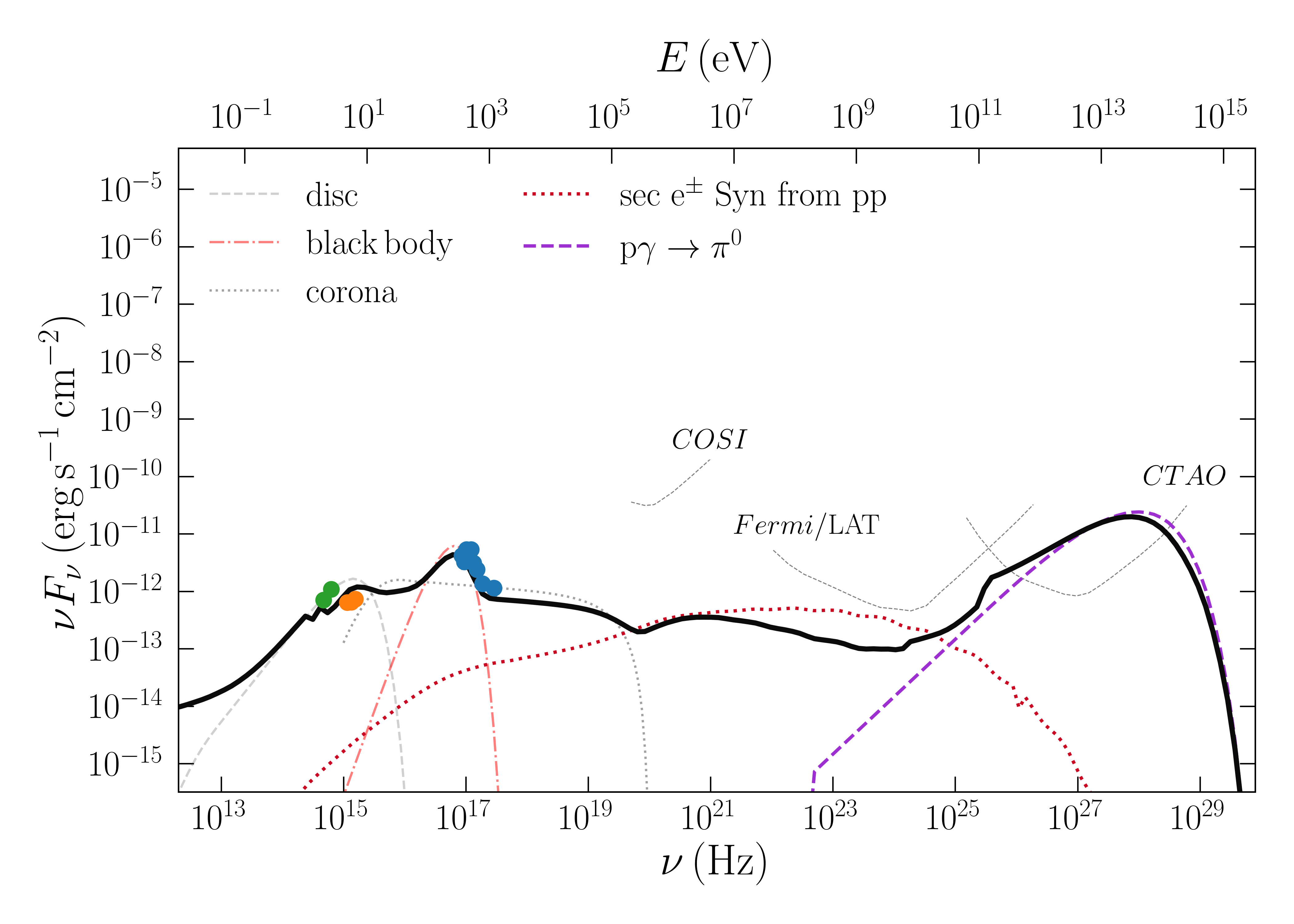}
    \vspace{-.2cm}
    \includegraphics[width=\columnwidth]{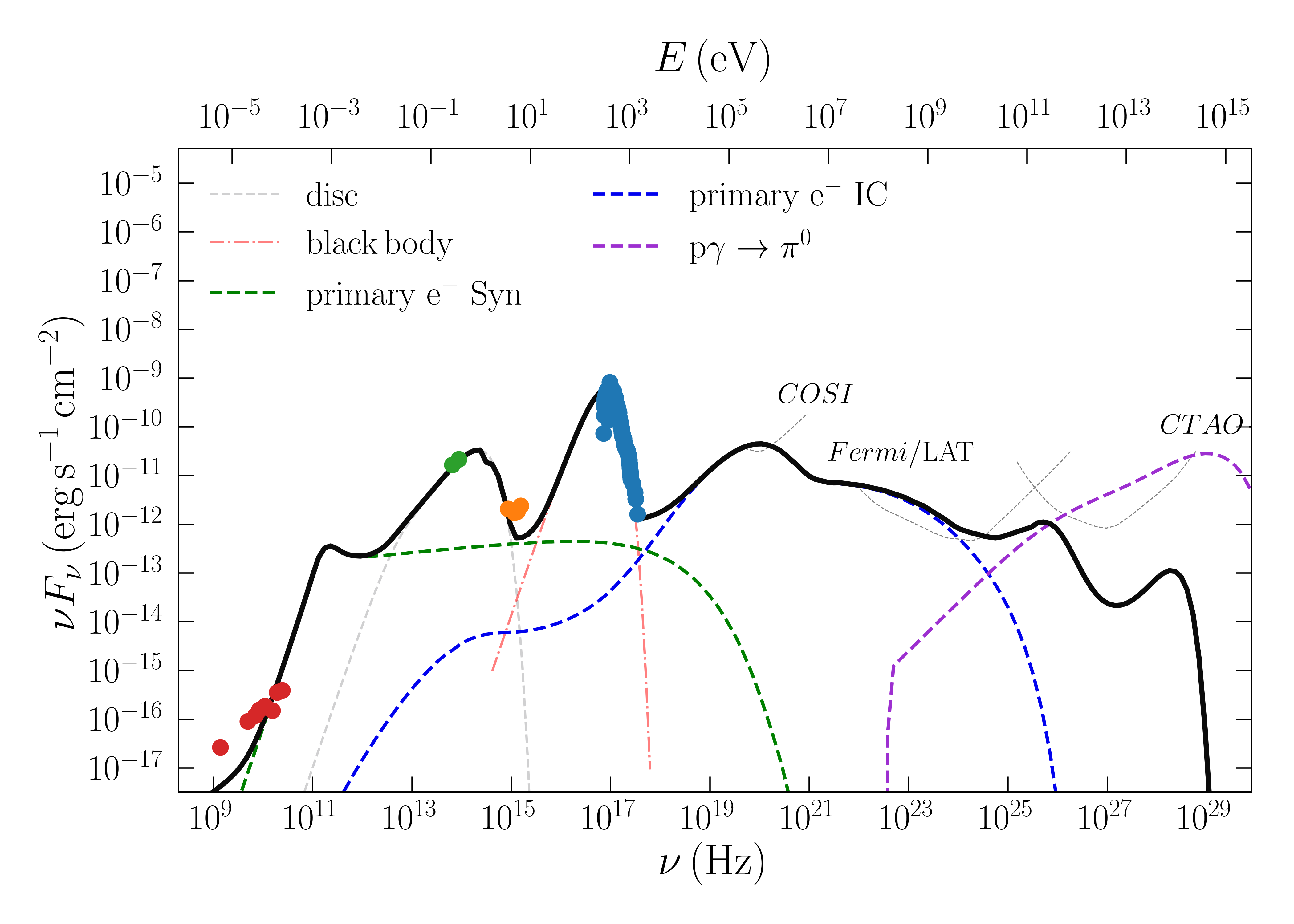}
    \caption{Spectral energy distributions of reference model A for \dsg (upper), \ehz (middle) and \ali (lower panel). The thick black line shows the total model emission after accounting for absorption. 
    The thermal disc, black-body, and coronal components are shown by dashed grey, dash-dotted red, and dotted grey curves, respectively. 
    Hadronic \gr components are shown by red dashed for synchrotron from secondary electrons from $\pi^{\pm}$ decay, and purple dashed for $p\gamma\rightarrow\pi^0$.
    For leptonic components, the dashed green and blue curves show the synchrotron and inverse-Compton emission of primary electrons, respectively.
    The thin grey curves indicate the COSI, Fermi/LAT, and CTAO reference sensitivities.}
    \label{fig: sed model A of all sources}
\end{figure}

Figure~\ref{fig: sed model A of all sources} shows the SEDs of the reference model A for each source. These models are overplotted on the multiwavelength data to assess whether the chosen physical regime is broadly compatible with the observed emission, but the curves should not be interpreted as unique best-fit SEDs. In each panel, the thick black line is the total model spectrum. The thermal emission from the disc is shown by a dashed grey curve, the black-body component by a dash-dotted red curve, and the coronal emission, when present, by a dotted grey curve. The thin grey curves mark the COSI, Fermi/LAT, and CTAO reference sensitivities indicated in the panels. The remaining curves show the rest of the components that dominate each model: neutral-pion decay from $p\gamma$ interactions by a dashed purple line and synchrotron emission from secondary pairs by dotted red curves. The dashed green and blue curves show the primary electron synchrotron and inverse-Compton components, respectively.

The reference Model~A of \dsg produces a high-energy SED in which the TeV band is dominated by the $p\gamma\rightarrow\pi^0$ component. The same target photon field that enables photohadronic interactions also attenuates part of the intrinsic \gr emission, producing a structured escaping spectrum. The model reaches the CTAO band without overproducing the optical/UV or X-ray data, and it therefore provides a useful reference case for a TDE that is simultaneously \gr and neutrino bright. 

\ehz model~A reaches the CTAO band through a high-energy component dominated by $p\gamma\rightarrow\pi^0$ emission. The cold corona supplies an additional target photon field for photohadronic interactions, while the hard proton distribution keeps a large fraction of the non-thermal proton power near the energies relevant for pion production. The same photon field also shapes the escaping spectrum through internal attenuation, so that the total SED is not a simple power law at high energies. This model provides our reference case for a TeV-bright \ehz-like TDE.

\ali model~A produces a broad, high-energy SED that is potentially relevant for COSI and Fermi/LAT, while remaining below the CTAO benchmark. The MeV--GeV emission is mainly shaped by the primary inverse-Compton component and the onset of the hadronic component, whereas the highest-energy part of the spectrum is associated with $p\gamma\rightarrow\pi^0$ emission. Because there is no active corona, the target photon field is less compact than in the cold-corona models of \ehz. The model can therefore produce a sizeable neutrino flux without generating a CTAO-bright TeV component.

The low-frequency data points provide an additional constraint on the normalisation of the non-thermal particle population. In all models, the radio emission can be broadly accounted for by synchrotron radiation from primary electrons. We therefore use the radio data mainly as a consistency check on the magnetic-field strength and the normalisation of the accelerated electrons. However, this interpretation is not unique. The observed radio emission of \ali has been discussed in terms of outflow or jet-like activity \citep{Alexander2016ASASSN14liRadio, vanVelzen2016ASASSN14liRadio}, and in our models additional synchrotron channels can contribute at low frequencies.

In Appendix~\ref{App: model A detailed seds}, we show all the components of all processes for each baseline model of all three sources. Further, in Appendix~\ref{app: section of all models of all sources}, we show all the SEDs for each model for the three sources discussed here. 

To avoid relying only on visual comparisons with the sensitivity curves, we compute for each model a proximity ratio 
\begin{equation}
R_X = \max \left[ \frac{F_{\rm model}(E)}{F_{X,\rm sens}(E)}\right],
\end{equation}
where 
$F_{\rm model}$ is the predicted flux, $F_{\rm sens}$ is the sensitivity flux at energy $E$,
$X$ denotes COSI, Fermi/LAT, or CTAO, and the maximum is taken only over the energy range covered by the corresponding reference sensitivity curve. We tabulate these values for the energy values where the ratio becomes the maximum in Table~\ref{tab:detectability_ratios}. These ratios are not formal detection significances; they provide a compact way to summarize which
models approach or exceed the adopted benchmark curves.

\begin{table*}
\caption{Peak-flux and maximum proximity ratios for the model predictions relative to the reference sensitivity curves.
\tablefoot{
For each electromagnetic facility, $R_X$ is the maximum value of $F_{\rm model}(E)/F_{X,\rm sens}(E)$ over the energy interval covered by the corresponding reference sensitivity curve. Values $R_X>1$ indicate that the model exceeds the reference sensitivity at least at one energy. These ratios are used only as proximity diagnostics and are not formal detection significances. The flux is in units of $10^{-12}\, \mathrm{erg\,cm^{-2}\,s^{-1}}$.  
}
}
\label{tab:detectability_ratios}
\renewcommand{\arraystretch}{1.1} % Increases row height by 50%
\begin{tabular}{lllll|lll|lll}
\toprule
\multirow{3}{*}{Source}              & \multirow{3}{*}{Model} & \multicolumn{9}{c}{Instrument}                                                      \\
                                     &                        & \multicolumn{3}{c}{COSI} & \multicolumn{3}{c}{Fermi/LAT} & \multicolumn{3}{c}{CTAO} \\
                                     &                        & $E_\mathrm{at\,max}$ & $F_\mathrm{at\, max}$ &  $R_\mathrm{max}$   & $E_\mathrm{at\,max}$ & $F_\mathrm{at\, max}$ &  $R_\mathrm{max}$    & $E_\mathrm{at\,max}$ & $F_\mathrm{at\, max}$ &  $R_\mathrm{max}$   \\
\midrule
\multirow{5}{*}{\dsg} & A & 0.26 MeV & $90.35$ & 2.69 & 13.4 GeV & $14.82$ & 2.641 & 4.6 TeV & $50.82$ & 59.03 \\
                                     & B & 0.26 MeV & $90.56$ & 2.697 & 30.7 GeV & $7.085$ & 5.557 & 3.5 TeV & $60.61$ & 71.42 \\
                                     & C & 0.26 MeV & $90.49$ & 2.694 & 13.35 GeV & $3.253$ & 5.798 & 3.5 TeV & $48.45$ & 57.10 \\
                                     & D & 0.26 MeV & $90.91$ & 2.707 & 7.7 GeV & $4.136$ & 9.003 & 4.6 TeV & $16.98$ & 19.72 \\
                                     & E & 0.46 MeV & $0.603$ & $1.9\times10^{-2}$ & 1.5 GeV & $1.131$ & 1.787 & 70 TeV & $0.008$ & 50.51 \\
                                     \hline
\multirow{5}{*}{\ehz} & A & 0.46 MeV & $0.289$ & $9.1\times10^{-3}$ & 7.7 GeV & $0.574$ & 1.249 & 6 TeV & $14.91$ & 16.11 \\
                                     & B & 0.46 MeV & $1.054$ & $3.3\times10^{-2}$ & 7.7 GeV & $1.594$ & 3.471 & 4.6 TeV & $11.08$ & 12.87 \\
                                     & C & 0.46 MeV & $0.835$ & $2.6\times10^{-2}$ & 70.1 GeV & $3.613$ & 1.26 & 6 TeV & $19.57$ & 21.14 \\
                                     & D & 0.26 MeV & $1.097$ & $3.3\times10^{-2}$ & 7.7 GeV & $4.884$ & 10.63 & 6 TeV & $25.67$ & 27.72 \\
                                     & E & 0.26 MeV & $0.140$ & $4.1\times10^{-3}$ & 7.7 GeV & $4.422$ & 9.626 & 0.16 TeV & $1.012$ & 0.21 \\
                                     \hline
\multirow{4}{*}{\ali} & A & 0.35 MeV & $44.07$ & 1.41 & 0.83 GeV & $2.387$ & 2.898 & 6 TeV & $6.194$ & 6.69 \\
                                     & B & 0.46 MeV & $0.541$ & $1.7\times10^{-2}$ & 4.4 GeV & $0.763$ & 1.555 & 6 TeV & $3.915$ & 4.23 \\
                                     & C & 0.35 MeV & $1.120$ & $3.6\times10^{-2}$ & 5.8 GeV & $0.690$ & 1.461 & 6 TeV & $5.933$ & 6.41 \\
                                     & D & 0.35 MeV & $10.16$ & 0.33 & 0.27 GeV & $2.738$ & 1.94 & 4.6 TeV & $0.02$ & $0.024$\\
\bottomrule
\end{tabular}
\end{table*}

\subsection{Muon-neutrino predictions}
\label{Sec: neutrinos of all model A of all three sources}
\begin{figure}[!ht]
    \centering
    \vspace{-.4cm}
        \includegraphics[width=\columnwidth]{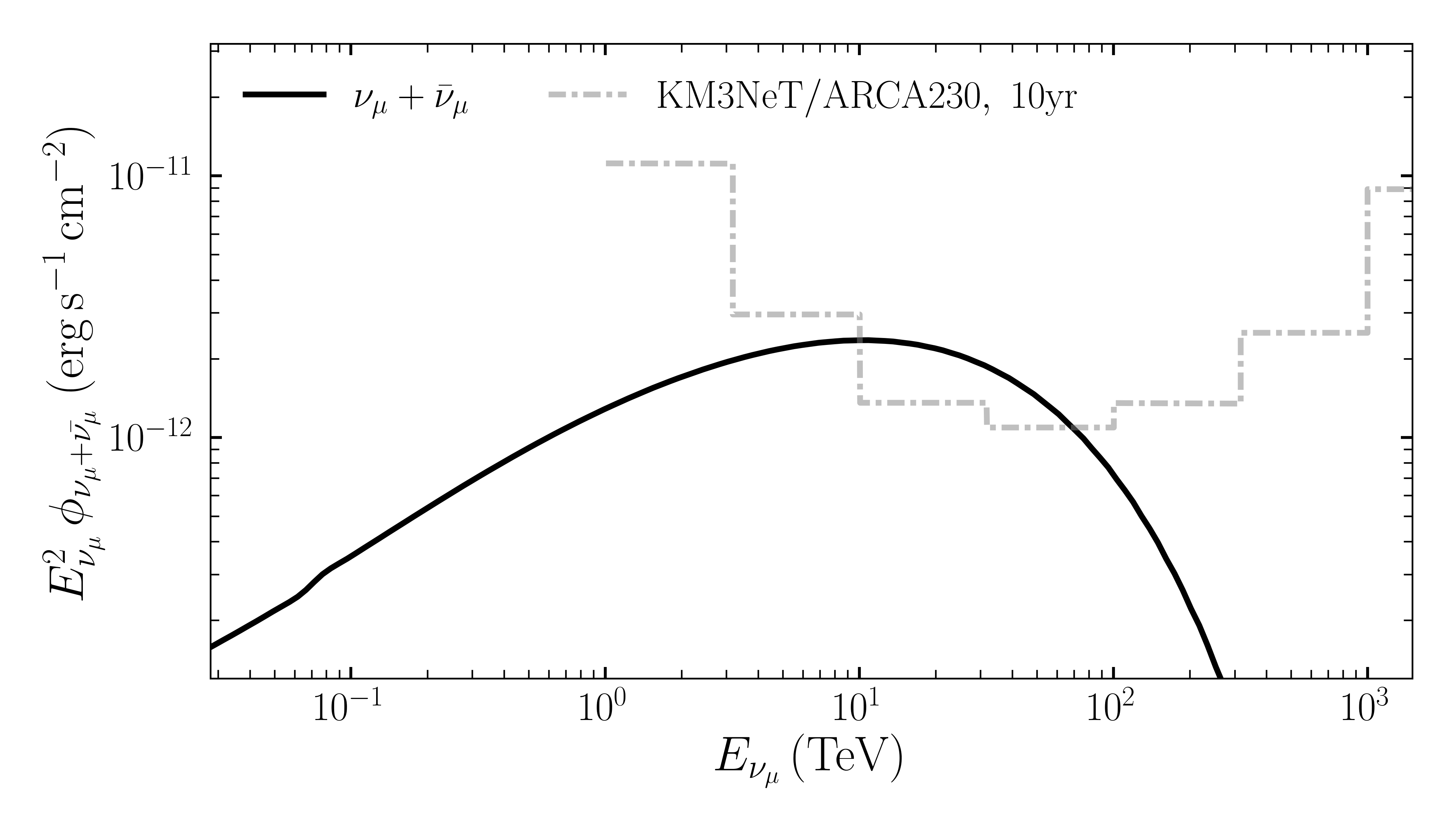}
    \vspace{-.3cm}
        \includegraphics[width=\columnwidth]{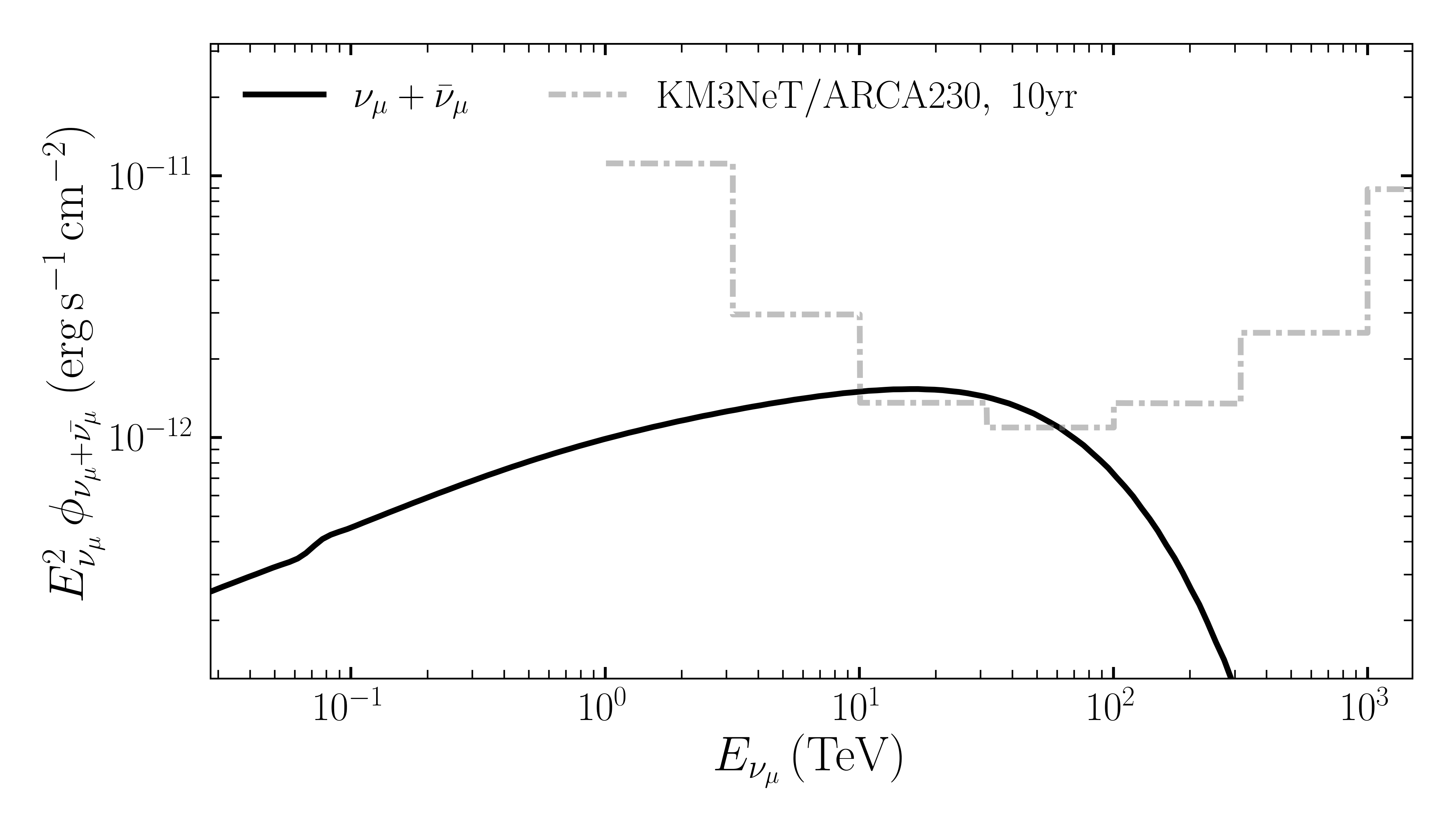}
        \vspace{-.2cm}
        \includegraphics[width=\columnwidth]{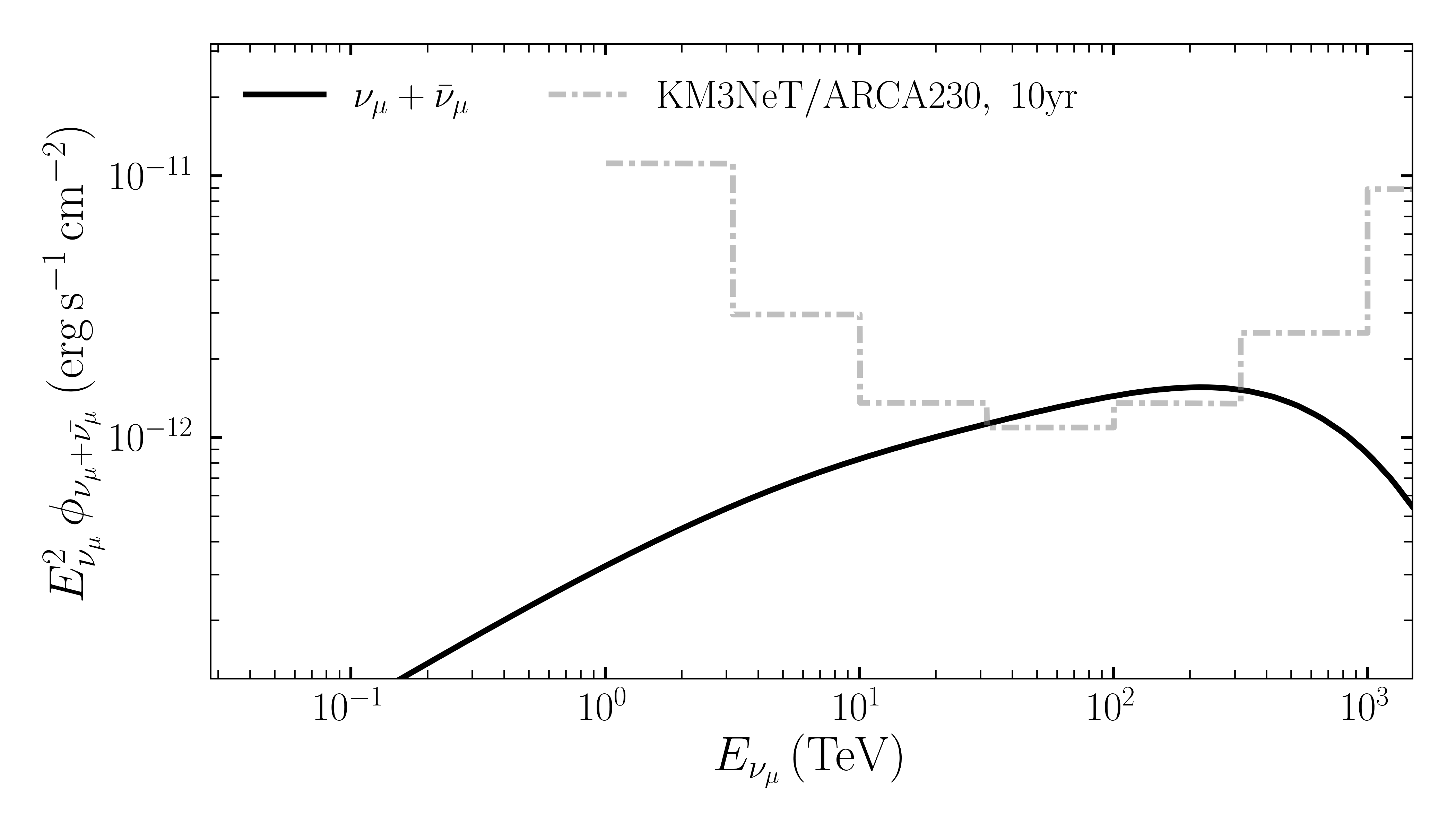}
    \caption{Muon-neutrino plus antineutrino spectra predicted for the baseline model A for \dsg (upper), \ehz (middle) and \ali (lower panel). The black curves show $E_{\nu_\mu}^2\phi_{\nu_\mu+\bar{\nu}_\mu}$, and the grey dash-dotted curves show the nominal ten-year KM3NeT/ARCA230 point-source sensitivity}
    \label{fig: neutrinos for all fiducial models of all three sources}
\end{figure}

The corresponding $\nu_\mu+\bar{\nu}_\mu$ spectra of the reference models are shown in Fig.~\ref{fig: neutrinos for all fiducial models of all three sources}. The plotted quantity is $E_{\nu_\mu}^2\phi_{\nu_\mu+\bar{\nu}_\mu}$, allowing a direct comparison with the KM3NeT/ARCA230 reference sensitivity.  \dsg model~A peaks in the multi-TeV range and approaches the nominal ARCA230 sensitivity around the energy interval where the detector is most sensitive. This is why we choose Model~A as the reference high-energy configuration for \dsg: it provides a neutrino-bright solution without relying on the largest $\gamma_{p,\rm min}$ or the largest non-thermal proton fraction explored here. The comparison with the ARCA230 curve should be interpreted conservatively.  The grey line is a steady point-source sensitivity for a ten-year exposure, whereas the TDE emission is expected to last for a finite flare duration.  Therefore, a model curve lying near the sensitivity is not by itself a detection claim.  Instead, the figure identifies which spectral shapes and flux levels are worth evaluating with the full event-count calculation, including the source declination, active time window, and flavour-dependent effective area.

For \dsg we also compute the expected number of IceCube $\nu_\mu+\bar{\nu}_\mu$ track events by folding the model neutrino spectra with the declination-dependent IceCube effective area over the adopted active interval. This provides a direct comparison with the single event IC191001A, which arrived approximately 154 days after the optical/UV peak of \dsg. We emphasise, however, that the present calculation is time-averaged over the selected MJD window and is not designed to model the temporal evolution of the neutrino production site. Consequently, our models can test whether the overall neutrino fluence is of the right order to produce an IceCube event, but they do not provide a physical explanation for the observed delay. A time-dependent treatment of the evolving radiation field, particle injection, and internal opacity would be required to address this question.

The corresponding $\nu_\mu+\bar{\nu}_\mu$ spectrum for \ehz model~A shows a broad multi-TeV neutrino spectrum that approaches the nominal KM3NeT/ARCA230 sensitivity near the energy range where the detector is most sensitive. This behaviour follows naturally from the same hard proton population and cold coronal photon field that produce the CTAO-band emission.

The corresponding $\nu_\mu+\bar{\nu}_\mu$ spectrum for the \ali Model~A
peaks at higher energies than the baseline models of \dsg and \ehz, with most of the neutrino power emerging from tens to hundreds of TeV. This reflects the hard proton distribution and the comparatively transparent no-corona environment: protons can reach high energies, but the photon field does not produce a strong CTAO-band counterpart.

In Appendix~\ref{app: section of all models of all sources} we show the neutrino spectra for all models of all three sources. In Table~\ref{tab:neutrino event counts}, we gather neutrino counts per source for both IceCube and KM3NeT/ARCA over the adopted active interval of each source.

\begin{table}
\centering
\caption{Expected neutrino event counts for IceCube and KM3NeT/ARCA. The values are computed by folding the predicted neutrino spectra with the corresponding detector effective areas over the adopted active duration of each source.}
\label{tab:neutrino event counts}
\renewcommand{\arraystretch}{1.1} % Increases row height by 50%
\begin{tabular}{lccc}
\hline
Source & Model & IceCube & KM3NeT/ARCA \\
\multirow{5}{8em}{\dsg}
 & A &  0.31 & 0.36 \\
 & B &  0.32 & 0.38 \\
 & C &  0.25 & 0.30 \\
 & D &  0.29 & 0.34 \\
 & E &  $3.0\times10^{-6}$ & $4.0\times10^{-6}$ \\
\hline
\multirow{5}{8em}{\ehz}
 & A &  0.05 & 0.06 \\
 & B &  0.08 & 0.10 \\
 & C &  0.06 & 0.08 \\
 & D &  0.09 & 0.11 \\
 & E &  $8.0\times10^{-5}$ & $6.0\times10^{-5}$ \\
\hline
\multirow{4}{8em}{\ali}
 & A &  0.36 & 0.41 \\
 & B &  0.40 & 0.46 \\
 & C &  0.45 & 0.51 \\
 & D &  $7.0\times10^{-4}$ & $8.0\times10^{-4}$ \\
\hline
\end{tabular}
\end{table}

\section{Discussion}\label{Sec: discussion}
In this work, we aim to map the regions of parameter space in which a TDE can become visible in  COSI, \fermi, CTAO, and KM3NeT. This is important because the same non-thermal proton population produces both neutrinos and secondaries, but the escaping \grs are strongly regulated by the local radiation field. In a transparent source, the neutral-pion component can reach the GeV-to-TeV band and may be observable by \fermi and/or CTAO. In a more compact or photon-rich source, the same intrinsic hadronic power can be reprocessed by $\gamma\gamma$ absorption, shifting part of the observable counterpart toward the X-ray or MeV--GeV range. Therefore, a non-detection in the TeV band does not necessarily imply a neutrino-quiet TDE; it may instead indicate that the source is opaque in TeV \grs.

A controlled comparison in which the same leptonic and hadronic parameter grid is applied to all three sources would be needed to fully disentangle intrinsic source properties from modelling choices. The present work instead identifies physically motivated detectability regimes for each source. Such a uniform-grid study, combined with time-dependent light-curve modelling, will be pursued in future work.

\subsection{Accelerated-proton parameter phase space}
\label{Sec: discussion parameter space}
Across the three sources, the most important particle parameters are those that define the shape of the non-thermal proton distribution. Hard proton spectra place a larger fraction of the injected proton power close to the photomeson threshold and therefore naturally enhance the multi-TeV neutrino and high-energy \gr components. This is why the reference hard-proton models of \dsg and \ehz can approach the KM3NeT/ARCA reference sensitivity without requiring the most extreme values of the non-thermal proton fraction. Softer proton spectra can still be viable, but only if the proton power is shifted toward higher energies by increasing $\gamma_{p,\rm min}$, by increasing the fraction of dissipated power injected into non-thermal protons, or by increasing the acceleration efficiency so that protons reach the $p\gamma$ regime before escaping, such as \dsg Models~B, C, and \ehz Model~C.

The second controlling quantity is the compactness and spectrum of the target photon field. A disc or corona provides the soft photons needed for $p\gamma$ interactions and can therefore raise the neutrino luminosity. At the same time, the same photon field increases the internal opacity of the source. This produces a natural tension: efficient neutrino production favours dense photon targets, whereas CTAO detectability requires that at least part of the GeV-to-TeV cascade can escape. The \ehz models show this most clearly. For Model~A, we use a cold coronal photon field that is efficient in both $p\gamma$ interactions and internal reprocessing, while Model~B, for which we remove the corona, becomes more transparent to GeV-to-TeV photons. The comparison suggests that CTAO detections should favour sources or phases in which the high-energy emission region is not deeply embedded in the densest part of the optical/UV/X-ray radiation field, such as \ehz.

The third controlling quantity is the magnetic field, which we probe indirectly through the plasma beta in the \ali models. For the cases of \dsg and \ehz, we assume a plasma in energy equipartition, namely $\beta=1$. A lower plasma beta strengthens the magnetic field and can enhance synchrotron emission from protons and secondary pairs. This changes the electromagnetic counterpart without necessarily producing a brighter TeV signal.  In \ali Model~C, for example, proton synchrotron and secondary-pair emission can contribute in the radio-to-infrared and GeV bands, while the source remains quiet for CTAO. This demonstrates that a magnetically enhanced hadronic model may be more visible through radio, MeV, or GeV diagnostics than through TeV \grs.

\subsection{Comparison between the sources}
\label{Sec: discussion source comparison}
The three sources occupy different regions of the same parameter space. \dsg is the most direct benchmark for a possible neutrino-associated optical TDE, but our results support a cautious interpretation. Several combinations of parameters can place the predicted $\nu_\mu+\bar\nu_\mu$ spectrum near the KM3NeT/ARCA reference curve. However, these combinations are not unique: a hard proton distribution, a large $\gamma_{p,\rm min}$, a large non-thermal proton fraction, or efficient acceleration can all lead to similar neutrino levels.  This degeneracy means that the neutrino prediction cannot be inferred from the bolometric TDE luminosity alone. It depends on how efficiently the TDE channels power into protons above the pion-production threshold and on how transparent the source is to the associated cascade emission.

Within the source-tailored parameter space explored here, \ehz provides the clearest example of a potentially TeV-bright TDE. The reference Model~A combines a hard proton spectrum with a cold coronal photon field, producing a strong $p\gamma$ component and a multi-TeV neutrino spectrum. The no-corona hard-proton Model~B is even more transparent and is therefore the most favorable case for simultaneous \fermi, CTAO, and KM3NeT searches. The soft-proton models remain viable only after the low-energy part of the proton distribution is suppressed or the non-thermal proton fraction is increased. Thus, compared to \dsg, \ehz more clearly separates two physical routes to detectability: a photon-rich route that enhances neutrino production, and a transparent route that favors escaping GeV-to-TeV photons.

Within the \ali parameter sequence explored here, the source occupies a different regime. In the explored parameter space, the neutrino-bright models do not naturally become CTAO-bright. The most promising electromagnetic counterpart is instead in the MeV-to-GeV band, where primary inverse-Compton emission, pion-induced cascades, and secondary-pair synchrotron can all contribute. This makes \ali a useful reminder that neutrino-bright TDEs need not be TeV-bright TDEs. The radio data provide an additional constraint, but they do not uniquely select a leptonic or hadronic solution. Primary-electron synchrotron is sufficient to account for the radio points at the level needed in our exploratory modelling, yet previous studies have interpreted the radio emission of \ali in terms of either an outflow or compact jet-like activity \citep{Alexander2016ASASSN14liRadio,vanVelzen2016ASASSN14liRadio,Krolik2016ASASSN14li,PashamVanVelzen2018}. In our models, magnetically enhanced configurations can also allow proton synchrotron or secondary pairs to contribute at low frequencies. Rather than a dedicated relativistic jet calculation, we use \ali to test whether a compact non-thermal shock can explain the radio-to-high-energy SED within X-ray and \gr constraints. Primary electron synchrotron emission in our reference model broadly matches the radio data, though this represents a consistency check rather than a unique physical origin. Future radio light curves and spectra are therefore essential for separating the location and physical origin of the radio-emitting plasma.

\subsection{\texorpdfstring{\gr}{gamma-ray} and neutrino detections}
The possible association of \dsg with IceCube-191001A motivated many of these theoretical developments \citep{Stein:2020xhk}. Additional proposed associations, such as AT2019fdr and AT2019aalc, strengthened the motivation for considering accretion flares as neutrino sources \citep{Reusch:2021ztx,vanVelzen:2021zsm}. At the same time, the statistical status of these associations remains uncertain \citep{IceCube:2025uzh}. Updated neutrino reconstructions and catalogues can change the spatial consistency of individual events, and recent IceCube updates motivate treating the proposed associations as suggestive rather than established \citep{IceCube:2025uzh}.  Our results are therefore best interpreted as a detectability study rather than as a claim that any of the modelled sources must have produced a specific IceCube event.

The absence of robust GeV detections from optically selected TDEs is another important point. Searches with \fermi have found no significant GeV emission from samples of candidate neutrino-associated TDEs and from broader TDE samples, leading to upper limits that strongly constrain transparent hadronic scenarios \citep{Peng2022FermiTDEs}. Time-dependent cascade calculations have also shown that \fermi limits can constrain the emission radius and the maximum energy of the accelerated protons, and that the corresponding expected neutrino event numbers can be small for many configurations \citep{YuanWinter2023}. Our models follow the same physical logic: the neutrino-bright cases are allowed only if the high-energy photons are either emitted during a short observing window, shifted into the MeV band, or absorbed and reprocessed inside the source. The \ali models illustrate this particularly well, because they can be relevant for KM3NeT and Fermi/LAT while remaining below the CTAO benchmark.

This motivates a special role for the MeV band. Several TDE neutrino models predict that hidden or partially hidden hadronic activity should be accompanied by hard X-ray or soft \gr emission rather than by an easily detectable TeV counterpart \citep{Murase:2020lnu,YuanWinter2023}. COSI will operate in the soft \gr range and is designed to explore the MeV sky with improved sensitivity, spectroscopy, and sky coverage \citep{Tomsick:2023aue}. In the context of our models, a COSI detection would be especially valuable because it would probe the cascade and primary inverse-Compton components that sit between the thermal X-ray/UV emission and the GeV-to-TeV band. Conversely, a COSI non-detection during a nearby TDE would help rule out the more luminous hidden-cascade configurations, even if \fermi and CTAO do not detect the source.

Finally, CTAO and KM3NeT provide complementary tests. CTAO will improve transient very-high-energy \gr sensitivity and will be important for rapid follow-up of multi-messenger alerts \citep{Carosi2021CTA}. KM3NeT/ARCA will provide strong sensitivity to TeV-to-PeV neutrinos from the southern sky \citep{KM3NeT:2024arca}. The models presented here show that the best candidates for joint CTAO--KM3NeT detection are not necessarily the most luminous optical TDEs, but the ones in which a hard proton population coexists with a moderately transparent emission region. In contrast, sources with high internal opacity may be better targets for COSI and \fermi, even when their neutrino flux is high.

\section{Summary and conclusions}
\label{Sec: summary and conclusions}
We have explored the conditions under which TDEs can reveal their hidden non-thermal engines through MeV, GeV, and TeV \gr emission and high-energy neutrinos. Our goal was not to obtain a unique best-fitting model for each event, but to identify the physical regimes in which TDEs become promising targets for COSI, \fermi, CTAO, and KM3NeT. To this end, we applied a lepto-hadronic emission framework to three representative sources: \dsg, a neutrino-motivated TDE; \ehz, a prompt optical/UV outburst with high-energy detectability potential; and \ali, a radio--X-ray bright event with strong constraints on the low-frequency synchrotron component.

The main conclusions of this work are as follows.
\begin{enumerate}
\item \textit{TDEs become high-energy sources only in restricted regions of parameter space.}  The key parameters are the proton spectral index, the minimum Lorentz factor of the non-thermal protons, the fraction of dissipated power injected into non-thermal particles, the acceleration efficiency, the compactness of the target photon field, and the magnetic field strength. Hard proton spectra and/or large values of $\gamma_{p,\rm min}$ move more proton power above the pion-production threshold and therefore favour both multi-TeV neutrino production and hadronic \gr emission.

\item \textit{Neutrino-bright does not necessarily mean TeV-bright.}  Dense optical/UV/X-ray photon fields increase the efficiency of $p\gamma$ interactions and can enhance the neutrino luminosity. However, the same fields also increase the internal $\gamma\gamma$ opacity and can reprocess the associated \gr emission into lower-energy cascade radiation. The observational outcome therefore depends on the competition between proton cooling, particle escape, and photon attenuation. A TeV non-detection should not be interpreted automatically as evidence for a neutrino-quiet TDE.

\item \textit{\dsg provides the benchmark for a neutrino-motivated TDE.}  In the explored models, several different parameter combinations produce $\nu_\mu+\bar\nu_\mu$ spectra close to or above the KM3NeT/ARCA reference curve.  Models with hard protons, enhanced $\gamma_{p,\rm min}$, or efficient acceleration can also reach the \fermi and CTAO benchmark sensitivities. The leptonic comparison model remains GeV-bright but neutrino-quiet, showing that a GeV counterpart alone would not uniquely imply efficient hadronic acceleration.

\item \textit{\ehz is the clearest TeV-orientated source in our sample.}  The hadronic models of \ehz can become CTAO- and KM3NeT-relevant when the proton spectrum is hard or when the low-energy part of the proton distribution is suppressed. The comparison between cold-corona and no-corona cases highlights the central tension of TDE multimessenger emission: photon-rich environments favour neutrino production, while more transparent environments favour escaping GeV-to-TeV photons.

\item \textit{\ali probes a different regime.}  In contrast to \dsg and \ehz, \ali is most naturally bright in the MeV-to-GeV band among the compact shock models explored here. Its reference Model~A is relevant for COSI, \fermi, and KM3NeT, while the source remains at most marginal for CTAO. The radio data are broadly consistent with primary-electron synchrotron emission, but they do not uniquely determine the emission channel. In magnetically enhanced models, proton synchrotron and synchrotron emission from secondary pairs can also contribute at radio-to-infrared frequencies.

\item \textit{The MeV band is a key diagnostic of hidden hadronic activity.}  If TeV photons are absorbed inside the source, a fraction of the hadronic power can emerge through electromagnetic cascades at hard-X-ray and MeV energies. COSI will therefore probe a regime that is complementary to \fermi and CTAO. In this sense, the absence of a TeV signal and the presence of an MeV excess would be a powerful clue that the source is partially hidden rather than intrinsically weak.

\item \textit{Benchmark proximity is not a detection claim.}  We quantified the models using the maximum ratio between the predicted flux and the reference sensitivity curve in each band. This provides a compact way to compare models, but it does not replace a time-dependent detectability calculation. For transient sources, the relevant quantities are the active duration, observing cadence, background level, source position, and, for neutrinos, the flavour-dependent effective area and expected event number.

\end{enumerate}

Overall, our results suggest that TDEs can appear in several distinct multimessenger regimes. Some are transparent enough to shine in GeV--TeV \grs and neutrinos; some are hidden enough to produce neutrinos and MeV cascades without a strong TeV counterpart; and some are predominantly leptonic, producing GeV emission with little associated neutrino flux.  The same source class can therefore look very different to COSI, \fermi, CTAO, KM3NeT, and radio telescopes, depending on where and when the non-thermal particles are accelerated. Further studies will be essential for determining whether TDEs are rare, spectacular multimessenger engines, a subdominant contribution to the diffuse neutrino sky, or both.

\begin{acknowledgements}
We are grateful for the very fruitful comments of the anonymous reviewer. 
D.K. acknowledges support by Tamkeen under the NYUAD Research Institute grant CG014. 
D.A.L. and I.L. were funded by the European Union ERC-2022-STG - BOOTES - 101076343. Views and opinions expressed are however those of the author(s) only and do not necessarily reflect those of the European Union or the European Research Council Executive Agency. Neither the European Union nor the granting authority can be held responsible for them. This research was partially funded by the Deutsche Forschungsgemeinschaft (DFG, German Research Foundation) as part of the DFG Research Unit FOR5195 – project number 443220636.

\end{acknowledgements}

\bibliographystyle{aa_url} % use this to include links to the refs
\bibliography{tdejet} % your references Yourfile.bib

\begin{appendix}

\section{Detailed radiative-component decompositions}
\label{App: model A detailed seds}
This appendix shows the full component decompositions of the reference Model~A used for each source. These figures are not separate fits; they expose the individual radiative channels that are grouped or omitted in the simplified SED panels shown in the main text. In all panels, the thick black curve shows the total model spectrum. The primary electron synchrotron and inverse-Compton components are shown by green and blue dashed curves, respectively. The $pp$ channel is represented by the red dashed $\pi^0$-decay component and by the red dotted synchrotron emission from secondary pairs. The $p\gamma$ channel is represented by the purple dashed $\pi^0$-decay component and by the purple dotted synchrotron emission from secondary pairs. When visible, proton synchrotron radiation is shown by a magenta dash-dotted curve. The thermal disc, black-body, and coronal photon fields are shown by grey dashed, red dash-dotted, and grey dotted curves, respectively.

\subsection{\dsg}
\label{App: AT2019dsg detailed sed}
\begin{figure}[!ht]
    \centering
    \includegraphics[width=\columnwidth]{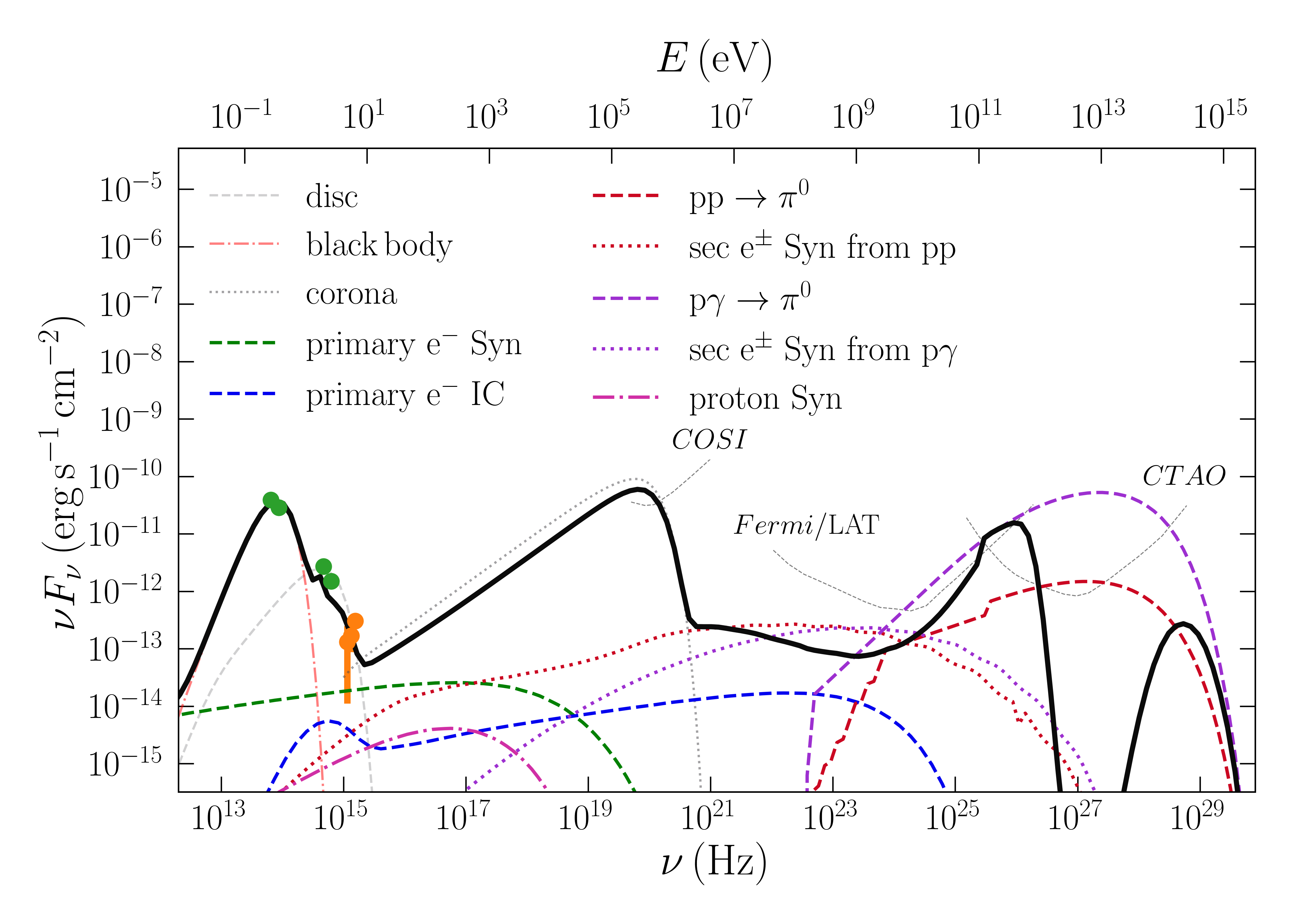}
    \caption{Full radiative-component decomposition of the reference \dsg Model~A.  The line styles identify the primary leptonic emission, $pp$ components, $p\gamma$ components, proton synchrotron radiation, and the thermal/coronal photon fields.}
    \label{fig: at2019dsg modelA all components}
\end{figure}

Figure~\ref{fig: at2019dsg modelA all components} shows the full decomposition of the reference \dsg Model~A. The high-energy emission is dominated by the photohadronic $p\gamma\rightarrow\pi^0$ component in the CTAO band, while the $pp$ channel and secondary-pair synchrotron emission remain subdominant.  The decomposition therefore clarifies the connection between the TeV $\gamma$-ray component and the multi-TeV neutrino spectrum discussed in the main text.

\subsection{\ehz}
\label{App: AT2019ehz detailed sed}
\begin{figure}[!ht]
    \centering
    \includegraphics[width=\columnwidth]{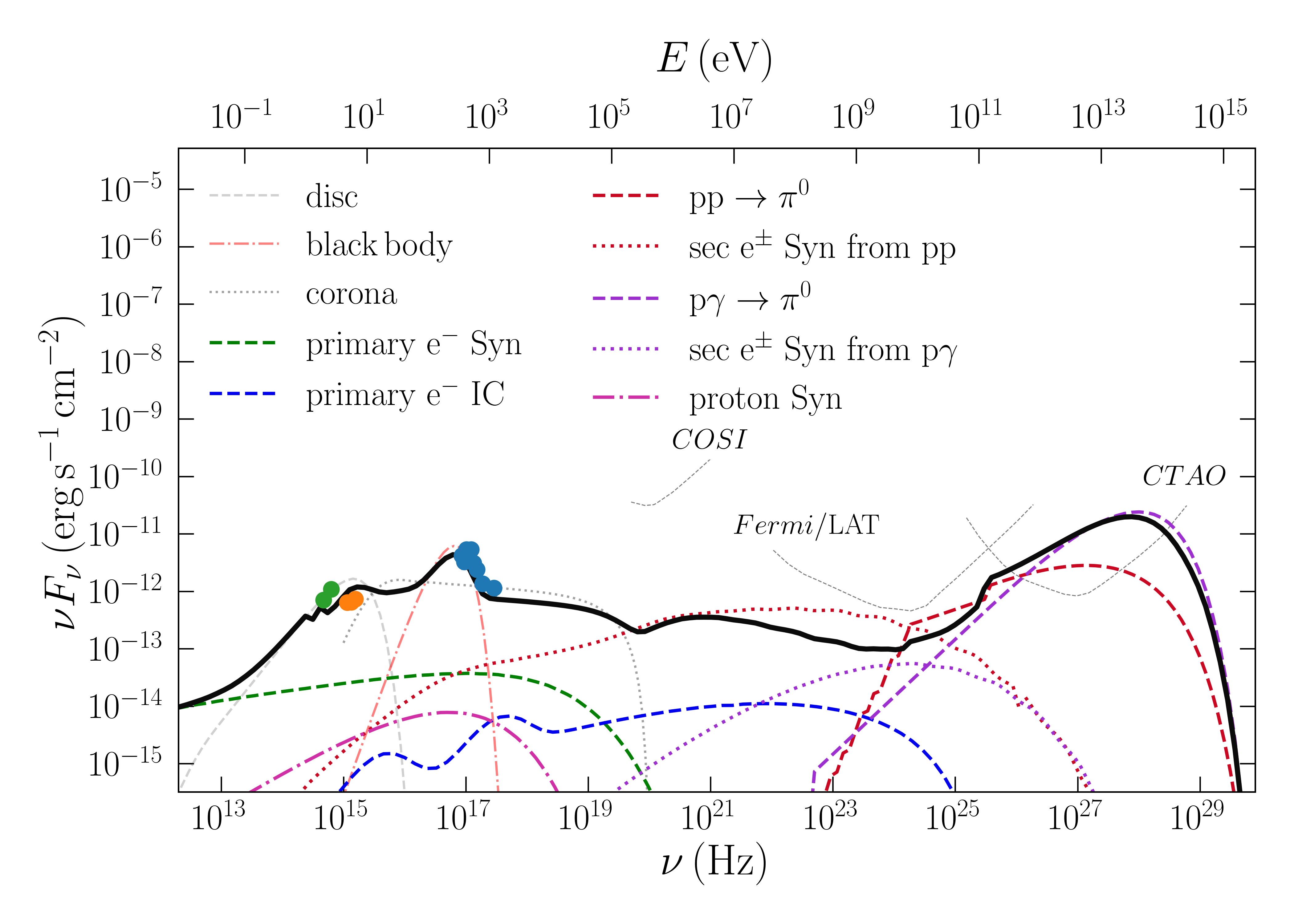}
    \caption{Full radiative-component decomposition of the reference \ehz Model~A.  The line styles identify the primary leptonic emission, $pp$ components, $p\gamma$ components, proton synchrotron radiation, and the thermal/coronal photon fields.}
    \label{fig: at2019ehz modelA all components}
\end{figure}

Figure~\ref{fig: at2019ehz modelA all components} shows the full decomposition of the reference \ehz Model~A.  The cold corona supplies an additional target photon field for photohadronic interactions and also contributes to the internal opacity.  The CTAO-band emission is mainly associated with $p\gamma\rightarrow\pi^0$ decay, while the secondary-pair components trace the electromagnetic cascade initiated by hadronic interactions and photon--photon absorption.

\subsection{\ali}
\label{App: ASASSN14li detailed sed}
\begin{figure}[!ht]
    \centering
    \includegraphics[width=\columnwidth]{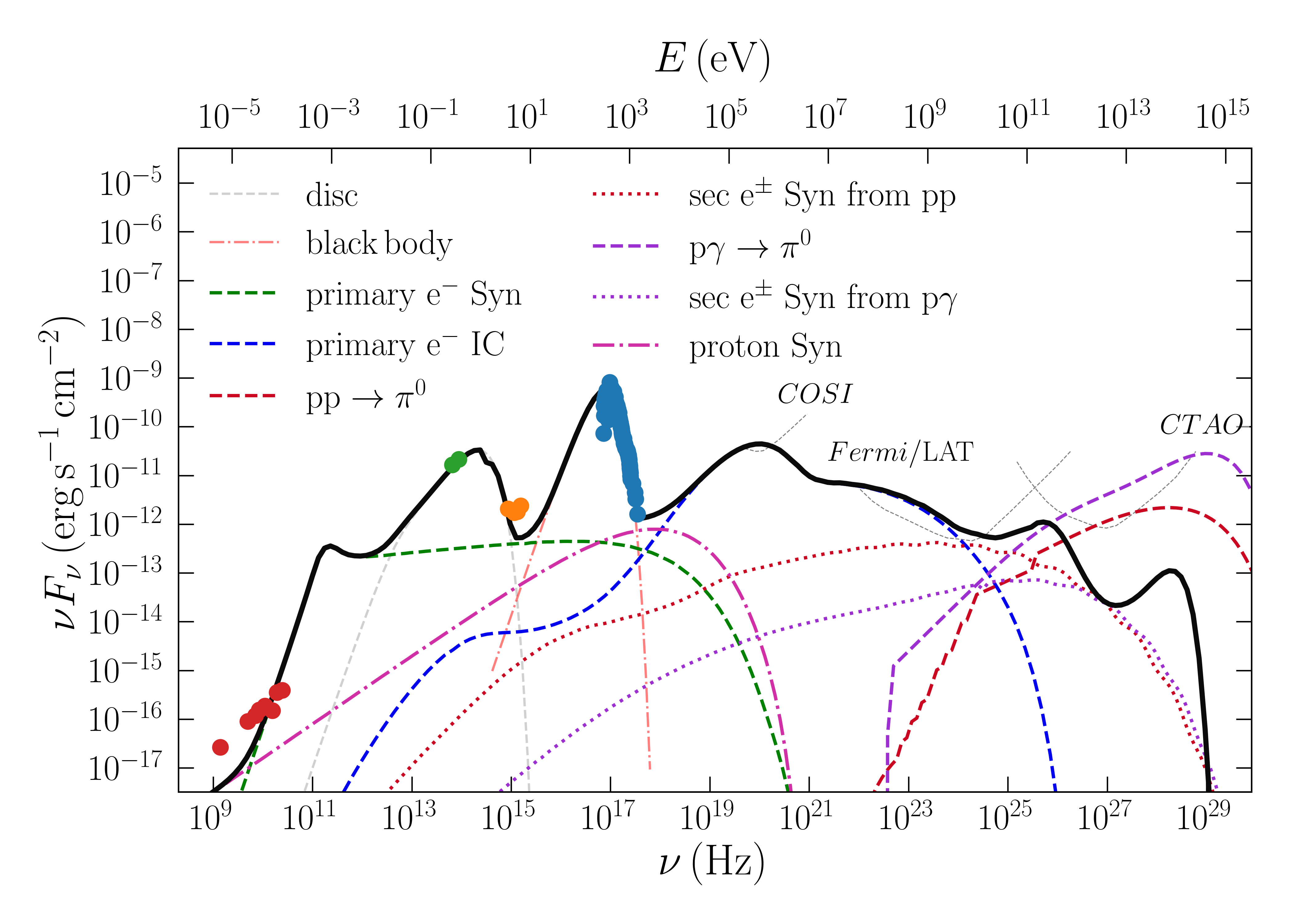}
    \caption{Full radiative-component decomposition of the reference \ali Model~A. The line styles identify the primary leptonic emission, $pp$ components, $p\gamma$ components, secondary-pair synchrotron radiation, and the thermal photon fields.}
    \label{fig: asassn14li modelA all components}
\end{figure}

Figure~\ref{fig: asassn14li modelA all components} shows the full decomposition of the reference \ali Model~A.  This source differs from \dsg and \ehz because the neutrino-bright configurations remain comparatively faint in the CTAO band.  The decomposition is useful for separating the primary inverse-Compton component, which shapes the MeV--GeV emission, from the hadronic components that contribute to the high-energy tail and to the neutrino output.  It also illustrates how internal photon--photon absorption and subsequent cascading can redistribute power away from the TeV band.

\section{Extended analysis of all models of \dsg, \ehz and \ali}
\label{app: section of all models of all sources}
\subsection{\dsg}
\subsubsection{High-energy spectral energy distributions}
\label{app: subsection for seds for all dsg models}
\begin{figure*}[!ht]
    \centering
        \includegraphics[width=0.48\textwidth]{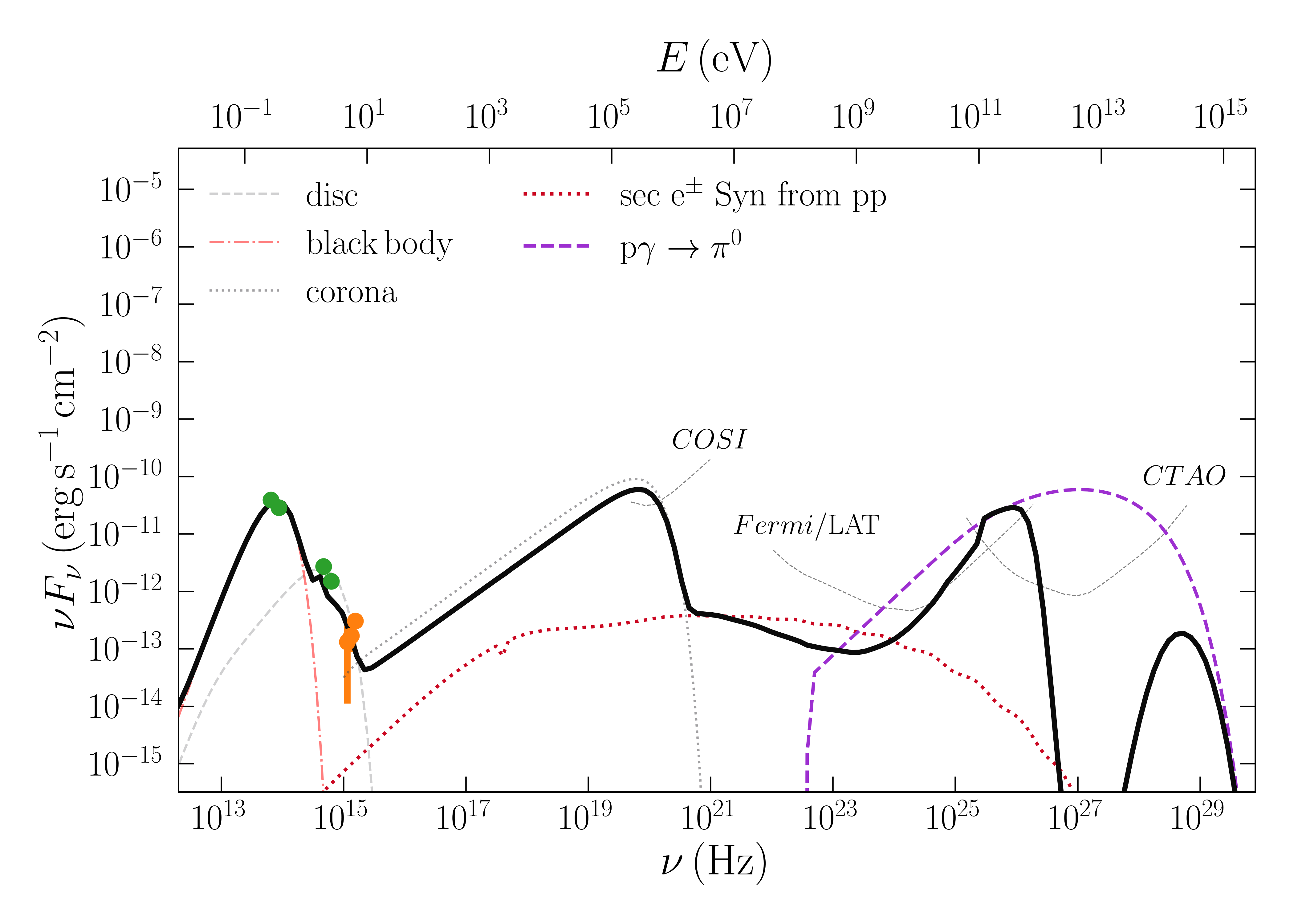}
        \includegraphics[width=0.48\textwidth]{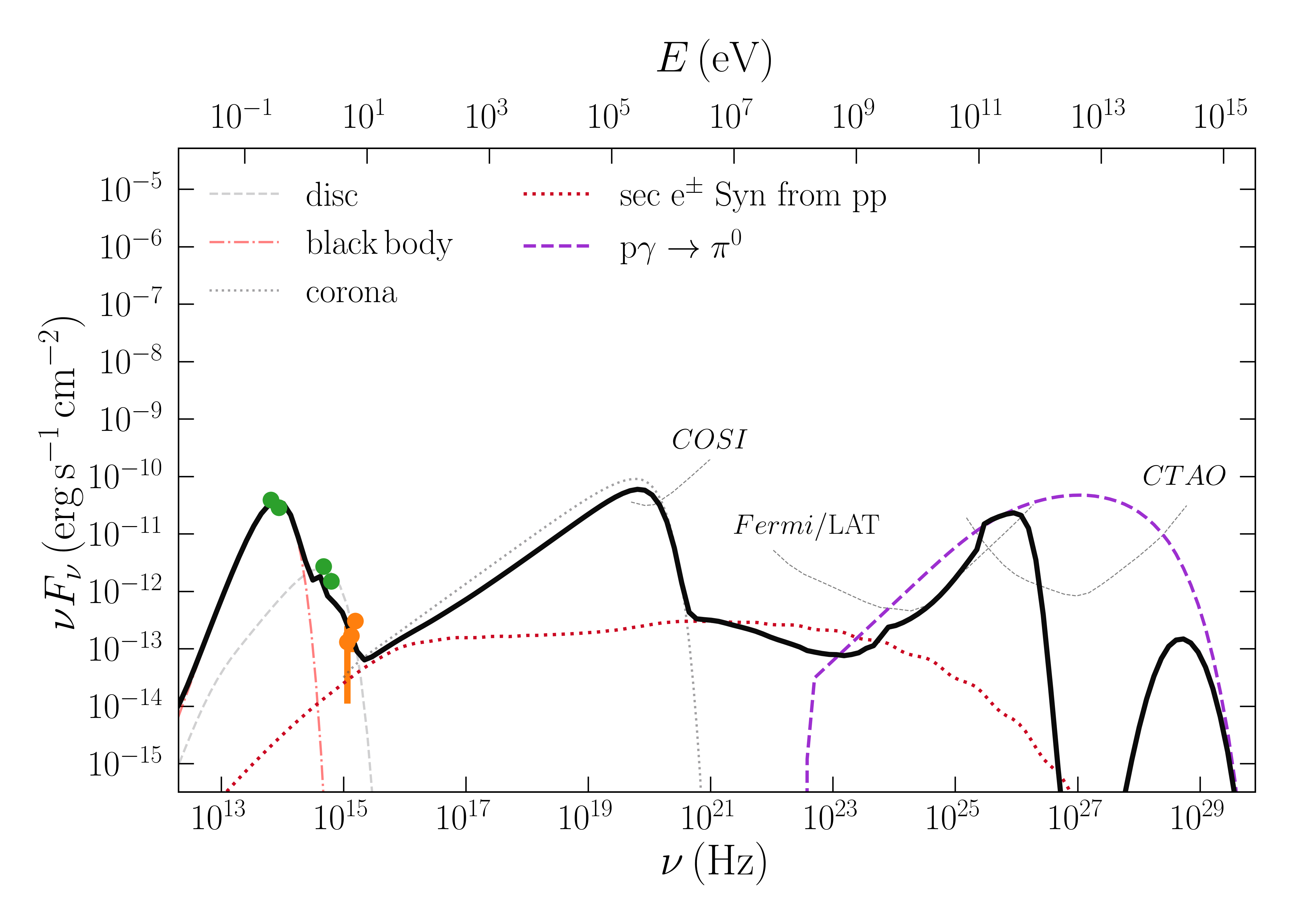}
        \includegraphics[width=0.48\textwidth]{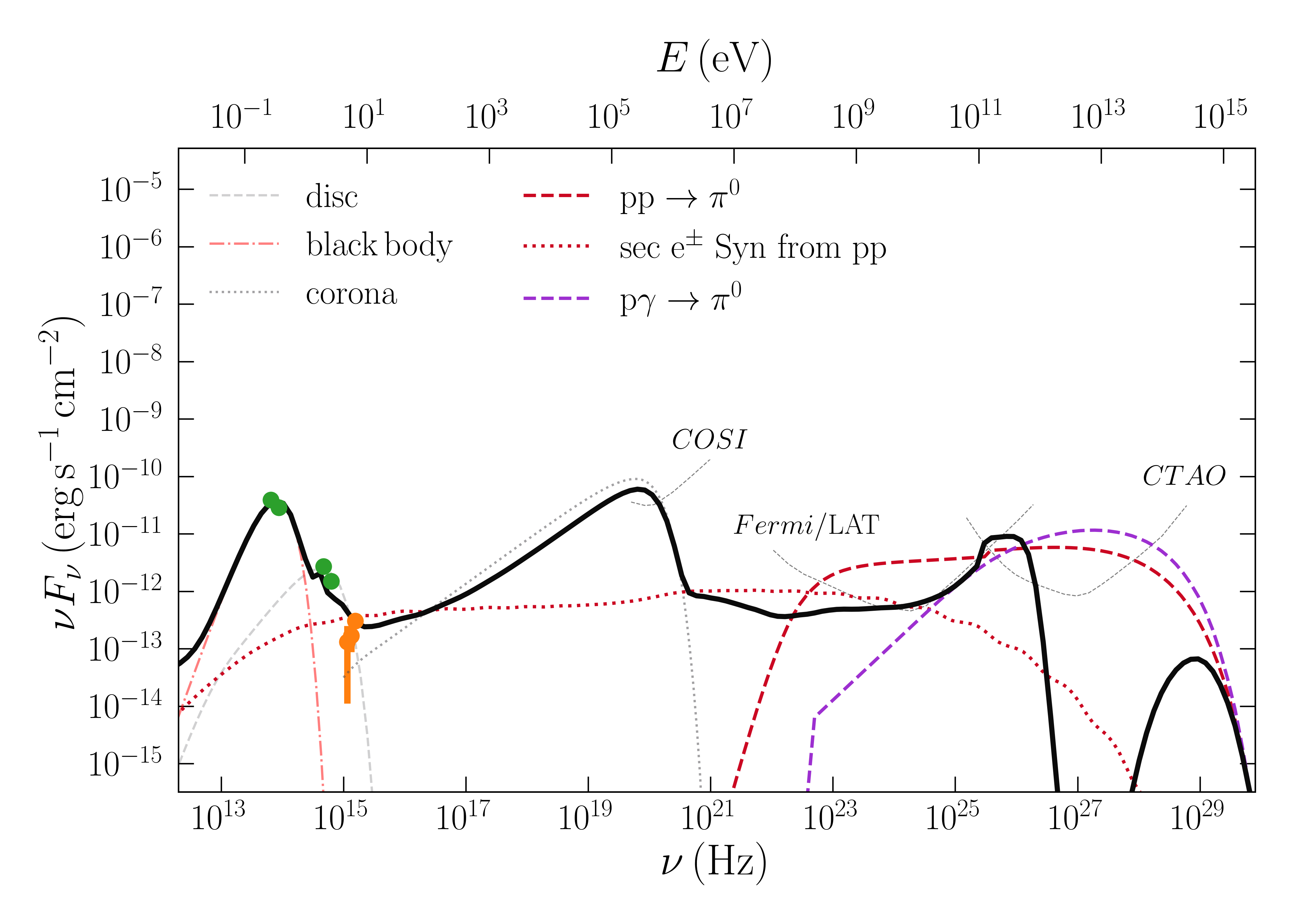}
        \includegraphics[width=0.48\textwidth]{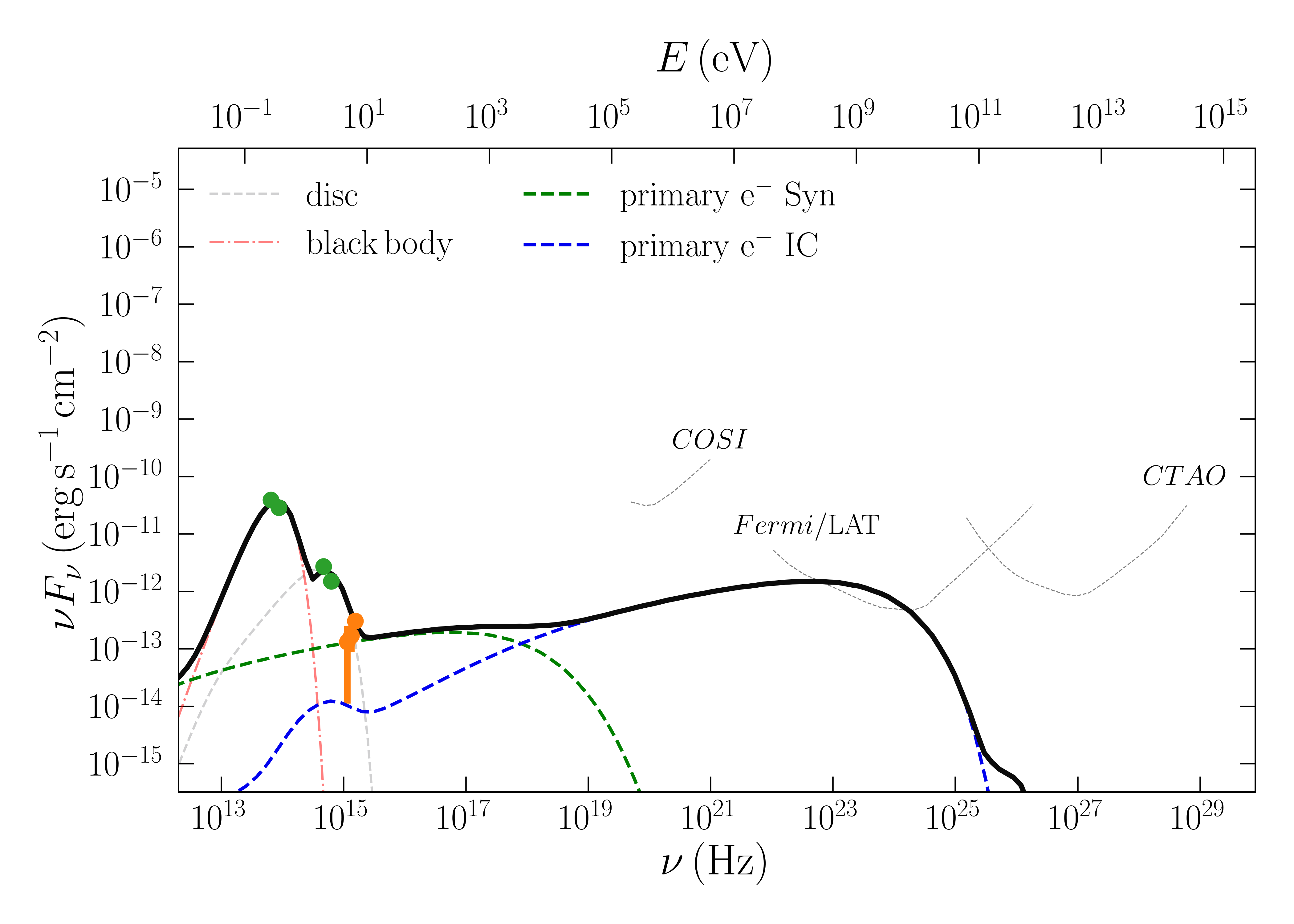}
    \caption{Spectral energy distributions of the four exploratory Models~B--E for \dsg. The thick black line shows the total model emission after accounting for absorption. The thermal disc, black-body, and coronal components are shown by dashed grey, dash-dotted red, and dotted grey curves, respectively. Hadronic \gr components are shown by red and purple curves: dashed red for $pp\rightarrow\pi^0$, purple dashed for $p\gamma\rightarrow\pi^0$, and dotted curves for synchrotron radiation from secondary pairs. In Model~E the green and blue dashed curves show the synchrotron and inverse-Compton emission of primary electrons. The thin grey curves indicate the COSI \citep{Tomsick:2023aue}, Fermi/LAT, and CTAO reference sensitivities.}
    \label{fig: appendix at2019dsg seds}
\end{figure*}

Figure~\ref{fig: appendix at2019dsg seds} shows the SEDs of the four \dsg models (B--E). In each panel, the thick black line is the total model spectrum. The thermal emission from the disc is shown by a grey dashed curve, the black-body component by a red dash-dotted curve, and the coronal emission, when present, by a grey dotted curve. The thin grey curves mark the COSI, Fermi/LAT, and CTAO reference sensitivities indicated in the panels. The remaining curves show the high-energy components that dominate each model: neutral-pion decay from $pp$ interactions is shown by a red dashed line, neutral-pion decay from $p\gamma$ interactions by a purple dashed line, and synchrotron emission from secondary pairs by dotted curves. In the leptonic Model~E, the green dashed and blue dashed curves show the primary electron synchrotron and inverse-Compton components, respectively.

Model~B demonstrates that, for a softer proton spectrum with $p_{\rm p}=2$, a much larger $\gamma_{p,\rm min}$ is needed to concentrate the non-thermal proton power near energies relevant for pion production. In this case, both $pp$ and $p\gamma$ channels contribute in the CTAO range, but the requirement of a high low-energy cutoff makes the model less natural than Model~A.  Model~C reaches a similar goal by increasing the fraction of power placed into non-thermal protons.  This boosts the secondary synchrotron and pion-decay components while keeping $\gamma_{p,\rm min}$ lower than in Model~B. Model~D shows that even a low $\gamma_{p,\rm min}$ can yield an observable high-energy component if the acceleration is more efficient and the proton distribution extends to the photohadronic regime. Finally, Model~E illustrates a different detectability regime: the GeV emission is produced mainly by the primary electron inverse-Compton component, while the hadronic and neutrino outputs are negligible.  This model is therefore relevant for Fermi/LAT searches but is not expected to be a promising KM3NeT source.

\subsubsection{Muon-neutrino predictions}
\label{app: Sec: neutrinos for all dsg models}
\begin{figure}[!ht]
    \centering
        \includegraphics[width=0.48\textwidth]{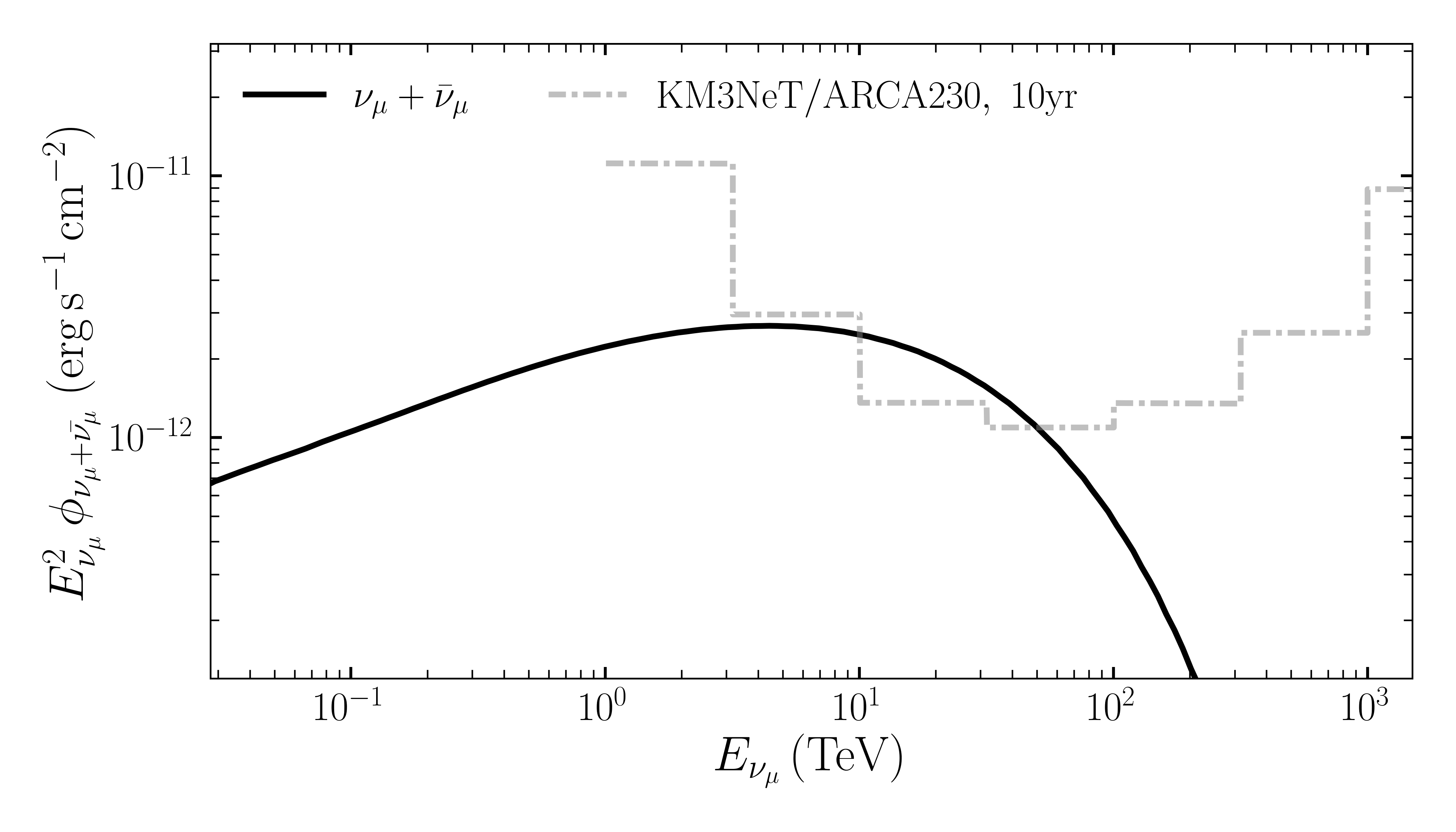}
    \vspace{-0.2cm}
        \includegraphics[width=0.48\textwidth]{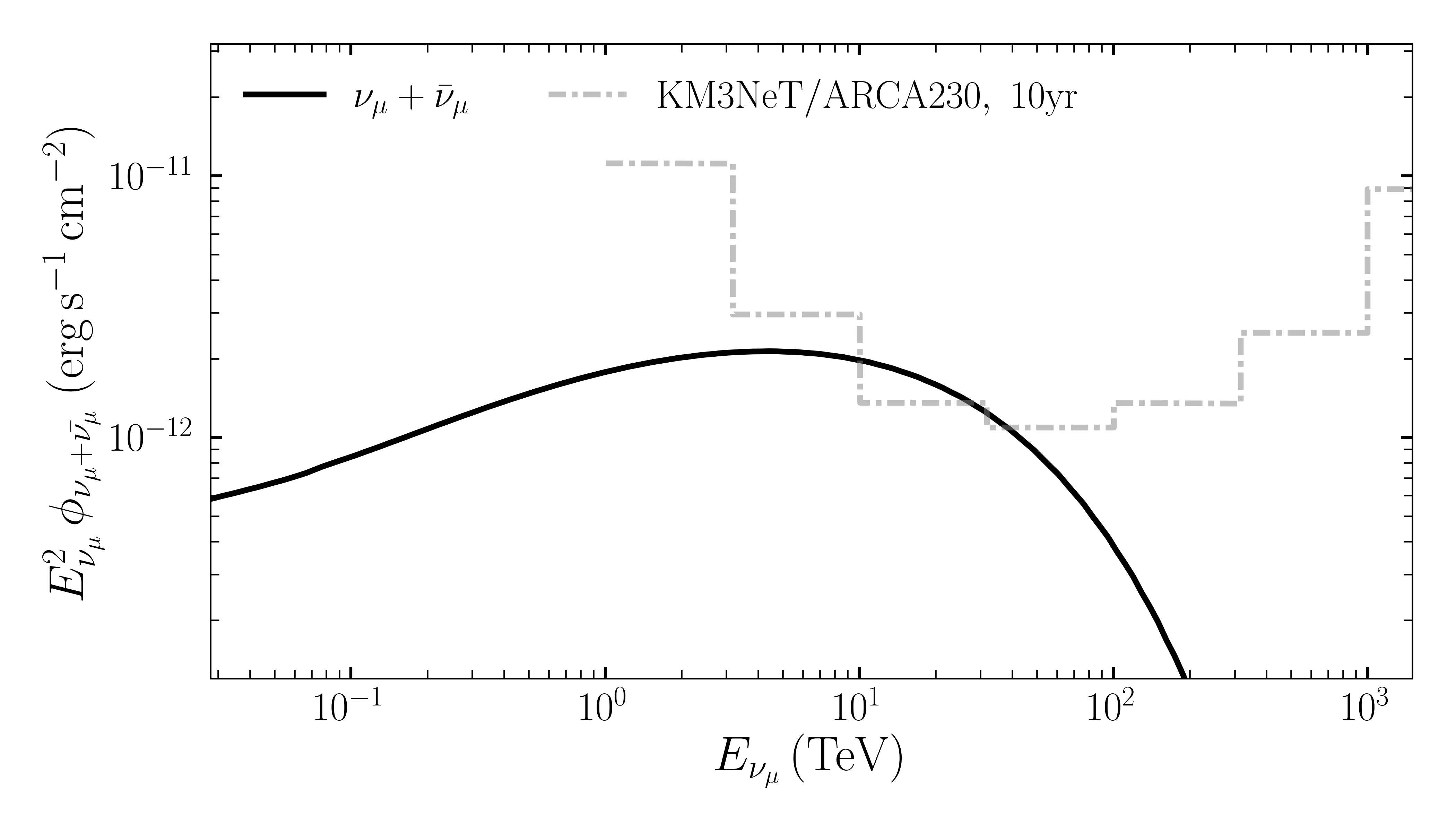}
    \vspace{-0.2cm}
        \includegraphics[width=0.48\textwidth]{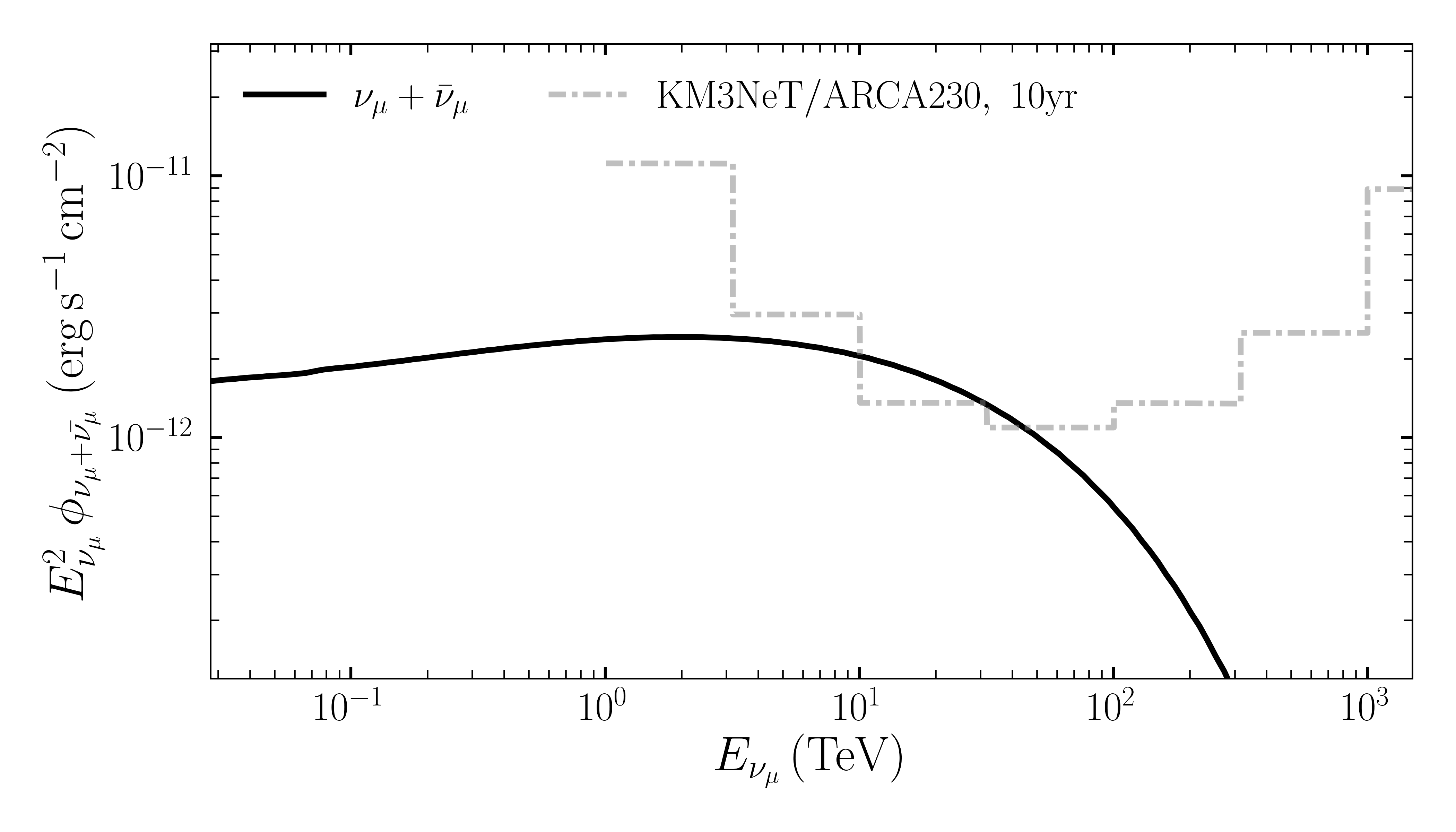}
    \caption{Muon-neutrino plus antineutrino spectra predicted for Models~B--D of \dsg. The black curves show $E_{\nu_\mu}^2\phi_{\nu_\mu+\bar{\nu}_\mu}$, and the grey dash-dotted curves show the nominal ten-year KM3NeT/ARCA230 point-source sensitivity \citep{KM3NeT:2024arca}. Model~E is not shown because its neutrino flux is negligible on this scale.}
    \label{fig: appendix at2019dsg neutrinos}
\end{figure}

The corresponding $\nu_\mu+\bar{\nu}_\mu$ spectra are shown in Fig.~\ref{fig: appendix at2019dsg neutrinos}.  The plotted quantity is $E_{\nu_\mu}^2\phi_{\nu_\mu+\bar{\nu}_\mu}$, allowing a direct comparison with the KM3NeT/ARCA230 reference sensitivity. Model~B reaches a similar level to Model~A only after increasing $\gamma_{p,\rm min}$ to 200, emphasising that soft proton spectra are less efficient at producing a detectable TeV neutrino signal unless the proton power is large. Model~C instead boosts the signal through the large non-thermal proton fraction, while Model~D shows that efficient acceleration can partly compensate for the smaller $\gamma_{p,\rm min}$. The neutrino flux of Model~E is far below the range displayed in Fig.~\ref{fig: appendix at2019dsg neutrinos}, consistent with its interpretation as a GeV \grs, rather than neutrino, scenario.

The comparison with the ARCA230 curve should be interpreted conservatively.  The grey line is a steady point-source sensitivity for a ten-year exposure, whereas the TDE emission is expected to last for a finite flare duration.  Therefore, a model curve lying near the sensitivity is not by itself a detection claim.  Instead, the figure identifies which spectral shapes and flux levels are worth evaluating with the full event-count calculation, including the source declination, active time window, and flavour-dependent effective area.

\subsubsection{Particle distributions and maximum energies in the reference model}
\label{app: Sec: AT2019dsg model A diagnostics}
\begin{figure*}[!ht]
    \centering
    \subfigure[Electron distribution.]{
        \includegraphics[width=0.48\textwidth]{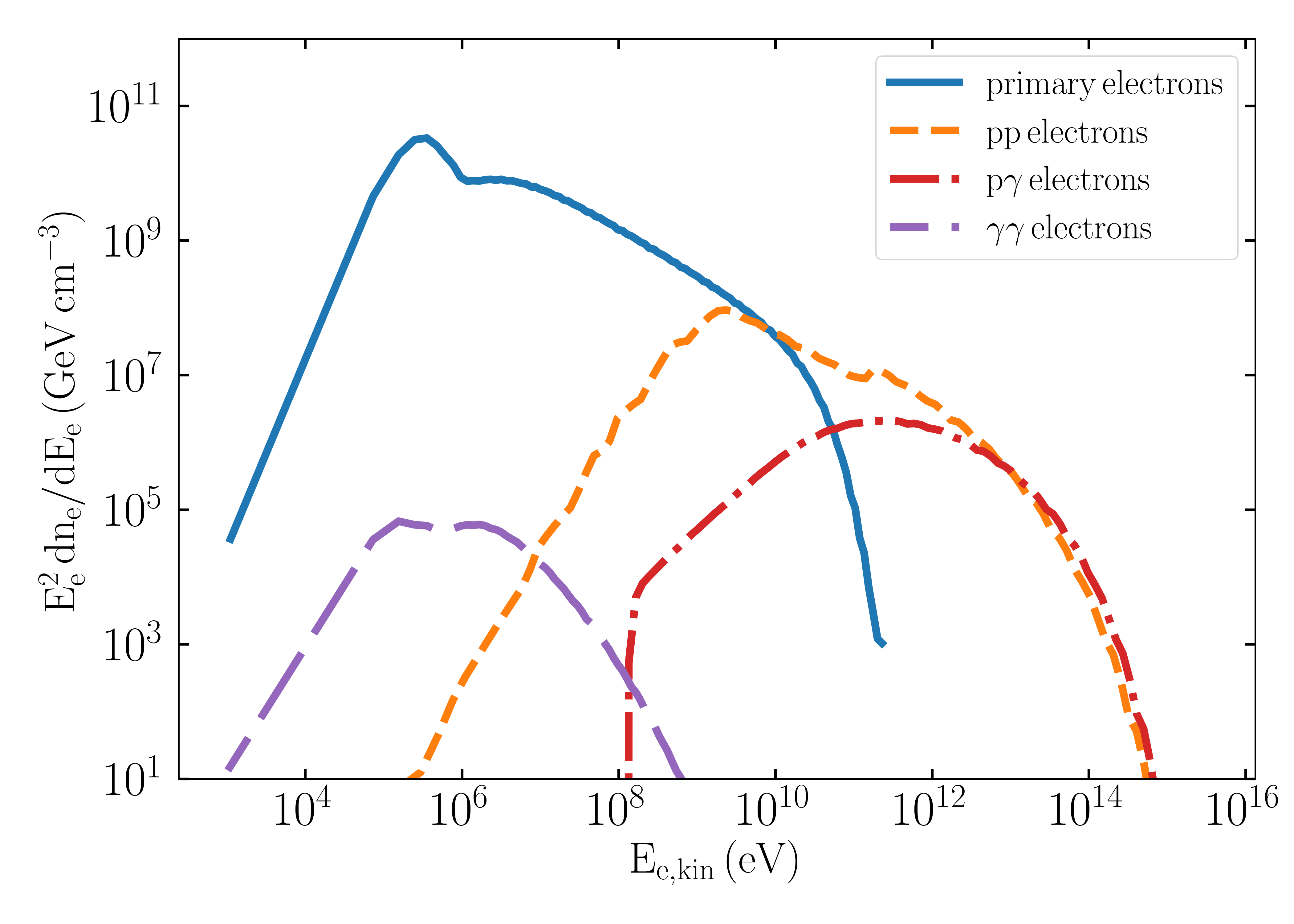}}
    \subfigure[Proton distribution.]{
        \includegraphics[width=0.48\textwidth]{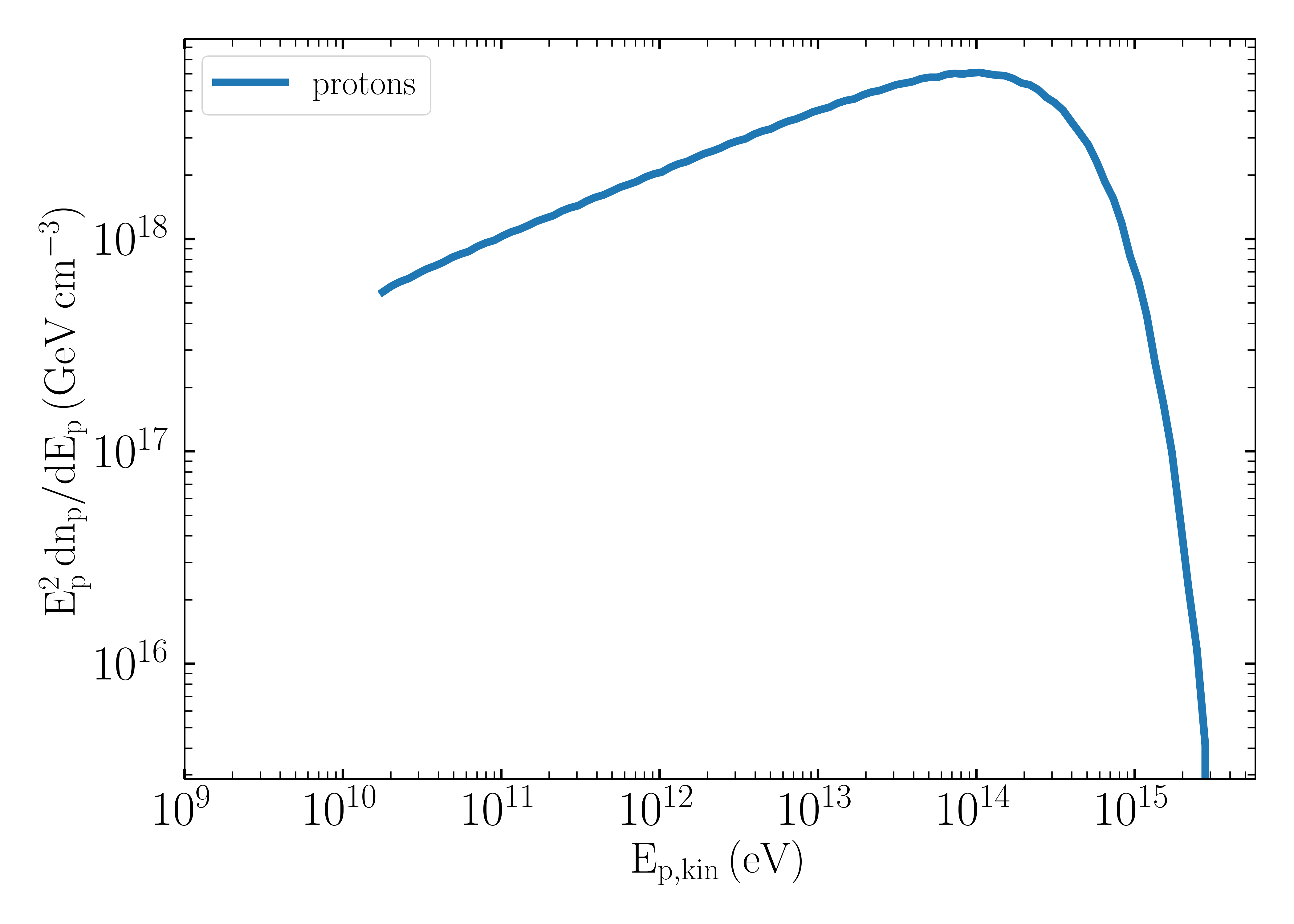}}
    \vspace{0.2cm}
    \subfigure[Electron timescales.]{
        \includegraphics[width=0.48\textwidth]{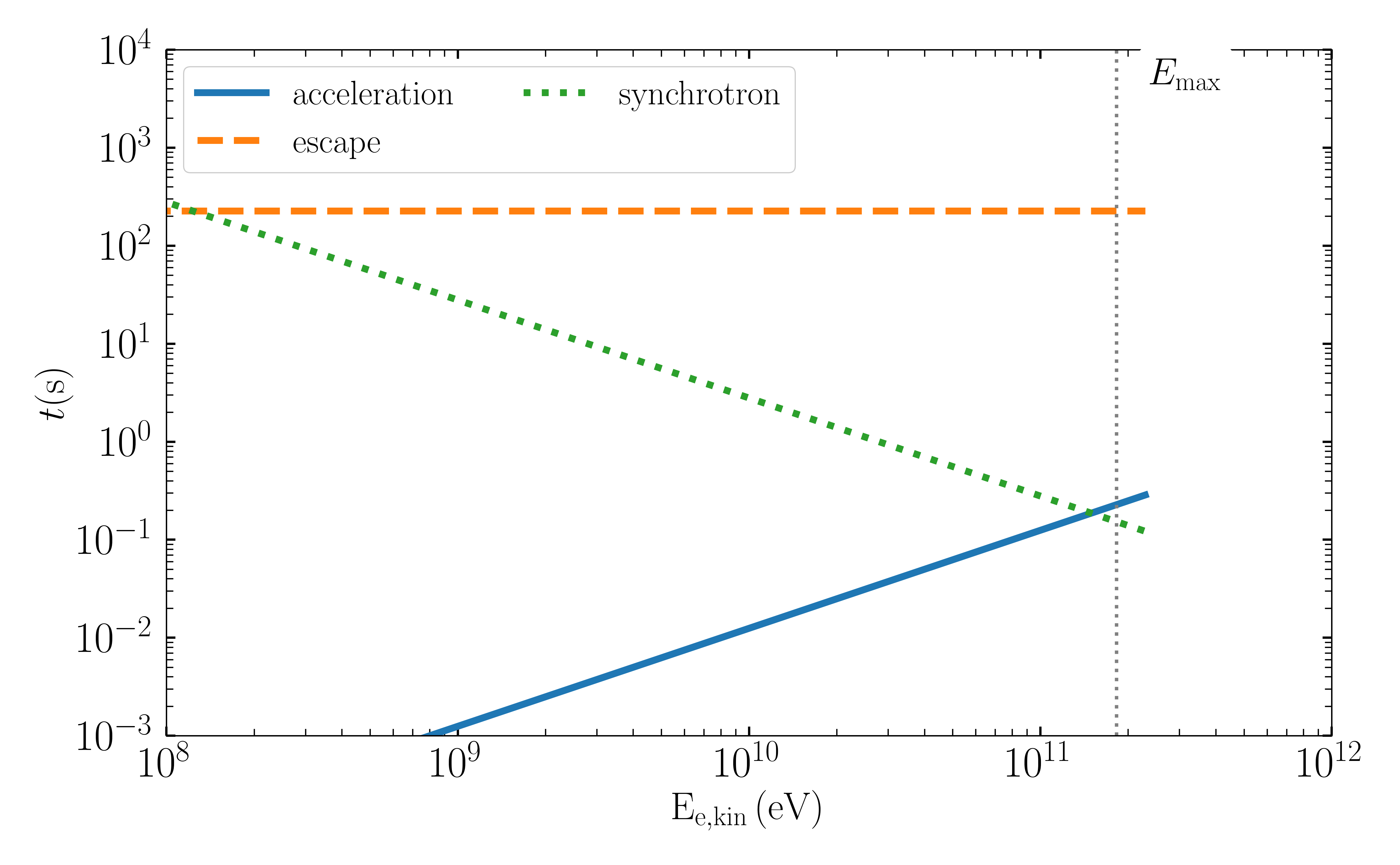}}
    \subfigure[Proton timescales.]{
        \includegraphics[width=0.48\textwidth]{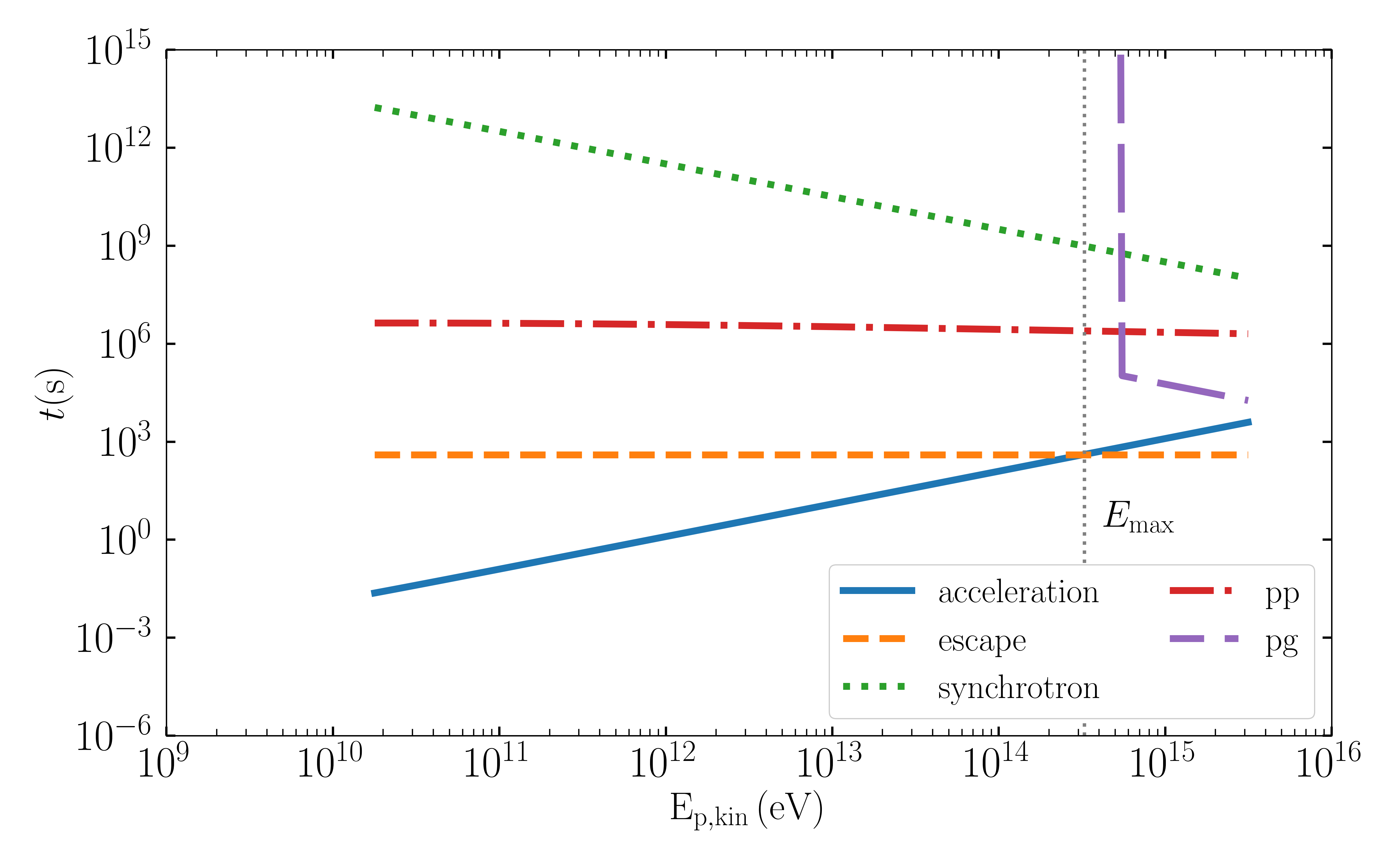}}
    \caption{Particle distributions and acceleration/cooling timescales for the reference \dsg Model~A.  The top panels show $E_i^2\,dn_i/dE_i$ for electrons and protons.  Primary electrons are shown in blue, secondary pairs from $pp$ interactions in orange, secondary pairs from $p\gamma$ interactions in red, and pairs from $\gamma\gamma$ absorption in purple.  The bottom panels show the acceleration, escape, synchrotron, $pp$, and $p\gamma$ timescales, as labelled in the panels.  The vertical grey dotted line marks the adopted maximum energy.}
    \label{fig: appendix at2019dsg modelA diagnostics}
\end{figure*}

To illustrate how the high-energy emission is produced, Fig.~\ref{fig: appendix at2019dsg modelA diagnostics} shows the particle distributions and acceleration/cooling timescales for Model~A. The electron distribution is dominated by primary electrons at low kinetic energies. Secondary pairs produced in $pp$ interactions populate the GeV-to-TeV kinetic-energy range, while the $p\gamma$ channel contributes at higher energies once the proton distribution crosses the photomeson threshold. The $\gamma\gamma$ pairs form at lower electron energies and trace the internal absorption of high-energy photons. The proton distribution is broad and hard, as expected for $p_{\rm p}=1.7$, and cuts off in the PeV range.

The timescale panels show why this model can generate both TeV gamma rays and neutrinos. For electrons, the maximum energy is determined by the balance between acceleration and synchrotron cooling; escape is slower and therefore does not set the electron cutoff. For protons, synchrotron and $pp$ cooling are inefficient over most of the relevant energy range. The maximum energy is instead controlled mainly by the competition between acceleration and escape, with $p\gamma$ losses becoming important only close to the highest energies.  Thus the model is not a full proton calorimeter, but it accelerates protons efficiently enough to enter the photohadronic regime before they escape. This explains why the same configuration can produce a CTAO-relevant \gr component and a neutrino spectrum close to the KM3NeT/ARCA benchmark.

The five models therefore map out three qualitatively different regimes.  First, hard proton spectra with moderate $\gamma_{p,\rm min}$ can place both the \gr and neutrino components near the sensitivity of CTAO and KM3NeT, as in Model~A. Second, soft proton spectra can reach comparable fluxes, but only by increasing either the $\gamma_{p,\rm min}$ or the fraction of non-thermal power injected into protons, as in Models~B and C. Third, GeV \gr detectability does not imply neutrino detectability: Model~E can approach the Fermi/LAT band through primary electron inverse-Compton emission while remaining neutrino quiet. This separation between GeV-bright and neutrino-bright configurations will be useful when comparing \dsg to \ehz below.

\subsection{AT2019ehz}
\label{app: Sec: AT2019ehz results}
\subsubsection{High-energy spectral energy distributions of \ehz}
\label{app: appendix of all five models of ehz}
\begin{figure*}[!ht]
    \centering
        \includegraphics[width=0.48\textwidth]{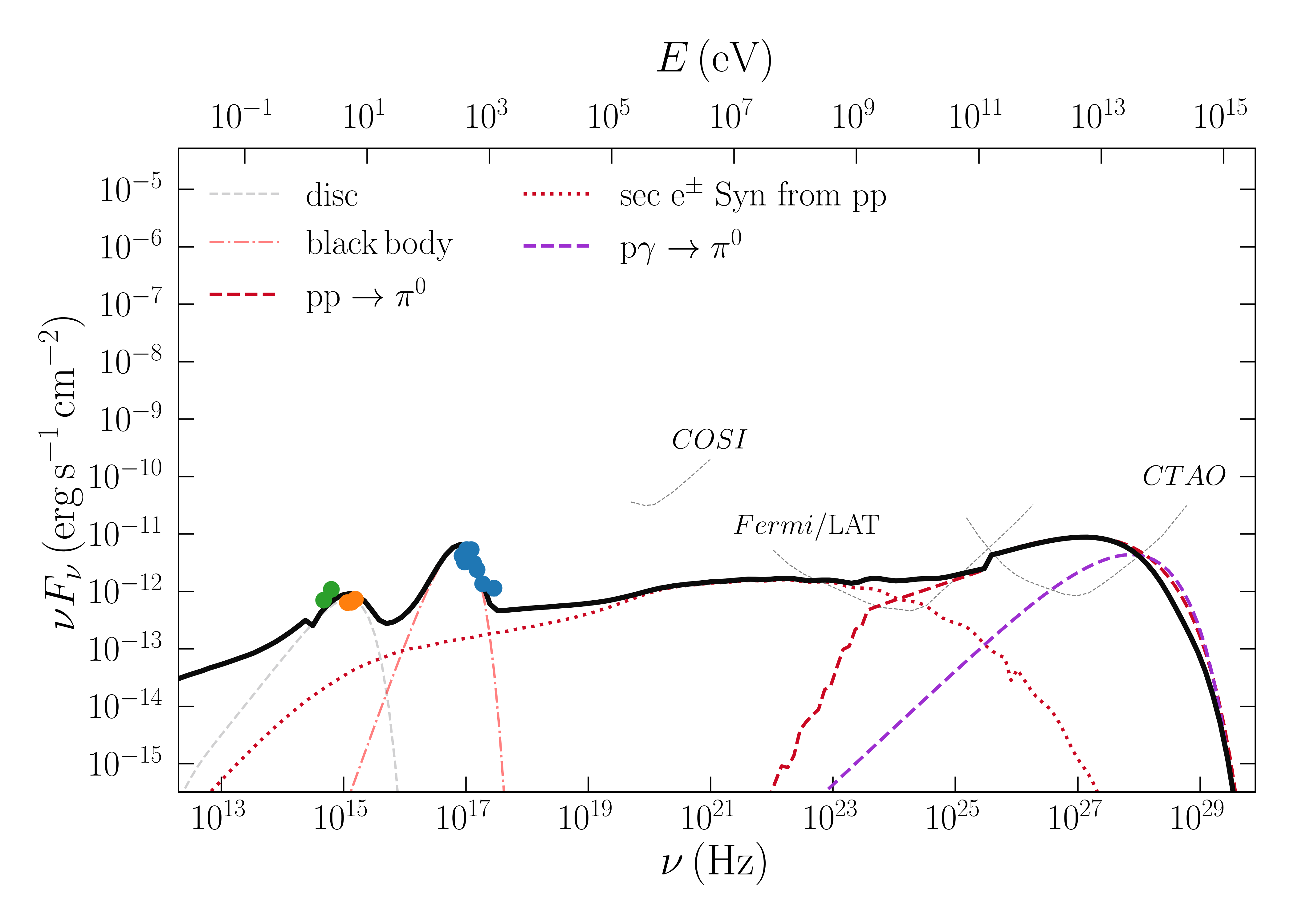}
    \vspace{0.2cm}
        \includegraphics[width=0.48\textwidth]{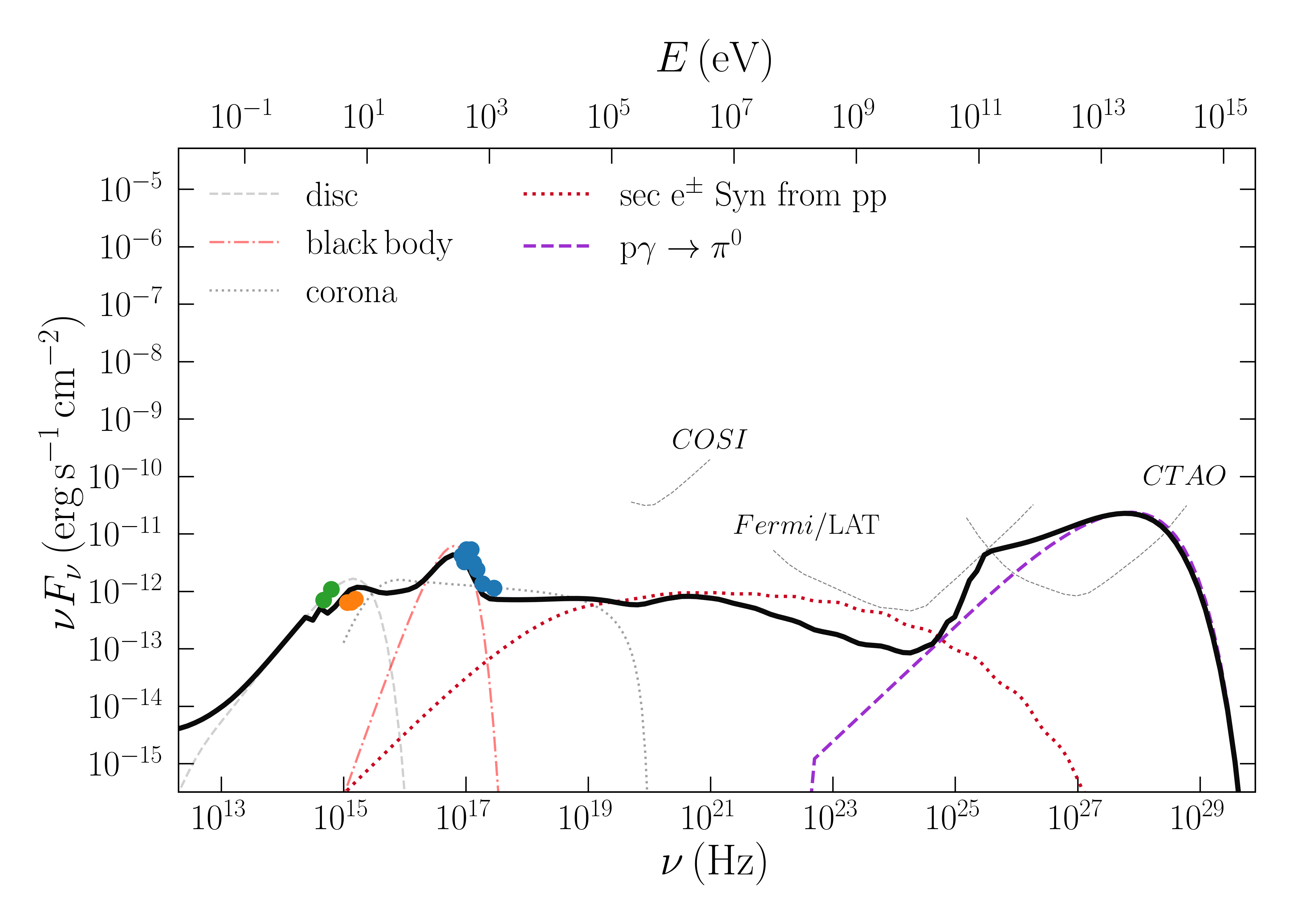}
        \includegraphics[width=0.48\textwidth]{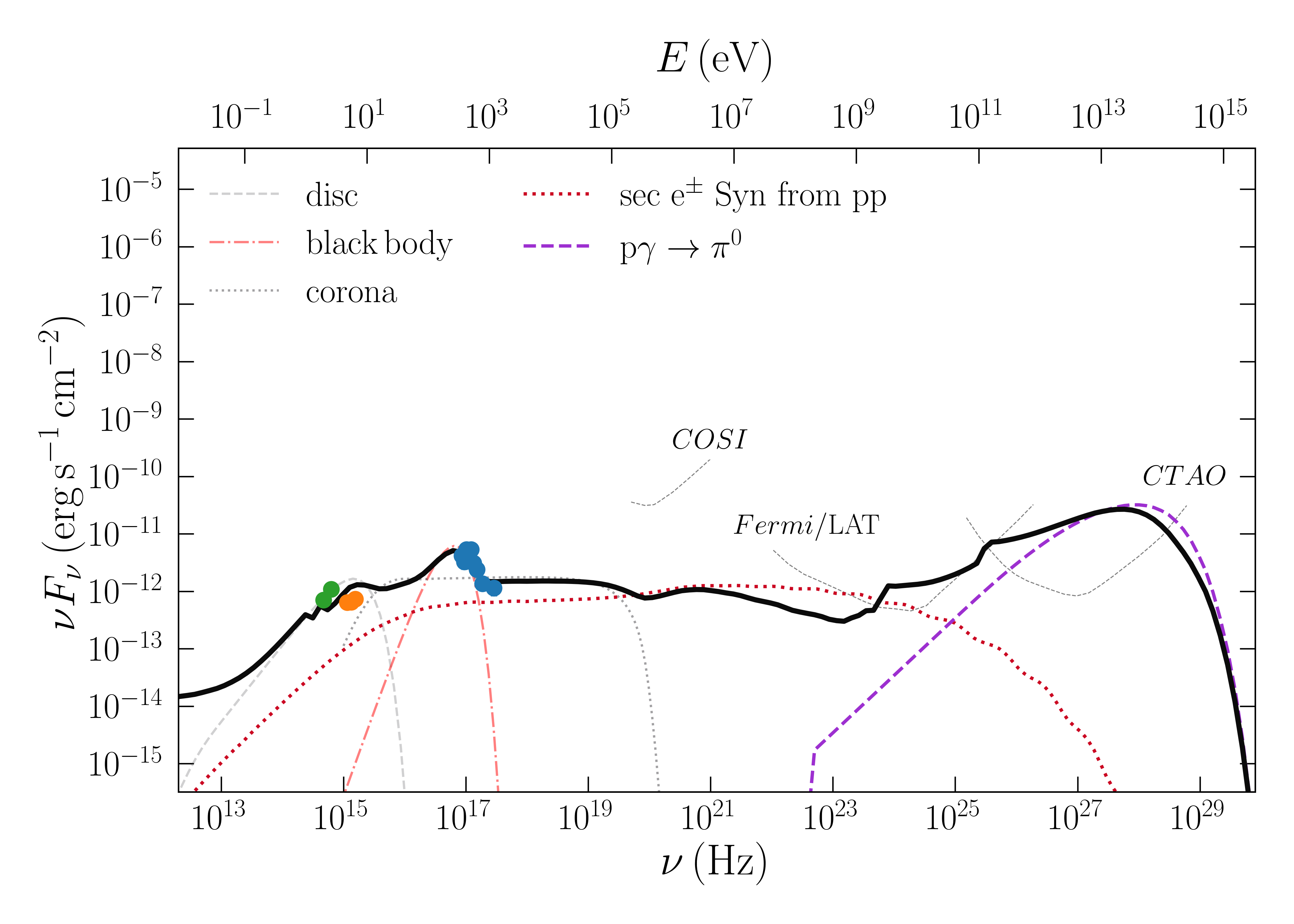}
    \vspace{0.2cm}
        \includegraphics[width=0.48\textwidth]{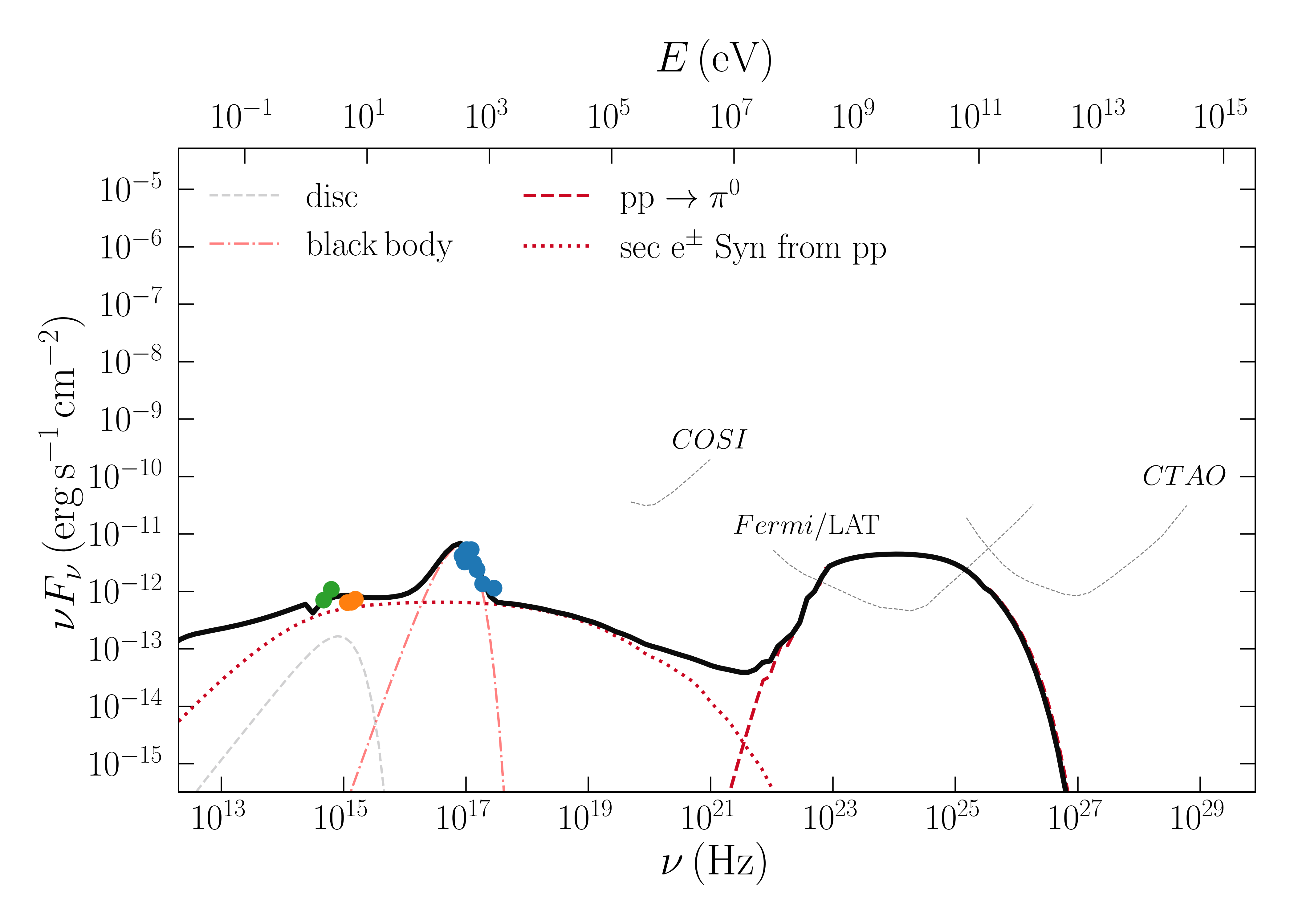}
    \caption{Spectral energy distributions of the four exploratory Models~B--E for \ehz. The thick black line shows the total model emission.  The thermal disc, black-body, and coronal components are shown by grey dashed, red dash-dotted, and grey dotted curves, respectively. Hadronic \gr components are shown by red and purple curves: red dashed for $pp\rightarrow\pi^0$, purple dashed for $p\gamma\rightarrow\pi^0$, and dotted curves for synchrotron radiation from secondary pairs.  The thin grey curves indicate the COSI \citep{Tomsick:2023aue}, Fermi/LAT, and CTAO reference sensitivities.}
    \label{fig: appendix at2019ehz seds}
\end{figure*}

The SEDs of the four \ehz models (B--E) are shown in Fig.~\ref{fig: appendix at2019ehz seds}.  The line styles follow those used for \dsg. The thick black curve shows the total model spectrum, the grey dashed curve shows the thermal disc, the red dash-dotted curve shows the additional black-body component, and the grey dotted curve shows the corona when present. The hadronic components are shown by red and purple curves: red dashed for $pp\rightarrow\pi^0$, purple dashed for $p\gamma\rightarrow\pi^0$, and dotted curves for synchrotron radiation from secondary pairs. In panels where primary leptonic components are shown, the green dashed and blue dashed curves represent primary electron synchrotron and inverse-Compton emission. The thin grey curves indicate the COSI, Fermi/LAT, and CTAO reference sensitivities.

Model~B shows the effect of removing the corona while keeping a hard proton spectrum. The reduced coronal opacity allows a larger fraction of the high-energy photons to escape, and the combination of $\eta=4$ and $f_{\rm sc}=10^{-3}$ makes the model relevant across the Fermi/LAT and CTAO bands.  This model is therefore the most direct multimessenger showcase in the \ehz set, but we do not use it as the reference case because it relies on a more transparent environment. Model~C demonstrates that a soft proton spectrum can still produce CTAO-relevant emission if the low-energy part of the proton distribution is suppressed by increasing $\gamma_{p,\rm min}$ to 500. Model~D reaches the high-energy band through a different route: it keeps a soft proton spectrum but increases the non-thermal proton fraction to $\varepsilon_p=0.8$, boosting both the secondary-pair synchrotron and pion-decay components. Model~E is qualitatively different. Its high-energy emission is concentrated mainly in the Fermi/LAT range and the infrared band receives a visible contribution from synchrotron radiation of secondary pairs produced in $pp$ interactions.  This case therefore illustrates that a GeV-detectable TDE does not necessarily have to be a strong TeV or neutrino source.

\subsubsection{Muon-neutrino predictions}
\label{app: Sec: AT2019ehz neutrinos}

\begin{figure*}[!ht]
    \centering
        \includegraphics[width=0.32\textwidth]{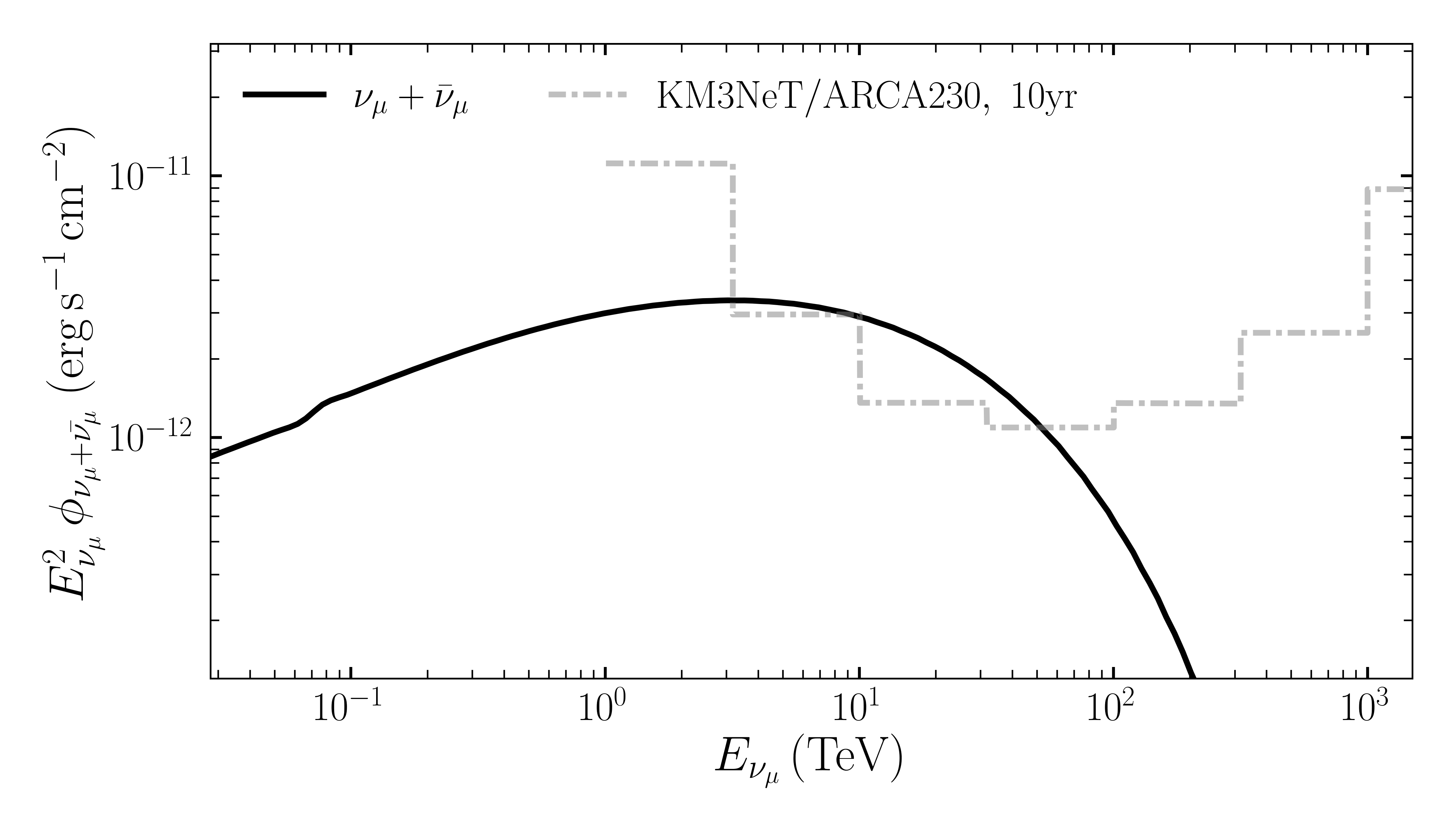}
    \vspace{0.2cm}
        \includegraphics[width=0.32\textwidth]{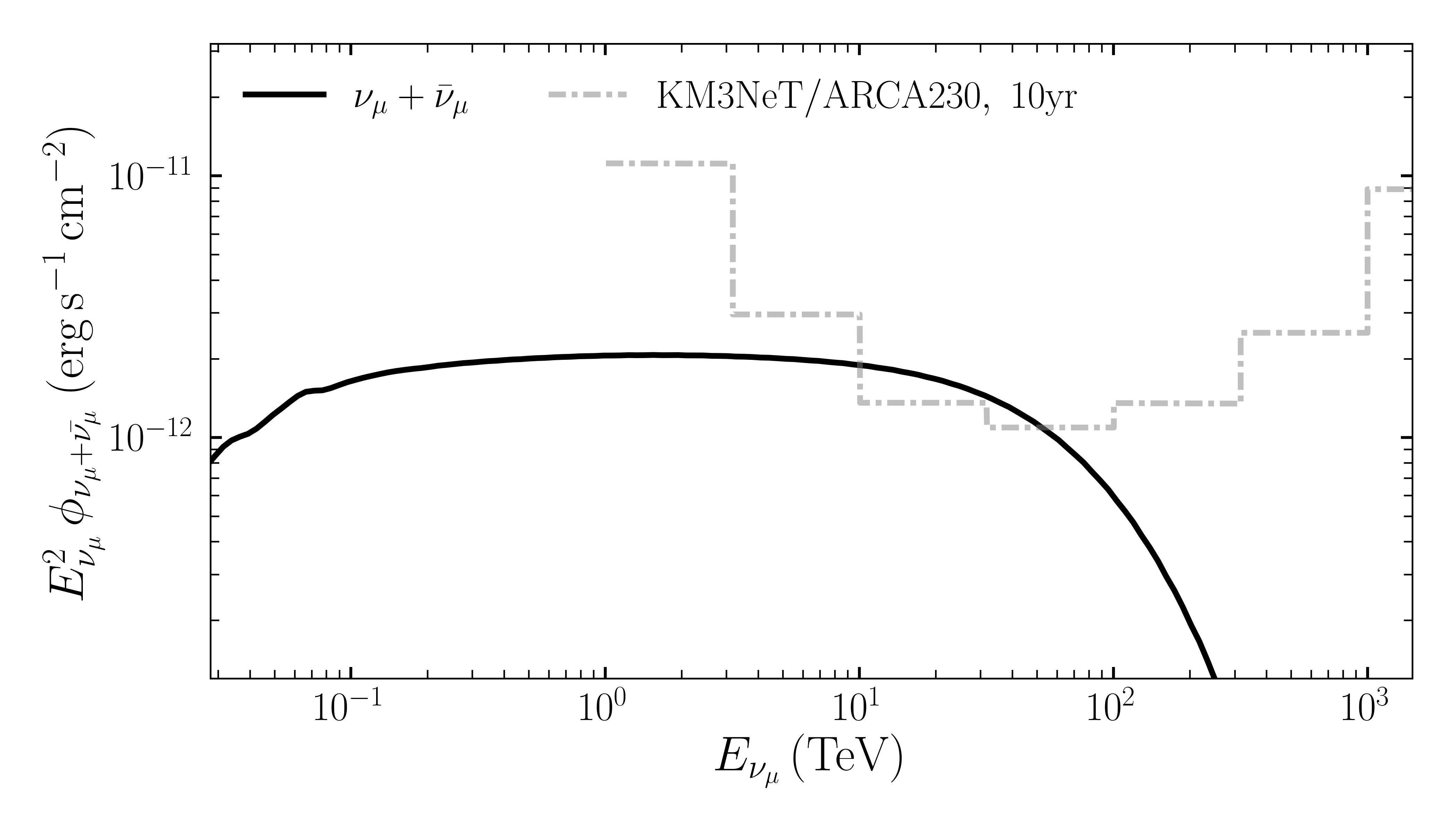}
        \includegraphics[width=0.32\textwidth]{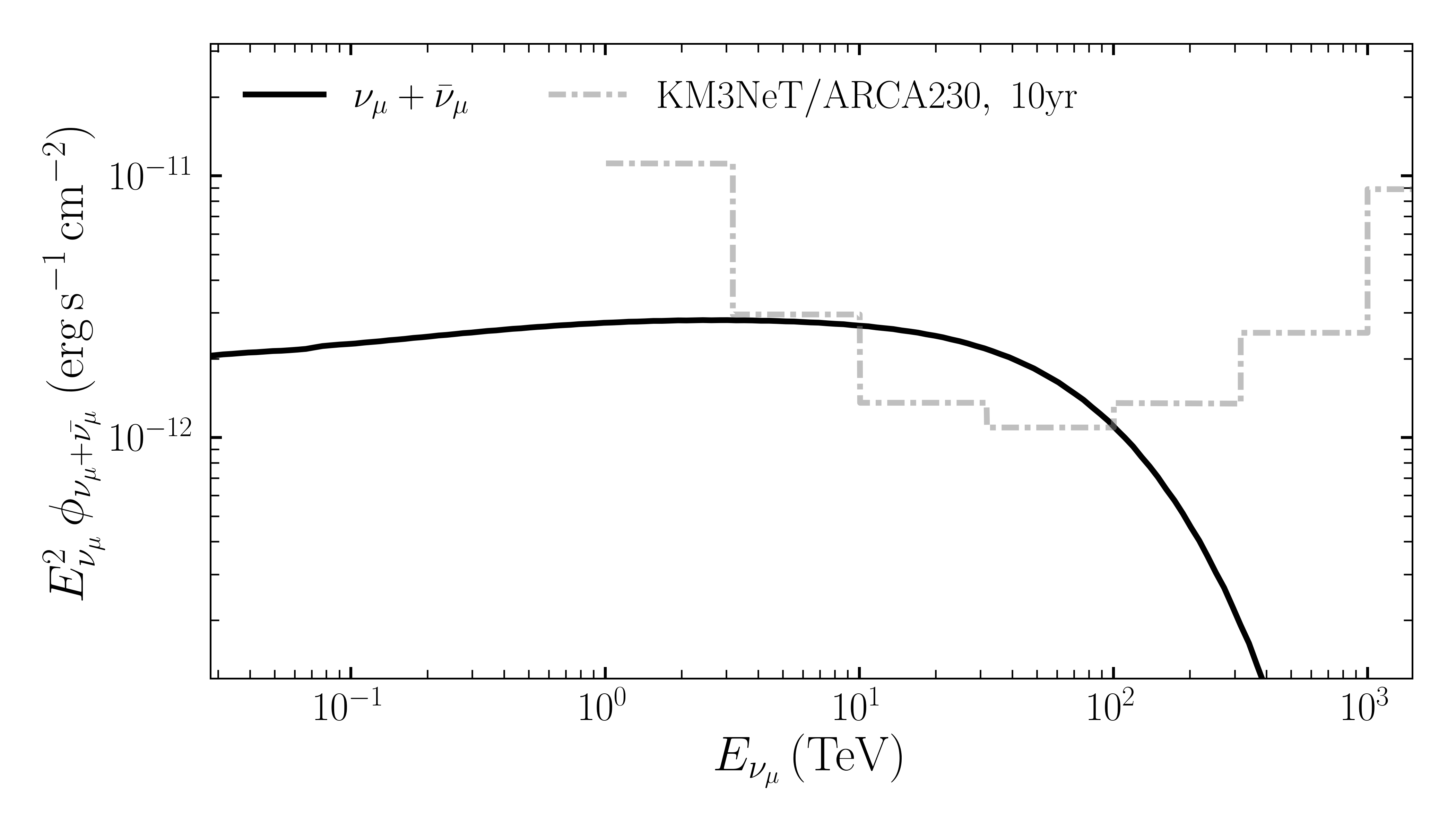}
    \caption{Muon-neutrino plus antineutrino spectra predicted for Models~B--D of \ehz.  The black curves show $E_{\nu_\mu}^2\phi_{\nu_\mu+\bar{\nu}_\mu}$, and the grey dash-dotted curves show the nominal ten-year KM3NeT/ARCA230 point-source sensitivity \citep{KM3NeT:2024arca}.  Model~E is not shown because its neutrino flux is negligible on this scale.}
    \label{fig: appendix at2019ehz neutrinos}
\end{figure*}

The corresponding $\nu_\mu+\bar{\nu}_\mu$ spectra for Models~B--D are shown in Fig.~\ref{fig: appendix at2019ehz neutrinos}. Model~B gives a comparable, and in parts higher, neutrino flux despite the absence of a corona.  In this case, the hard proton spectrum and efficient acceleration allow the source to remain neutrino bright while also becoming more transparent to escaping GeV-to-TeV photons. Models~C and D show that soft proton spectra can also approach the KM3NeT benchmark, but only under more specific conditions. In Model~C, the large value of $\gamma_{p,\rm min}$ removes low-energy protons that do not efficiently contribute to pion production and concentrates the injected power closer to the relevant energy range. In Model~D, the same goal is achieved by increasing the non-thermal proton fraction. Model~E is not shown in Fig.~\ref{fig: appendix at2019ehz neutrinos}, because its high-energy output is optimised for the GeV band and the corresponding neutrino flux is negligible on this scale.

As in the case of \dsg, the comparison with the grey ARCA230 curve should be used only as a guide.  The grey line is a ten-year steady point-source sensitivity, whereas the neutrino emission from a TDE is transient.  The final detectability of each model must therefore be evaluated through the event-count calculation using the source position, the assumed active time window, and the flavour-dependent effective area.

\subsection{\ali}
\label{Sec: ASASSN14li results}
\subsubsection{High-energy spectral energy distributions}
\label{app: Sec: ASASSN14li SEDs}
\begin{figure}[!ht]
    \centering
        \includegraphics[width=\columnwidth]{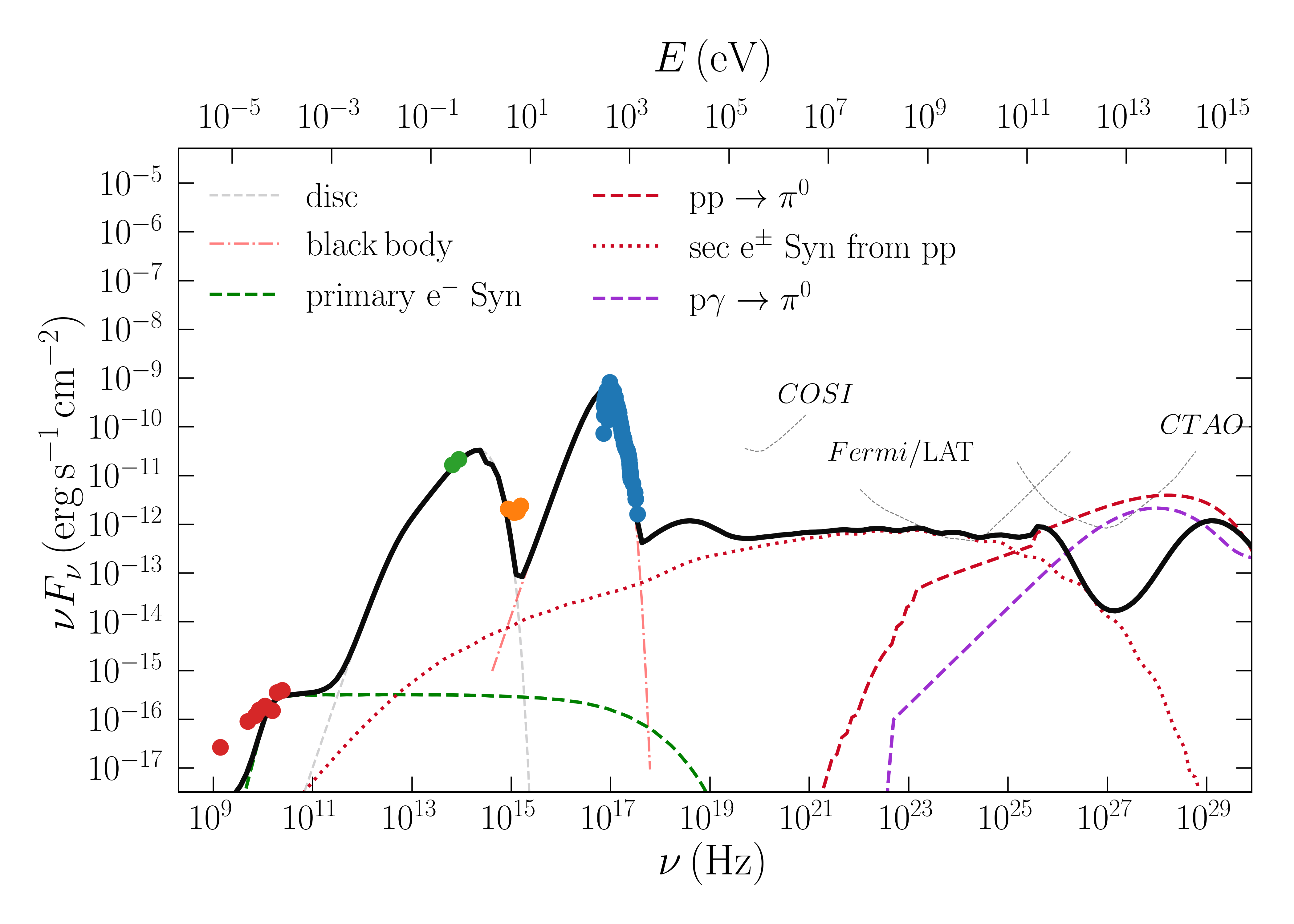}
    \vspace{0.2cm}
        \includegraphics[width=\columnwidth]{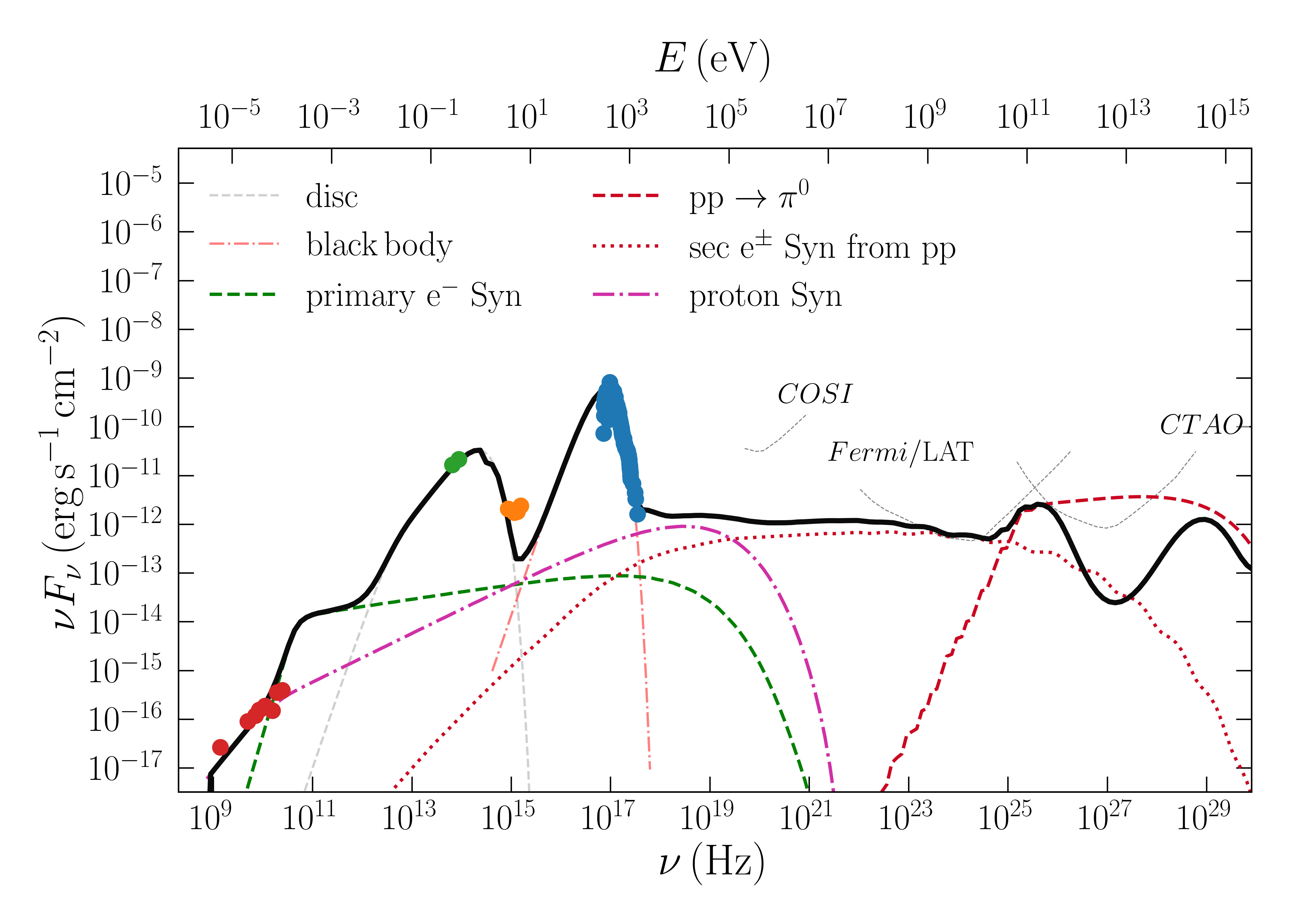}
        \includegraphics[width=\columnwidth]{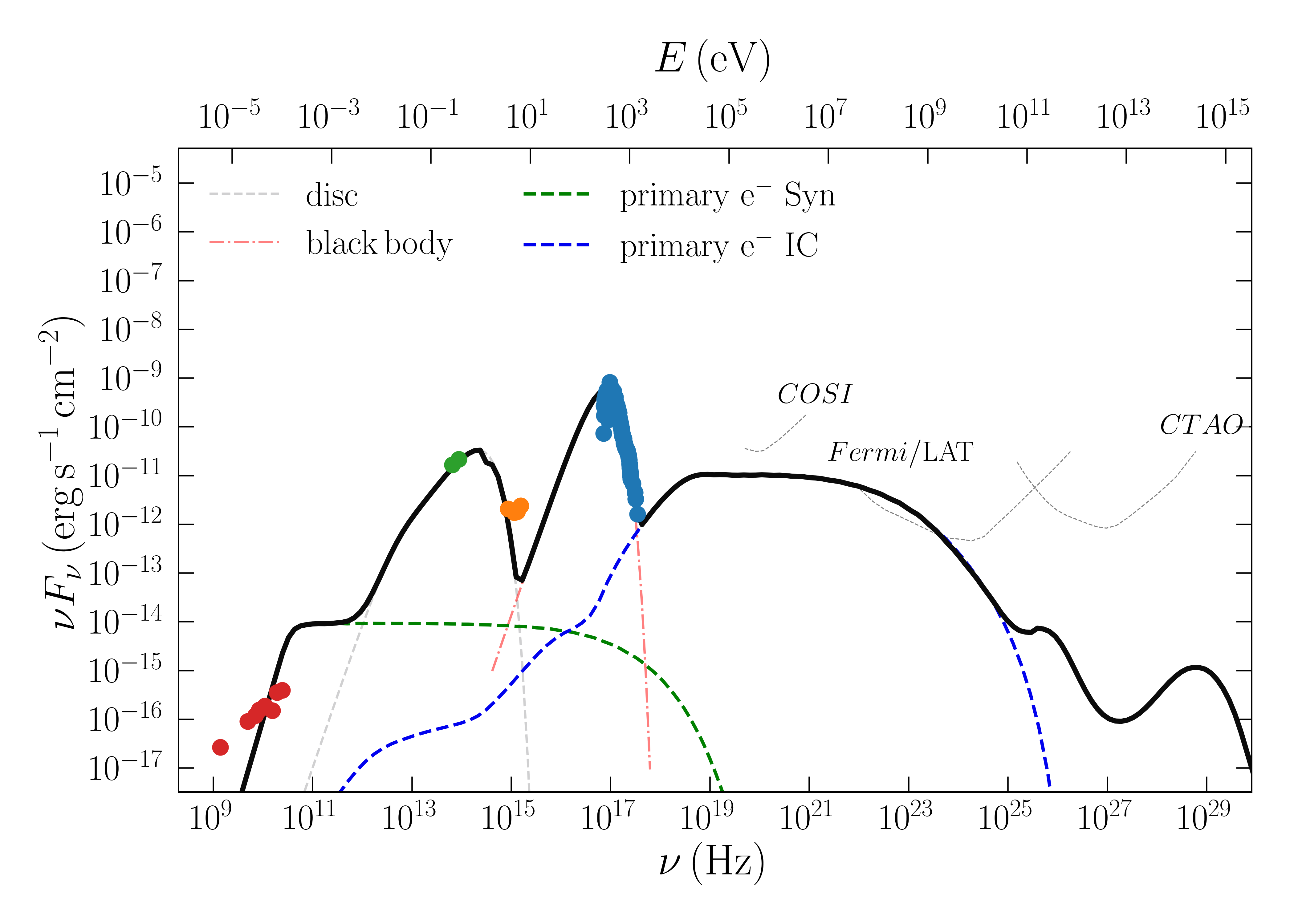}
    \caption{Spectral energy distributions of the three exploratory Models~B--D for \ali.  The thick black line shows the total model emission.  The thermal disc and black-body components are shown by grey dashed and red dash-dotted curves, respectively. Primary electron synchrotron and inverse-Compton emission are shown by green and blue dashed curves when visible.  Hadronic \gr components are shown by red and purple curves: red dashed for $pp\rightarrow\pi^0$, purple dashed for $p\gamma\rightarrow\pi^0$, and red dotted for synchrotron radiation from secondary pairs produced in $pp$ interactions. The radio data are broadly consistent with primary electron synchrotron, although additional synchrotron components, such as proton synchrotron in Model~C, may also contribute. The thin grey curves indicate the COSI \citep{Tomsick:2023aue}, Fermi/LAT, and CTAO reference sensitivities.}    
    \label{fig: appendix asassn14li seds}
\end{figure}

The SEDs of the three (B--D) \ali models are shown in Fig.~\ref{fig: appendix asassn14li seds}.  The line styles follow those used for \dsg and \ehz.  The thick black curve shows the total model spectrum. The grey dashed and red dash-dotted curves show the thermal disc and black-body components, respectively.  Primary electron synchrotron and inverse-Compton emission are shown by green and blue dashed curves when visible. Hadronic components are shown in red and purple: red dashed curves show $pp\rightarrow\pi^0$, purple dashed curves show $p\gamma\rightarrow\pi^0$, and red dotted curves show synchrotron emission from secondary pairs produced in $pp$ interactions. The thin grey curves indicate the COSI, Fermi/LAT, and CTAO reference sensitivities.

Model~B tests whether a large non-thermal proton fraction can compensate for a substantially lower jet power.  The high-energy SED is again hadronically dominated, but the total flux remains below the CTAO reference curve. Model~C explores a different extreme: the jet power and $\gamma_{p,\rm min}$ are increased, and the plasma beta is reduced to $\beta=0.1$, corresponding to a more magnetically dominated emitting region. This boosts the hadronic contribution and produces a broad GeV-to-TeV component, but even this configuration does not give a robust CTAO detection in our setup. Finally, Model~D is qualitatively different. Its high-energy output is dominated by primary electron inverse-Compton emission, while the hadronic normalization is weak. This model can therefore be relevant for Fermi/LAT searches, but it is not expected to be a neutrino-bright source.

\subsubsection{Muon-neutrino predictions}
\label{app: Sec: ASASSN14li neutrinos}
\begin{figure}[!ht]
    \centering
        \includegraphics[width=\columnwidth]{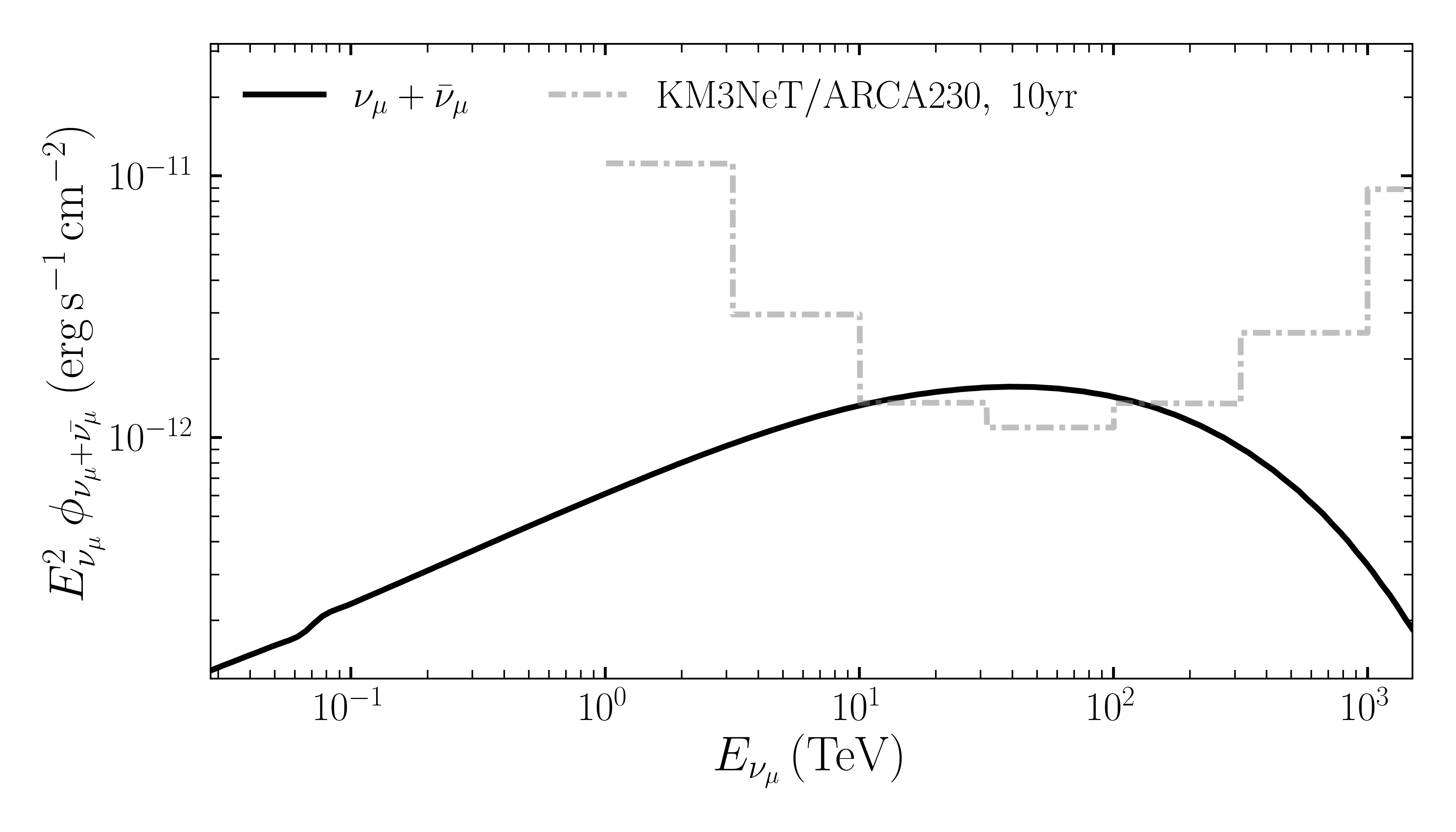}
        \includegraphics[width=\columnwidth]{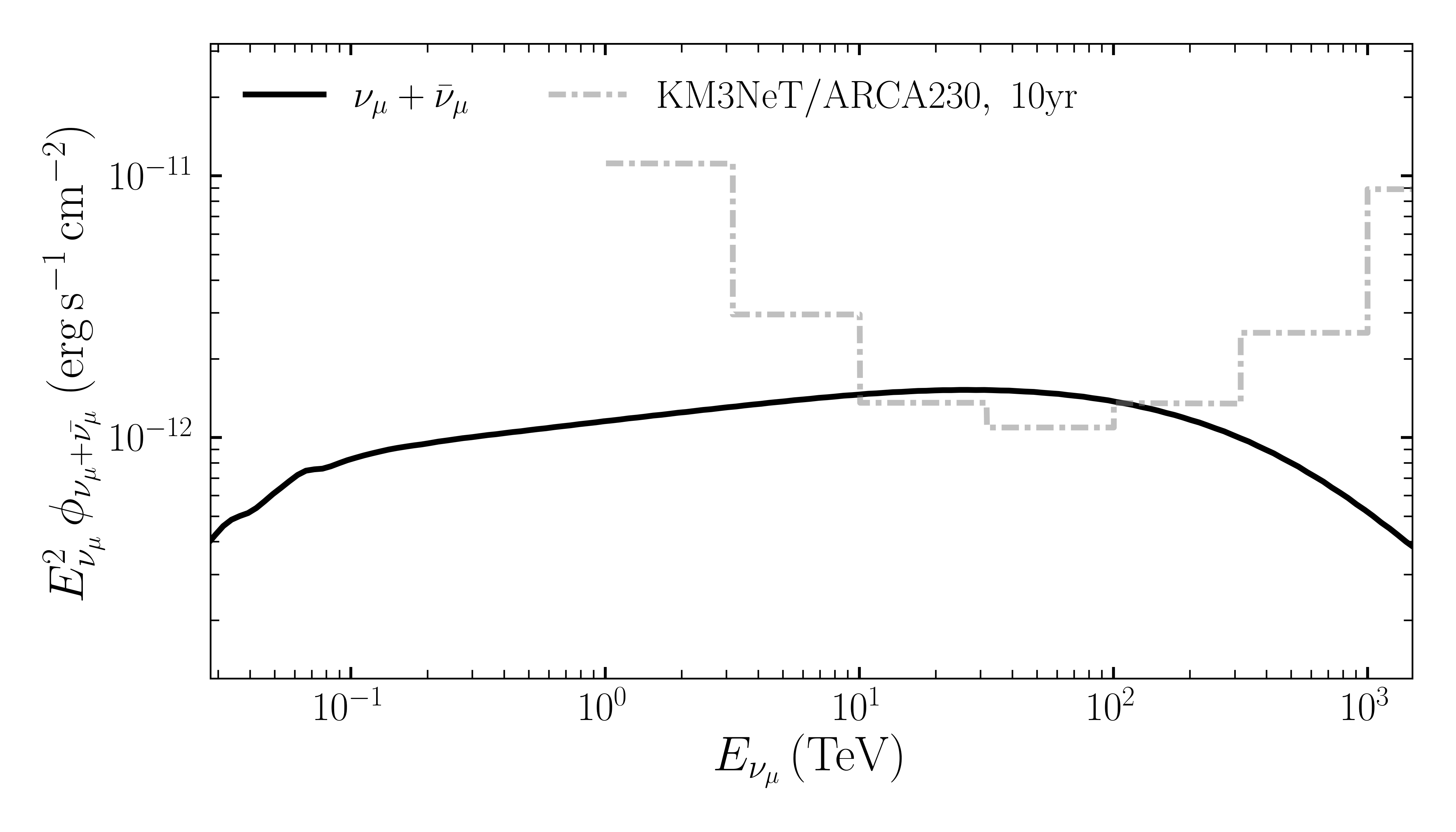}
    \caption{Muon-neutrino plus antineutrino spectra predicted for the hadronic \ali Models~B--C. The black curves show $E_{\nu_\mu}^2\phi_{\nu_\mu+\bar{\nu}_\mu}$, and the grey dash-dotted curves show the nominal ten-year KM3NeT/ARCA230 point-source sensitivity \citep{KM3NeT:2024arca}. Model~D is not shown because it is a lepton-dominated, neutrino-quiet configuration.}
    \label{fig: appendix asassn14li neutrinos}
\end{figure}

The corresponding $\nu_\mu+\bar{\nu}_\mu$ spectra for the hadronic Models~B--C are shown in Fig.~\ref{fig: appendix asassn14li neutrinos}. Model~B gives a comparable neutrino flux even though its jet power is lower, because a large fraction of the non-thermal proton power is placed into the power-law component. Model~C shows that a softer proton spectrum can also approach the KM3NeT/ARCA230 benchmark if the hadronic minimum Lorentz factor is made very large and the magnetic energy density is increased. The lepton-dominated Model~D is not shown in Fig.~\ref{fig: appendix asassn14li neutrinos}. Its weak hadronic normalization makes the predicted neutrino flux negligible on the scale of the plotted models. The \ali sequence therefore provides a useful contrast with \dsg and \ehz: in the parameter space explored here, neutrino-bright configurations do not automatically imply CTAO-bright gamma-ray emission. Instead, the detectable high-energy signature is more likely to appear in the MeV--GeV band and in neutrinos, while TeV gamma-ray detection remains challenging.

Photon-photon absorption is expected whenever the escaping high-energy photons propagate through a sufficiently dense soft-photon field. The threshold condition for pair production is \citep{Gould:1967zza,Murase:2020lnu}
\begin{equation}
E_\gamma \epsilon (1-\cos\theta) \geq 2(m_{\rm e}c^2)^2 ,
\end{equation}
for target photons of energy $\epsilon$, and, for head-on interactions, gives
\begin{equation}
E_{\gamma,\rm thr} \simeq
260\left(\frac{\epsilon}{1\,{\rm keV}}\right)^{-1}\,{\rm MeV}.
\end{equation}
Thus, the bright optical/UV and X-ray radiation fields in \ali can absorb photons from the GeV range upward, and TeV photons are well above the threshold for interactions with the X-ray field. The absorbed power is then redistributed through the electromagnetic cascade, helping to explain why the models can be relevant for COSI/Fermi-LAT and KM3NeT while remaining below the CTAO benchmark.

\end{appendix}

\end{document}